\documentclass[12pt,preprint]{aastex}

\shortauthors{Westfall {\it et al.}}
\shorttitle{Extended Structure of Sculptor}

\begin{document}

\title{Exploring Halo Substructure with Giant Stars VIII:\\ The Extended
  Structure of the Sculptor Dwarf Spheroidal Galaxy} 

\author{Kyle B. Westfall\altaffilmark{1,2}, Steven R.
  Majewski\altaffilmark{1,3,4}, James C. Ostheimer\altaffilmark{1,5}, Peter
  M. Frinchaboy\altaffilmark{1,3}, William E. Kunkel\altaffilmark{3,6},
  Richard J. Patterson\altaffilmark{1}, and Robert Link\altaffilmark{1,7}}

\altaffiltext{1}{University of Virginia, Department of Astronomy,
  Charlottesville, VA 22903-0818}
\altaffiltext{2}{Department of Astronomy, University of
  Wisconsin-Madison, 475 N. Charter St., Madison, WI 53706} 
\altaffiltext{3}{Visiting Astronomer, Cerro Tololo Inter-American
  Observatory, National Optical Astronomy Observatory, which is operated
  by the Association of Universities for Research in Astronomy, Inc., under
  cooperative agreement with the National Science Foundation. }
\altaffiltext{4}{Visiting Associate, The Observatories of the Carnegie
Institution of Washington, 813 Santa Barbara Street, Pasadena, CA 91101.}
\altaffiltext{5}{Present address: 1810 Kalorama Rd., NW, \#A3, Washington,
  D.C. 20009.}
\altaffiltext{6}{Las Campanas Observatory, Carnegie Institution of
  Washington, Casilla 601, La Serena, Chile.}
\altaffiltext{7}{Present address: Northrop Grumman Information Technology
- TASC, 4801 Stonecroft Blvd., Chantilly, VA 20151.}

\email{westfall@astro.wisc.edu, srm4n@virginia.edu,
  jco9w@alumni.virginia.edu, pmf8b@virginia.edu, kunkel@jeito.lco.cl,
  ricky@virginia.edu, robert.link@ngc.com}

\begin{abstract}

  We explore the spatial distribution of stars in the Sculptor dwarf
  spheroidal (dSph) galaxy over an area of 7.82 deg$^2$, including coverage
  of the central region but extending mostly south and east of the dSph
  core.  Two methods are used to identify stars that are most likely
  associated with the dSph, and these filtered samples of stars are used
  to map its spatial structure.  First, following the method of 
  previous contributions in this series, we utilize Washington $M,
  T_2+DDO51$ photometry to identify red giant branch (RGB) star candidates
  with approximately the same distance and metallicity as the Sculptor
  dSph.  Second, a prominent blue horizontal branch (BHB) population
  provides a fairly populous and pure sample of Sculptor stars having
  broadband colors unlike the bulk of the Galactic field star population.
  A spectroscopically observed subset of Sculptor candidate stars (147
  total stars: $\sim5\%$ of all Sculptor candidates, $\sim10\%$ of Sculptor
  giant candidates)  yields a systemic heliocentric velocity for the system
  of $v_{\rm hel}=110.43\pm0.79\,{\rm km\,s^{-1}}$, in good agreement with
  previous studies.  We also find a global velocity dispersion of
  $\sigma_v=8.8\pm0.6\,{\rm km\,s^{-1}}$ with slight indications of a 
  rise in the velocity dispersion past $\sim0.4r_{\rm lim}$.
  These spectra also provide a check on the reliability of our
  candidate Sculptor giant sample to $M \sim 19$: 94\% of the photometrically-selected
  Sculptor giant star candidates with follow-up spectroscopy are found to
  be kinematically associated with Sculptor, while four out of ten stars outside
  of our Sculptor giant star selection criteria that we tested
  spectroscopically appear to be velocity members of Sculptor.  These
  percentages are in agreement with results for an additional 22 Sculptor
  field stars with radial velocities in the literature.  All available
  velocities show that our methodology for picking Sculptor giants is both
  reliable and conservative.  Thus, these giant star samples should
  provide a reliable means to explore the structure of the Sculptor dSph.
  Nevertheless, considerable care has been taken to assess the level of
  background contamination in our photometric sample to ensure an
  accurately derived density profile of the Sculptor dSph to large radii.
  Multiple background assessments verify that we detect a considerable
  stellar density of Sculptor stars to the limits of our main survey area
  for both the RGB and BHB candidate samples.  While we find that a King
  profile of limiting radius $r_{\rm lim} = 79\farcm6$ fits the density
  profile of Sculptor well to $\sim60\arcmin$, beyond this, we identify a
  ``break'' in the profile and a clearly detected population of Sculptor
  stars following a $\Sigma\propto r^{-2}$ decline to more than
  2$r_{\rm lim}$.  This break population must signify either the presence
  of an extremely broad distribution of bound ``halo stars'' around the
  Sculptor dSph, or the presence of unbound tidal debris.  If the latter
  is true, we determine a fractional mass-loss rate of approximately
  $0.042\,{\rm Gyr^{-1}}$ for the Sculptor dSph.  Additional support for
  the notion that there is tidal disruption comes from the two-dimensional
  distribution of our Sculptor candidate stars: Both the RGB and BHB
  samples show increasingly elongated isodensity contours with radius that
  point to an apparent stretching reminiscent of what is seen in
  models of disrupting satellite galaxies.  Finally, we find that RGB stars
  that are more likely to be metal-poor (based on their color and
  magnitude) are significantly less centrally concentrated and therefore
  constitute the primary contributing stellar population to the likely
  tidally-stripped parts of the dSph.

\end{abstract}


\section{INTRODUCTION}

Dwarf spheroidal (dSph) galaxies were immediately recognized as mysterious
stellar systems upon Shapley's discovery of the first example of this
galaxy class, ``a stellar system of a new type'', in the constellation
Sculptor \citep{s38a,s38b,s39}.  As \citet{s38b} pointed out after his
discovery of a second example in the Fornax constellation soon afterward,
the dSph galaxies ``have some properties in common with globular clusters,
others with spheroidal galaxies, and still others (nearness and complete
resolution into stars) with the Magellanic Clouds.''  The uniqueness of
these systems was further born out by Thackeray's study of Sculptor's
variable star population, which indicated a ``physical dissimilarity''
between this new stellar system and a globular cluster \citep{thackeray50}.
However, to Shapley the density profile was the most remarkable aspect of
these diffuse, low-surface-brightness ``clusters''; indeed, he points out
that had not the discovery plate been abnormally sensitive and taken during
sky conditions that were also above average, ``the Sculptor cluster would
not have been found.''

It is fair to say that the density distribution (stellar number and mass
density) of dSph galaxies remains one of the most perplexing aspects of
these systems.  When the radial light profiles are converted to a stellar
mass density and compared to that suggested by the internal dynamics of
stars in these systems, very large dark matter contents are inferred
\citep[][see \citealt{mateo98} for a summary]{fl83, a83, har94, IH95}.
From these measurements, the very low mass dSph galaxies make up
the high end of the current mass-to-light ($M/L$) scale for distinct
stellar systems, with members having total $M/L$'s approaching 100
\citep[cf.][]{mateo98}, or even exceeding that value by almost an order of
magnitude \citep{kleyna01,kleyna02}.  Even on the low end of this scale,
with an apparent total $M/L \sim 10$ \citep{ad86, QDP, IH95}, the Sculptor
dSph is extraordinarily dominated by dark matter for such a low-mass
stellar system ($\sim10^{7}\,M_{\sun}$).

The physical extents of dSph galaxies also remain uncertain, but have
become a fulcrum on which recent debate regarding the true dark matter
content of dSph's lie.  The question is not only whether all of the light
has been fully accounted for \citep[e.g.,][]{oden01,pIV} but whether the
increasingly large extensions of the radial profiles found by various
observers on well-studied dSph examples are clues to something untoward in
our basic understanding of the true dark matter content/distribution
within, and the dynamical state of, these systems
\citep[e.g.,][]{kuhn93,kroupa97,gf98,pIX,Mu05b,Ma05b}.

The situation for the Sculptor dSph is more or less representative of the
history of attempting to understand the dSph galaxy class as a whole.  On
shallow images including only the top two magnitudes of its luminosity
function, \citet{s38a} originally showed the radius of the Sculptor system
to be at least $40\arcmin$, but with evidence for stars extending to as
much as a degree from the center.  Extending Shapley's starcount work,
\citet[][hereafter \citeauthor*{H61}]{H61} conducted a 2.2 deg$^2$ survey
using more plates of slightly shallower depth.  He determined a limiting
radius of $46\arcmin\pm3\arcmin$ for Sculptor, which was close to Shapley's
lower limit.  More importantly, \citeauthor*{H61} found a ``definite
limiting radius'' where the density of stars in Sculptor apparently reaches
zero.  Under the assumption of $M/L \sim 2$, this radius was consistent
with expectations for a system that is tidally limited by the Milky Way
(MW).  Unfortunately, density profiles that drop sharply to zero density
can be artificially created by overestimates of a subtracted background
level, and it would seem that Hodge's survey area was too limited to
contain true ``background'' regions free of Sculptor stars (as we show
later in this paper).

Subsequent studies of the distribution of Sculptor's substantial RR Lyrae
population found it to be much more extended than the quoted
\citeauthor*{H61} tidal limit of the galaxy: \citet{va73} states,
``Variable stars have been traced in the Sculptor system out to distances
of well over $60\arcmin$.'' Later \citet{ip79} fit a \citet{king62} model
to the distribution of RR Lyrae stars and found a King limiting radius
($r_{\rm lim}$) of $47\farcm6$, in good agreement with the
\citeauthor*{H61} measurement.  However, \citet{ip79} also found that
likely 10\% of Sculptor RR Lyraes are located {\it outside} this limiting
radius (which they interpret as a tidal radius) and that they extend to
$3\arcdeg$ from the center of the galaxy.  The dynamical nature of
individual stars (such as the above variable stars) or statistical stellar
overdensities beyond the ``limiting radii'' of dSphs is quite contentious,
but as a group these stars are often termed ``extratidal'', as was done by
\citet{ip79}.  While this is an often-adopted shorthand, note that (1) King
models may not necessarily apply to dSph galaxies, which have longer
relaxation timescales than the globular clusters for which the models were
intended, and, therefore, (2) some models of dwarf spheroidal structure may
include {\it bound} ``extratidal'' populations beyond a King profile.  To
allay confusion between the observed spatial position and the actual
dynamical state of stars found beyond the King limiting radius, we limit
use of the expression ``extratidal'' here, preferring instead the term
``break population'', which refers to a change in the slope of the radial
density law that these stars create.

The results from \citet{ip79} led others to reevaluate the spatial
distribution of the Sculptor dSph.  \citet{DKK} obtained photographic data
$\sim4$ magnitudes deeper than that presented by \citeauthor*{H61} and
found a limiting radius between $75\arcmin$ and $120\arcmin$; however,
\citeauthor*{DKK} note that this value is highly dependent on the adopted
background.  They could obtain the \citeauthor*{H61} limiting radius after
subtracting an excessively large background level (an effect no doubt
confounding the original \citeauthor*{H61} study); but, they could also
obtain a limiting radius that was ``essentially infinite'' by adopting
their lowest background estimate.  In a study that included multiple UK
Schmidt plates that reached much greater angular distances from the center
of Sculptor and $\sim2$ magnitudes deeper than the \citeauthor*{H61} data,
\citet{e88b} noted a similar sensitivity of his results to the adopted
background level.  In order to force the King model to fit all points to
the extent of his survey, the adopted background had to be 3$\sigma$ higher
than the calculated background, which forced {\it all} points with
$r>60\arcmin$ below the background level.  This study ultimately identified
a best-fit King model with $r_{\rm lim}=95\arcmin$ but noted the inability
of this fitted profile to fully describe the radial distribution of
Sculptor.  \citet[][hereafter \citeauthor*{IH95}]{IH95} used one UK Schmidt
plate of similar depth to those of \citet{e88b} and found
$r_{\rm lim}=76\farcm5$.  CCD data going approximately one magnitude deeper
but over slightly less than half the area of \citeauthor*{IH95} was
recently used by \citet[][hereafter \citeauthor*{W03}]{W03} who determined,
however, $r_{\rm lim}=44\arcmin$ using the theoretical \citet{king66}
profile.

In three of the four most recent studies, not only has the derived King
profile limiting radius grown, but the density of the presumed Sculptor
stars near and beyond this radius has been found to be significantly
($\gtrsim2\sigma$) in excess of densities predicted by the King model.  In
fact, the starcounts from \citet{e88b} show the overdense, ``extratidal''
population represents roughly 13\% of the whole of Sculptor, in agreement
with measurements from the RR Lyrae distribution \citep{ip79}.  Moreover,
\citeauthor*{W03} have claimed not only the existence of a break population
but also a hint of extratidal arms extending from Sculptor to the
northwest and southeast.  We address this and other previous results in
comparison with our own in more detail in \S5 and \S6.

An overdensity at large radii is not exclusive to the Sculptor dSph.  In
their study, which included eight of the now ten known local dSphs,
\citeauthor*{IH95} comment, ``There is a noticeable tendency for most of
the dSphs to show an excess of stars, with respect to the best-fitting King
model, at large radii.''  \citeauthor*{IH95} state that this excess density
is not likely caused by incorrect background determinations.  As previously
discovered by \citet{DKK} and \citet{e88b}, \citeauthor*{IH95} find an
unreasonably large background estimate would have to be adopted to
eliminate the excess density {\it in the case of every dSph}.
Subsequently, break populations were reported in other studies for Carina
\citep{ksh96,pII,pVI}, Ursa Minor \citep{kk00,md01,pIV}, Leo I
\citep{pX}, Draco \citep{wilk04}, and possibly Sextans \citep{GGRF92}.
Of course, the Sagittarius dSph (Sgr) \citep{igi95} is the clearest
indicator
that at least one satellite galaxy of the MW is experiencing
tidally-induced mass loss, which contributes a break population to its
radial profile.  This dSph has now been shown to have extratidal debris
wrapping more than $360\arcdeg$ around the Galaxy
\citep[e.g.,][]{ibata01,sgr1}.

Unfortunately, the existence or cause of the break populations in other
dSphs are not as clear as in the Sgr system.  For example, whether
a break population exists around the Draco dSph has been controversial and
highlights the importance of careful studies in the low-density regimes of
the radial density profile.  \citet{skh97}, \citet{kk00}, \citet{pp01}, and
\citet{wilk04} all report evidence for stars beyond the nominal Draco
limiting radius.  However, using Sloan Digital Sky Survey data,
\citet{oden01} find no tidal extensions and no evidence of a break in the
radial density profile, a result also found by \citet{pp02} \citep[who
  apparently reverse the previous conclusions of][]{pp01}.  In addition,
both \citet{oden01} and \citet{acm01} find that the radial distribution of
Draco is better fit by an exponential profile with no real limiting radius
as prescribed by the \citet{king62} profile, whereas a change in the slope
(from a ``break population") of the Draco density profile is obvious in
the data shown by \citet{wilk04}.  Similar controversy has been raised in
the case of the Carina dSph, for which \citeauthor*{IH95}, \citet*{ksh96}
and \citet{pII} all claim a break population but for which
\citeauthor*{W03} find no break population and \citet{morrison01} question
the \citet{pII} result.  A thorough reanalysis of the debate surrounding
the existence of a break population in Carina is given in
\citet[][hereafter \citeauthor*{pVI}]{pVI} and further proof that a Carina
break population exists is given in \citet[][hereafter
  \citeauthor*{Mu05b}]{Mu05b}.
 
These case histories suggest that the state of the dSph observational
record remains unsettled (or, at least, not universally accepted), and
this empirical ambiguity forestalls any meaningful (or, at least, complete)
interpretation of the physical state of these systems.  As discussed above,
 the observational difficulty lies in a proper accounting for the
low-density dSph regions against the usually overwhelming contamination by
foreground field stars and background galaxies.  As we have seen, the
physical extent of the Sculptor dSph has apparently grown with deeper
and/or more accurate studies conducted over larger areas.  To improve the
contrast of the true dSph members over the contaminating background
(hereafter the ``signal-to-background'' ratio or $S/B$), survey approaches
have progressed from the early simple star count analyses to searches in
selectively tuned subregions of color-magnitude space
\citep*[e.g.,][]{ksh96,gfiq95,oden01,rock02,pp02}.  For the dSphs, such
studies rely on deep photometry reaching to dSph-rich parts of the CMD,
such as the main sequence turn-off (MSTO), to boost the contributed dSph
signal.  However, to employ such a technique in the case of an expansive
and distant system like Sculptor, which in some previous estimates has an
$r_{\rm lim}$ as large as $1\fdg5$ and which has a MSTO near $V \sim 23.5$,
one anticipates needing a substantial amount of imaging (say $\sim16$
deg$^2$) on 4-m class telescopes or larger to confidently reach at least
twice the King limiting radius in all directions.  Even attempting such an
experiment on only a fraction of this area with the currently largest
mosaic CCD cameras is a daunting prospect.

In this series of papers we have taken an
alternative approach to high $S/B$ mapping of low-density, extended regions
of Galactic satellites by concentrating instead on {\it reducing the
background} in the problem.  Our technique employs more easily obtainable,
{\it relatively shallow} imaging in the Washington $M, T_2 + DDO51$ filter
system to isolate (typically low-metallicity) giant stars associated with
the dSph from the primary field contaminant, foreground dwarf stars from
the more metal-rich disk.  This methodology goes a long way toward the limit of zero
background, so that we are left with relatively pure samples of {\it bona fide}
dSph-associated giant stars.  A benefit of this approach is that we
identify the very stars needed for practical follow-up, spectroscopic
analysis.  So far in this series we have presented the methodology
\citep[][hereafter \citeauthor*{pI}]{pI} and applied it to determine the
distributions of giant stars in the Carina, Ursa Minor, Draco and Leo I dSphs
\citep[][hereafter \citeauthor*{pII}, \citeauthor*{pIV}, \citeauthor*{pIX} and
  \citeauthor*{pX}]{pII,pIV,pIX,pX}.  This technique has also been used
to study the M31 dSph galaxies And I, And II and And III \citep{ost02}.  A
preliminary report on our survey of Sculptor has been given in
\citet{westfall00}.  We have given a report of preliminary results on these
and other dSphs we are currently investigating, including Sculptor, in
\citet{m02} and \citet{m03}.

Since those summary reports and \citet{westfall00}, we have extended the
radial coverage of our Washington$+DDO51$ survey around Sculptor, improved
the critical assessment of the residual background noise, and included a
parallel analysis using Sculptor's horizontal branch stars.  We report here
evidence for an extended, King profile break population of stars
around the Sculptor dSph out to at least two limiting radii.
These results are found with independent analyses of both
blue horizontal branch (BHB) and red giant branch (RGB) star tracers.
These photometric mappings are backed up with spectroscopic verification of
radial velocity membership for a small subsample of selected Sculptor star
candidates (\S4).  Our findings here strengthen the claim that at least
some Galactic dSphs may be experiencing non-negligible stellar mass loss, most
likely tidally induced.  Thus, they provide important constraints
applicable to the ultimate goals of our overall, long-term project, which
are to (1) understand satellite disruption within the context of the
hierarchical build-up of the Galactic halo as predicted by Cold Dark Matter
models \citep[e.g.,][]{ps74,kw93,lc93,nfw96,nfw97,bkw01}, and explored in
simulations of halo substructure
\citep*[e.g.,][]{johnston98,bkw01,harding01,jsh02,hayashi03}, and (2)
determine the dynamical conditions of dSph galaxies within the context of
their inferred large $M/L$ values
\citep[e.g.,][; \citeauthor*{Mu05b}]{kuhn93,oh95,pp95,kroupa97,gf98}.

We present the photometric data and subsequent reduction methods in \S2.
In \S3, we describe our selection criteria and its application to our
photometric survey.  Spectroscopic measurements are presented in \S4 that
provide both a limited exploration of the dynamics of the dSph and an
estimate of the accuracy of our photometric selection technique.  Spatial
analyses of our photometrically-selected Sculptor candidate samples are
given in \S5 and we conclude with some discussion of our results in \S6.

\section{PHOTOMETRY}
        
A mosaic of the Sculptor dSph was created via CCD imaging undertaken over
the course of six observing runs at the Swope 1-m telescope located at the
Las Campanas Observatory.  Table \ref{tab:obs} lists the UT dates of these
observations, average seeing estimates, lunar illumination fraction, the
total number of Sculptor fields observed ($N_{\rm fields}$), and the number
of photometric fields ($N_{\rm phot}$).  Each individual pointing in the
survey area was observed once in the Washington $M$ and $T_2$ filters and
twice in the $DDO51$ filter with nominal exposure times of 240, 240, and
840 seconds, respectively.  Observations before the year 2001 were taken
with the $2048\times2048$ SITe \#1 CCD chip while those made after this
were taken with the $2048\times4096$ SITe \#3 CCD chip.  The SITe \#3 chip
has a hot pixel at $(x,y)\approx(808,3158)$ and, as a consequence, is
mounted in the dewar so that the good three quarters of the chip is
centered, while the poorer fourth quarter is somewhat vignetted.  In order
to maximize sky coverage, we have included the full, unvignetted portion of
the SITe \#3 chip frames in our reduction, which happens to include the hot
pixel\footnote{Due to its use in the latter stages of our observing
  program, fields completed with the SITe \#3 chip are primarily at large
  distances from the dSph core or are used to provide photometric anchors
  for calibration.  Inclusion of the hot pixel area will, {\it in this
    region of the CCD field}: (1) increase the photometric error of the
  objects detected, (2) hinder the detection of fainter sources, which
  effectively forces a brighter magnitude limit, and (3) cause an
  {\it underestimation} of the stellar density.  All of these effects are
  largely mitigated by applied limits in allowable photometric error
  (discussed later) and, in the end, only a small portion ($\lesssim5$\%)
  of the total survey area is affected.  In fact, the surface density
  profile is more affected by having unobserved regions in the elliptical
  annuli at large radii than by including the area affected by the hot
  pixel.}.  The total sky coverage is approximately $0.16$ deg$^2$
\citep[$0\farcs697$ per pixel,][]{cr91} for the SITe \#1 chip and $0.10$
deg$^2$ ($0\farcs435$ per pixel) for the SITe \#3 chip.

Individual pointings were arranged in a grid pattern with $\sim3\arcmin$
overlaps in order to get contiguous coverage of the Sculptor field yet
allow for calibration of non-photometric frames by overlapping, adjacent
photometric ones in a boot-strap process (see below).  With a previously
determined limiting radius of $r_{\rm lim}=76\farcm5$ (\citeauthor*{IH95}),
coverage of the entire galaxy to significantly large radii (e.g.,
$1.5r_{\rm lim}$ -- $2r_{\rm lim}$) would be very ambitious when one is
limited to 0.1 -- 0.16 deg$^2$ per pointing.  Therefore, to explore the
nature of the Sculptor radial profile to large radii, our survey strategy
focused on covering an area of approximately 1.28 deg$^2$ at the dSph
center, but with the majority of attention extended to $\sim 2r_{\rm lim}$
toward the east and south.  Even this ``lop-sided'' coverage over 7.82
deg$^2$ required nearly 100 separate pointings --- 59 from the SITe \#1
chip and 37 from the SITe \#3 chip --- and nearly 400 individual CCD
frames.  The survey includes four background ``control'' pointings taken
$5\arcdeg$ from the center of the galaxy in each cardinal direction.  These
control fields provide important tests of the stellar background near the
Sculptor dSph (\S5.1).

\begin{figure}
\begin{center}
\plotone{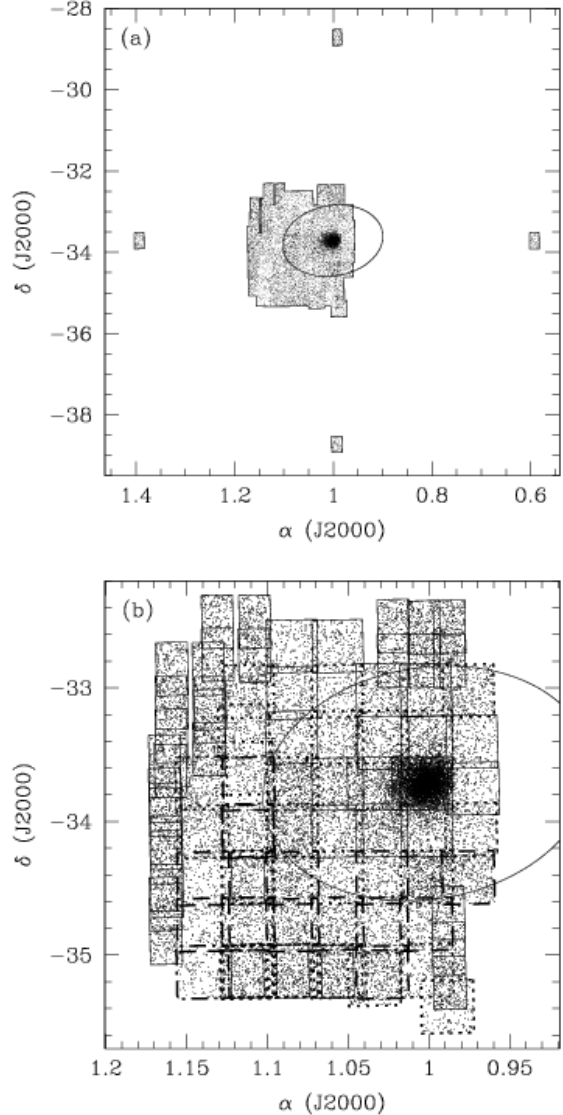} 
\caption{Spatial distribution of field boundaries and all photometered
  objects over (a) the entire survey area and (b) the contiguous inner
  region.  The ellipse centered on the core concentration of the dSph is
  the King limiting radius, $r_{\rm lim}=76\farcm5$, derived by
  \citeauthor*{IH95}.  The survey area extends to $\sim2r_{\rm lim}$ from
  the center to the south and east. In (a), only the outer boundary of the
  inner contiguous fields is shown.  In (b), individual pointing boundaries
  are shown with those taken under photometric conditions (54 out of 96)
  using {\it solid} boundaries, and fields requiring one (27), two (12),
  and three (3) boot-strapping steps from a photometric frame (explained in
  text) using {\it dotted}, {\it dashed}, and {\it dot-dashed} boundaries,
  respectively.  The varying depth of the fields (see Figure
  \ref{fig:histmaglim}) is due to differences in observing conditions.}
\label{fig:uncutfield}
\end{center}
\end{figure}

Figure \ref{fig:uncutfield} shows the distribution of all photometered
objects in both the full survey area including the four control pointings
and the main part of the survey.  Some pointings were later found not to
be at their nominal positions due to pointing errors at the telescope.
Subsequent maps in this paper will only show the inner regions though all
analyses were performed on both the main survey and the control fields.

Basic image reductions used the {\ttfamily imred.ccdred} package in
IRAF\footnote{IRAF is distributed by the National Optical Astronomy
  Observatories, which are operated by the Association of Universities for
  Research in Astronomy, Inc., under cooperative agreement with the
  National Science Foundation.}.  Most nightly observations contained bias
frames, dome flats, and sky flats for each of the three filters.  Those
nights without bias frames or dome flats were reduced using the most
appropriate calibration frames from the nearest adjacent night.  Those
nights with sky flats and/or uncrowded object images with high backgrounds
were combined and used to create illumination correction images.  Final
frames used for photometry had large-scale flux variations of at most
$\epsilon_{f}\lesssim2\%$, with the vast majority having $\epsilon_{f}
\lesssim1\%$, such that flat-fielding errors contribute a magnitude error
of $\epsilon_{m}\lesssim0.01$.

Point spread function(PSF)-fitted photometry of the object frames was done
with the stand-alone version of DAOPHOT II \citep{stetson92}.  Magnitudes,
magnitude errors, fitting errors (DAOPHOT parameters $\chi$ and sharp), and
$xy$-pixel coordinates were determined for each object by the stand-alone
task ALLSTAR using the best-fit PSF (determined from $\gtrsim20$ stars).
Frames from a single pointing were matched using the task DAOMASTER.  The
now single set of $xy$-pixel coordinates were converted to celestial
coordinates (J2000.0) using a reference list of stars from the USNO-A2.0
\citep{monet98} catalog imported into the IRAF task {\ttfamily tfinder}.

The Washington$+DDO51$ standard fields SA98, SA110, SA114, and NGC3680
\citep{geisler90,geisler96} were observed interspersed with object frames
during photometric nights.  Aperture corrections were obtained using the
program DAOGROW \citep{stetson90}.  Photometric transformations including
airmass, color (if necessary), and nightly zero-point terms were determined
as described in \citet{mkkb94}, utilizing the matrix inversion algorithm
of \citet{harris81} incorporated into a local code (see also discussion in
\citeauthor*{pIV}).  Five of the six observing runs had at least one night
for which photometric transformation coefficients could be derived.

All observed pointings were locked onto a single magnitude system in the
following way.  For photometric frames, the measured magnitude of each
individual star was calibrated using the nightly photometric transformation
coefficients as derived from our standard stars.  Stars observed in
multiple photometric frames were used to determine residual relative
magnitude zero-point offsets between these frames.  Zero point corrections
were applied to correct the photometric frames to a mean level between
frames, and these corrections were applied iteratively until the zero-point
shifts were all less than 0.001 magnitudes.  For the photometric frames,
initial offsets were $<0.04$ magnitudes in all frames and $<0.01$ in 87\%,
83\%, and 74\% of all $M$, $T_2$, and $DDO51$ frames, respectively.  In
each subsequent bootstrapping step, multiple non-photometric fields were
matched to the existing calibrated database, overlapping stars were used to
determine frame-by-frame zero-point offsets and color terms, and the
derived terms were used to convert the non-photometric magnitudes to the
calibrated system using an algorithm similar to that used for the
photometric transformation coefficients.  All 96 fields were matched and
calibrated for a final time using the frame-by-frame photometric or
bootstrapped transformation equations.  Again, zero-point shifts were
iteratively applied until all shifts were less than 0.001 magnitudes.  For
the entire database, initial zero-point offsets were always $<0.04$ and
$<0.01$ magnitudes in 88\%, 87\%, and 82\% of all $M$, $T_2$, and $DDO51$
frames, respectively.

Reddening values were found toward each individual star in our survey using
maps from \citet{sfd98}.  These reddening values have a mean of $E(B-V)
=0.0204$ and a standard deviation of 0.0044. Extinction and reddening
values in our observed $M$, $T_2$, and $DDO51$ bands were calculated
according to equations from \S2.2 of \citeauthor*{pI}.  Henceforth, all
magnitudes and colors given are extinction and reddening corrected and we
omit the ``dereddened'' subscript from our magnitudes and colors for
brevity (e.g., $M_0\equiv M$).

\begin{figure}
\begin{center}
\plotone{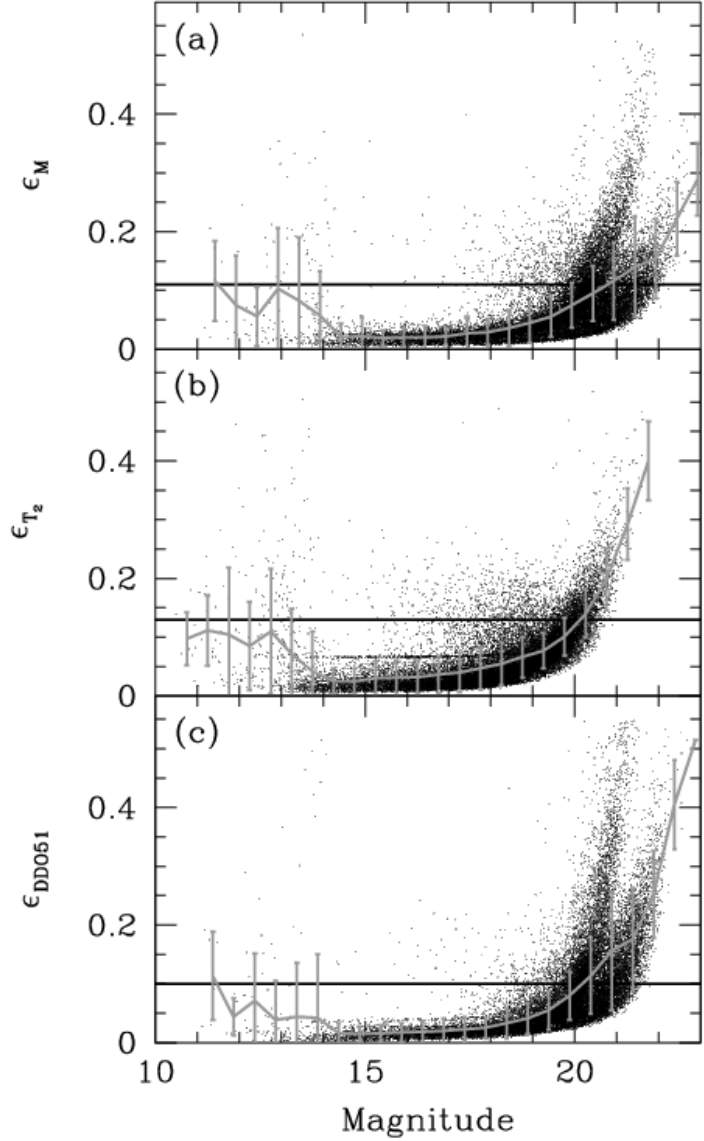} 
\caption{PSF measures of the magnitude error, $\epsilon_{X}$, with respect
  to the determined extinction corrected magnitude for the (a) $X=M$, (b)
  $X=T_{2}$, and (c) $X=DDO51$ filters.  The mean ({\it grey line}) and
  standard deviation ({\it error bars}) are shown to give a better sense
  of the statistical distribution of points.  The solid black lines of constant
  $\epsilon_{X}$ denote the adopted error limits.  Stars with measurement
  errors larger than these limits are discarded from any analysis.  Note
  that some pointings have photometry for which $\epsilon_{X}$ does not
  asymptote to zero.  This is most noticeable in the $T_2$ filter and is a
  product of a poorly converging PSF for some frames.}
\label{fig:magerr}
\end{center}
\end{figure}

\begin{figure}
\begin{center}
\plotone{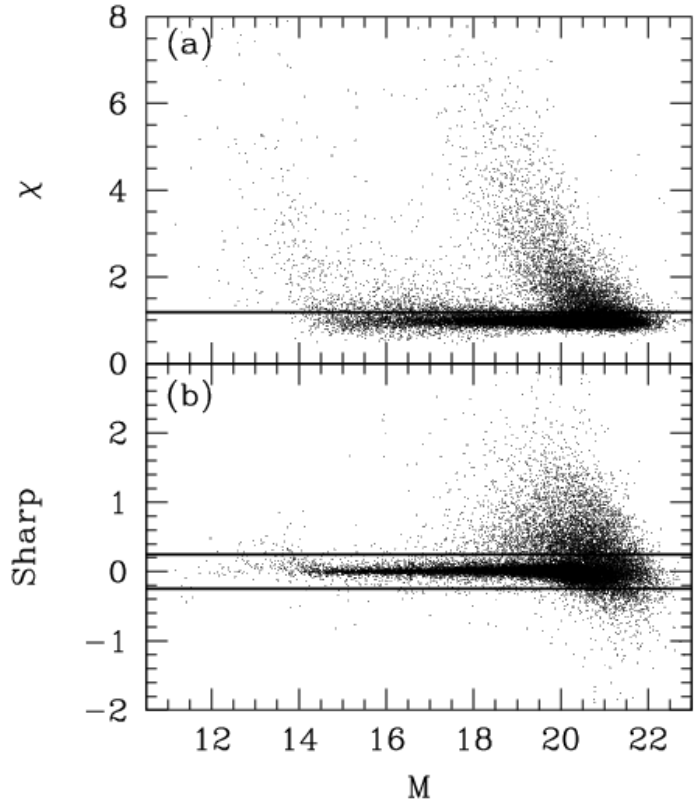} 
\caption{The PSF fitting parameters (a) $\chi$ and (b) sharp are plotted
  versus $M$ to show the stellar loci of good photometric measures.  The
  measured $\chi$ and sharp values for each CCD frame have been offset so
  that the stellar loci lie along $\chi=1$ and ${\rm sharp}=0$.  The solid
  lines denote the adopted range of acceptable ``stellar'' morphologies of
  these normalized parameters: (a) $\chi<1.18$ and (b)
  $-0.25<{\rm sharp}<0.25$.}
\label{fig:chirou}
\end{center}
\end{figure}

Figure \ref{fig:magerr} gives the photometric errors calculated for all
photometered objects in each filter in the survey area (Figure
\ref{fig:uncutfield}).  Any star having measurement errors larger than
$\epsilon_{M}=0.11$, $\epsilon_{T_{2}}=0.13$, and $\epsilon_{DDO51}=0.10$
is removed from our search for Sculptor stars.  (See \S4.6 for a discussion
of contamination of our ultimate sample due to photometric uncertainties.)
Through an analysis of the PSF-fitting parameters $\chi$ and sharp, the
sample is further limited to those objects with stellar morphologies.  The
stellar locus of $\chi$ and sharp values for each CCD frame was shifted to
bring them to common values of 1 and 0, respectively, so that a single set
of limits ($\chi<1.18$ and $-0.25<{\rm sharp}<0.25$) could be invoked
across the dataset to select acceptable stellar objects as shown in Figure
\ref{fig:chirou}.  These morphological limits eliminate most sample
contamination by galaxies (large $\chi$ at intermediate and faint
magnitudes), poorly imaged stars (large $\pm$sharp), and saturated stars
(large $\chi$ at bright magnitudes).

The varying conditions under which the survey was taken results in a
varying sensitivity across the survey (evident in the apparently
non-smoothly varying density of detected sources in Figure
\ref{fig:uncutfield}), and requires us to determine the limiting magnitude
of each field so that homogeneous samples can be created for structural
analysis of Sculptor (\S5).  Magnitude limits for each frame were set to
the mean magnitude value at the imposed error limits of the data.  A
histogram of the number of pointings per magnitude limit is shown in Figure
\ref{fig:histmaglim}.

\section{SELECTION OF SCULPTOR MEMBER STAR SAMPLES}

This section describes the creation of magnitude limited samples of likely
Sculptor giant star members.  Individual stars are subject to the following
criteria: First, the stars must have magnesium line/band strengths
consistent with those for metal-poor giant stars as gauged in the ($M -
T_{2}$, $M - DDO51$) diagram, hereafter referred to as the two-color
diagram (2CD). Second, those stars selected as giant stars must have
combinations of effective temperatures and apparent magnitudes consistent
with the RGB and/or red horizontal branch (RHB) of Sculptor in the ($M -
T_{2}$, $M$) color-magnitude diagram (CMD).  Ideally, a third criterion
would be that the stars must have heliocentric radial velocities,
$v_{\rm hel}$, consistent with the systemic radial velocity of the Sculptor
dSph.  However, we have to date spectroscopically observed only a small
sample ($\sim10$\%) of our selected Sculptor giant candidates.
Nevertheless, those spectroscopic data that we have obtained verify the
robustness of the other selection criteria toward selecting {\it bona
  fide} Sculptor giant stars (\S4.5).

A second sample of stars useful for exploration of the Sculptor morphology
are the BHB stars.  The near-complete isolation of this very blue
population from the bulk of foreground MW contaminants allows for a
qualitative check on the distribution the more numerous and brighter red
giant stars.  Stars satisfying either the RGB/RHB or BHB criteria are
then broken into various magnitude-limited samples that give different
sampling areas and sampling densities of the Sculptor field.

\subsection{The Two-Color Diagram}

The ($M - T_2$, $M - DDO51$) 2CD for stars in the area surveyed around the
Sculptor dSph is shown in Figure \ref{fig:2color}.  The separation of giant
and dwarf stars in the 2CD is based on the $DDO51$ filter measure of the
\ion{Mg}{1} triplet at 5150\AA\ and the MgH feature with bandhead at
5211\AA (\citeauthor*{pI}).  At a given stellar surface temperature, these
spectral features are primarily sensitive to stellar surface gravity and
secondarily to metallicity.  The $M$ broadband filter is used as a measure
of the continuum flux across the wavelength range of the magnesium
features so that the ($M - DDO51$) color index yields an effective measure
of the stellar surface gravity.  For K spectral type stars, giants have
larger ($M-DDO51$) due to weaker magnesium features.  The giant/dwarf star
separation becomes less pronounced and disappears at both earlier and later
spectral types.  To sort stars by effective temperature, we use the $T_{2}$
broadband filter to create the color index ($M - T_{2}$), which primarily
reflects stellar effective temperatures
\citep[\citeauthor*{pI}; ][]{bessell01}.

The secondary dependence of the magnesium features on metallicity also
allows for formulation of isometallicity loci for both dwarfs and giants
\citep[\citeauthor*{pI}; ][]{pb94}, as shown in Figure \ref{fig:2color}.
Thus, in principle we can tune our selection to giant stars of a given
metallicity.
The isometallicity loci (as displayed in Figure 2b of
\citeauthor*{pI}) are given for both dwarfs and giants; increasing metallicity
correlates with a decrease in ($M - DDO51$) index.  As in \citeauthor*{pI}, a shift
of -0.005 was applied in ($M-DDO51$) and -0.075  in ($M-T_2$) from the curves
of \citet{pb94}; this shift, which causes better coincidence of the \citet{pb94}
solar metallicity dwarf curve with our data from this and prior studies, can be
understood in light of the fact that the \citet{pb94} passbands were nonstandard
\citep[as discussed in][ and Paper I]{lb96}.  These
isometallicity loci are used to shape a 2CD giant selection region 
(Figure \ref{fig:2color}) that excludes dwarfs with [Fe/H]$>-2.0$ while selecting
the dominant populations of Sculptor giants, which have metallicities of
[Fe/H]$\sim-1.5$ and [Fe/H]$\sim-2.3$ \citep{kaluzny95,m99}.  The allowable
temperature range for stars selected to be Sculptor giant candidates with 
this initial criterion is set
to span from the blue edge of the MW field population redward to just
beyond the color of the tip of the Sculptor RGB.  Both diagonal edges of
the selection region are set approximately parallel to the dwarf locus but
offset to redder color.  This offset will help limit the number of dwarfs
accidentally landing within the selection region due to photometric errors.

Stars enclosed in the adopted 2CD giant region are only giant star
{\it candidates}, and, moreover, they are giant candidates not necessarily
specific to the Sculptor dSph.  Other objects that may fall within this
selection region could include (1) field dwarfs that are scattered into the
giant region by photometric error, (2) compact galaxies (typically
dominated by giant star light and possibly at redshifts where the Mg
features shift out of the $DDO51$ passband) that were not eliminated by the
morphological limits, (3) field giants, and (4) metal-poor subdwarfs with
[Fe/H] $\lesssim-2.0$.  An analysis of the level of contamination that we
expect is presented in \S4.5.

\subsection{The Color-Magnitude Diagram}

The ($M - T_2$, $M$)  CMD, is presented in Figure \ref{fig:cmd1}.  The RGB
and BHB sequences of Sculptor clearly stand out.  Below, we describe in
detail how Figures \ref{fig:2color} and \ref{fig:cmd1} together are used
to isolate these populations from the Galactic foreground contamination.

\subsubsection{Selection of Candidate Sculptor Stars from the Giant Star
               Sample}

Figure \ref{fig:cmd1}b shows that the Sculptor RGB is a predominant feature
of the stars passing the 2CD giant star selection criteria.  Application of
a second criterion defined in the CMD will further purify the Sculptor
giant sample.  Limit definitions for this selection are guided by two
methods illustrated in Figure \ref{fig:cmd2}.  First, based on stellar
proper motions found by \citet{schweitzer95}, those giant stars with a high
probability ($>80\%$) of being Sculptor member stars are used to define the
RGB locus to a magnitude of $M\sim20.5$ (Figure \ref{fig:cmd2}a); third
order polynomials have been defined to enclose this astrometric sample.  A
second third order polynomial is used to retain a small region of the CMD
centered at $(M - T_2, M) \approx (0.97, 19.9)$, where there is a
concentration of highly probable Sculptor giants.

Although the high probability proper motion members provide a useful means
to constrain the RGB for $M\lesssim20.5$, much of our survey extends
fainter than this.  Unfortunately, near the higher density dSph center
where the RGB is most clearly defined, our data are actually shallower than
in some of the more outlying fields of our survey.  However, deeper $B$ and
$V$ Sculptor photometry from \citep{m99} can be used as a substitute.  The
\citeauthor{m99} data (1) show a separation of the RHB and the RGB that is
actually more pronounced in ($B-V$, $V$) space than in ($M - T_2, M$)
space, and (2) extend to magnitudes required to define RGB bounding limits
suitable for our deepest $M$, $T_2$ data \citep[see Figure 1 of][]{m99}.
We isolate the RHB using
\begin{equation}
\begin{array}{c}
0.4<(B-V)<0.64 \\
-0.38(B-V)+20.35<V<-0.38(B-V)+20.6
\end{array}
\label{eq:rhb}
\end{equation}
and isolate the lower RGB by selecting
\begin{equation}
\begin{array}{c}
(B-V)<0.85 \\
V>20.05 \\
V>-6.67(B-V)+24.38.
\end{array}
\label{eq:lrgb}
\end{equation}
All stellar objects that have photometry from both this study and
\citet{m99} are shown in Figure \ref{fig:cmd2}b, regardless of magnitude
errors in either.  Guided by those stars satisfying Equations \ref{eq:rhb}
and \ref{eq:lrgb}, we extend the RGB/RHB selection to $(M - T_2) = 0.65$
between $20.0<M<20.47$ to include RHB stars and find it reasonable to
simply extend our third order polynomials to $M=21.0$ for the RGB fainter
than $M=20.5$.  The final adopted CMD selection criterion is applied to all
stars previously selected as giant star candidates in the 2CD over our
entire survey area (Figure \ref{fig:cmd2}c).

Note that in the RHB selection we are {\it not} attempting to include
{\it all} RHB stars in our survey.  Rather, we are only attempting to
include as many probable Sculptor stars {\it from our giant candidate
sample} as possible, after delineating that population by appealing to a
data set where the RHB is more distinct from the RGB.

\subsubsection{Selection of Candidate Sculptor BHB Stars}

The extreme blue colors of BHB stars relative to the MSTO of the MW field
population provides a unique opportunity to isolate a fairly pure sample of
Sculptor BHB candidates using only a CMD selection.  However, as seen by
comparing the BHB in Figures \ref{fig:cmd1}a and \ref{fig:cmd1}b,
imposition of the previously adopted magnitude error limits substantially
reduces the number of selected stars from the very blue end of the BHB,
primarily due to limitations of the $T_2$ photometry.  However, because the
BHB is {\it so} blue, for this CMD selection we can tolerate larger ($M -
T_2$) color errors (along the BHB) with little decrease in sample purity
(see Table \ref{tab:meanerr} for the mean color error of this sample).
Therefore we do not limit the selection of BHB stars by their photometric
error.  The Sculptor BHB candidate selection procedure is demonstrated in
Figure \ref{fig:hb}.

As with the RGB/RHB selection, we again limit the stars used for definition
of the bounding limits to those having Sculptor proper motion membership
probabilities greater than 80\% (Figure \ref{fig:hb}a).  In the definition
of the BHB selection, the three main considerations are: (1) enclosure of
the available blue extent of the BHB stars, (2) inclusion of RR Lyrae stars
(which, due to variability, will have a larger magnitude scatter than
non-variable HB stars), and (3) minimization of error-scattered
contaminants from the ``blue edge'' of the MW field dwarf population.  The
final selection criteria are applied to the entire survey as demonstrated
in Figure \ref{fig:hb}b.  For stars with $(M - T_2)>0.6$, the approximate
blue edge of the MW field population, error bars in both color and
magnitude are shown to demonstrate the low probability that a star from
this population might be scattered into the BHB selection region.

\subsection{Definition of Magnitude Limited Samples}

Figure \ref{fig:findb} shows the final CMD and spatial distribution of
objects selected from our survey to be candidate Sculptor members and those
objects pruned from further analyses.  Color and magnitude error bars are
included on individual candidate points
in Figure \ref{fig:findb}a to give a sense of the security of the
sample definitions with respect to their photometric errors.  We also include
average color and magnitude errors in 0.25 magnitude bins for the selected
Sculptor candidate sample and for the de-selected, ``field star'' sample.  These
average errors demonstrate that the quality of data between the field and 
candidate lists are comparable, but, in detail, the data outside the 
core of the galaxy are demonstrably of greater average quality 
than data taken near the Sculptor core.  From Figure
\ref{fig:findb}, we note: (1) the spatial extent of our selected BHB stars
roughly coincides with that of our selected RGB/RHB stars, (2) there seem
to be a significant number of stars that satisfy the RGB/RHB or BHB
selection criteria yet fall outside the King limiting radius as determined
by \citeauthor*{IH95} (these constitute a ``break population'' in the
radial profile, as shown below), (3) the distribution of some of our
selected stars fall along field boundaries due to the increased magnitude
depth in overlap regions, (4) there does not seem to be an inordinate
{\it excess} of stars included in our selected sample based on the fact
that there does not seem to be a {\it deficit} of stars within the RGB
boundary of the pruned star CMD in Figure \ref{fig:findb}c, but, on the
contrary, (5) our selection is conservative in order to be {\it reliable}
at the expense of being {\it complete}.  That we are missing some Sculptor
giant/HB stars is most obvious by the greater density of stars at the base
of the RGB in Figure \ref{fig:findb}c, and by a residual concentration of
stars near the spatial center in the deselected sample shown in Figure
\ref{fig:findb}d.  Sculptor stars may be omitted by the joint 2CD and CMD
selection criteria due to: (1) the decreasing sensitivity of the 2CD to the
giant/dwarf discrimination on the lower RGB and sub-giant branch, (2) RHB
stars that fall outside the 2CD and/or CMD selections, (3) asymptotic giant
branch or RGB stars that fall outside the RGB/RHB selection at bright
magnitudes, (4) RR Lyrae stars in variability phases that place them
outside our ``BHB'' selection, and (5) stars with photometric errors that
scatter the data points beyond our selection limits.

\begin{figure}
\begin{center}
\plotone{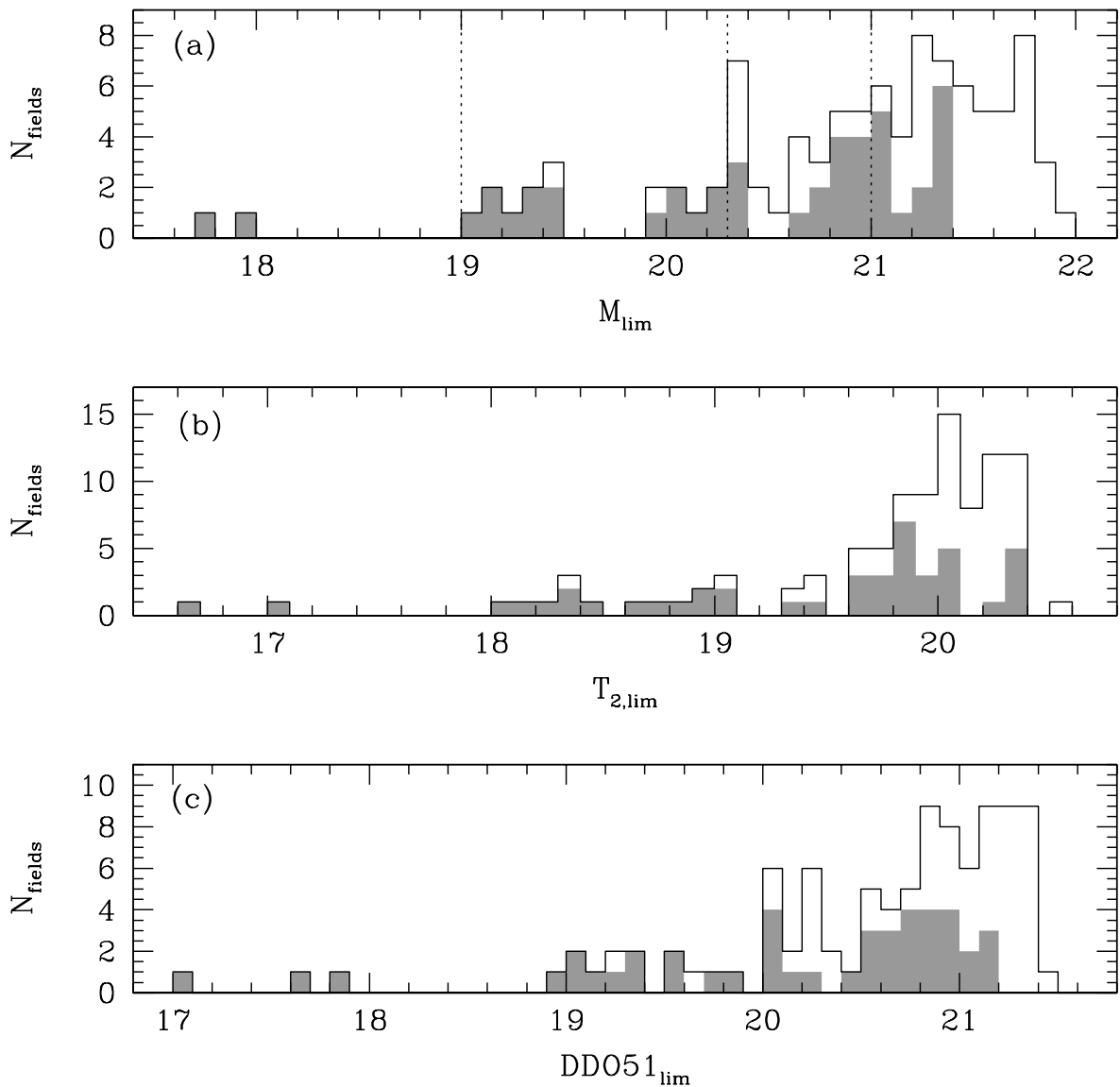} 
\caption{A histogram of the limiting magnitudes of all the pointings in the
  survey for (a) $M$, (b) $T_2$ and (c) $DDO51$.  The grey regions give the
  number of non-photometric fields in each bin.  Magnitude limits are
  determined by the mean magnitude of objects near the error limits of each
  frame.  The magnitude limits that will define some of our analysis
  samples (\S3.3) are shown by vertical dotted lines in the $M$ diagram. }
\label{fig:histmaglim}
\end{center}
\end{figure}

\begin{figure}
\begin{center}
\plotone{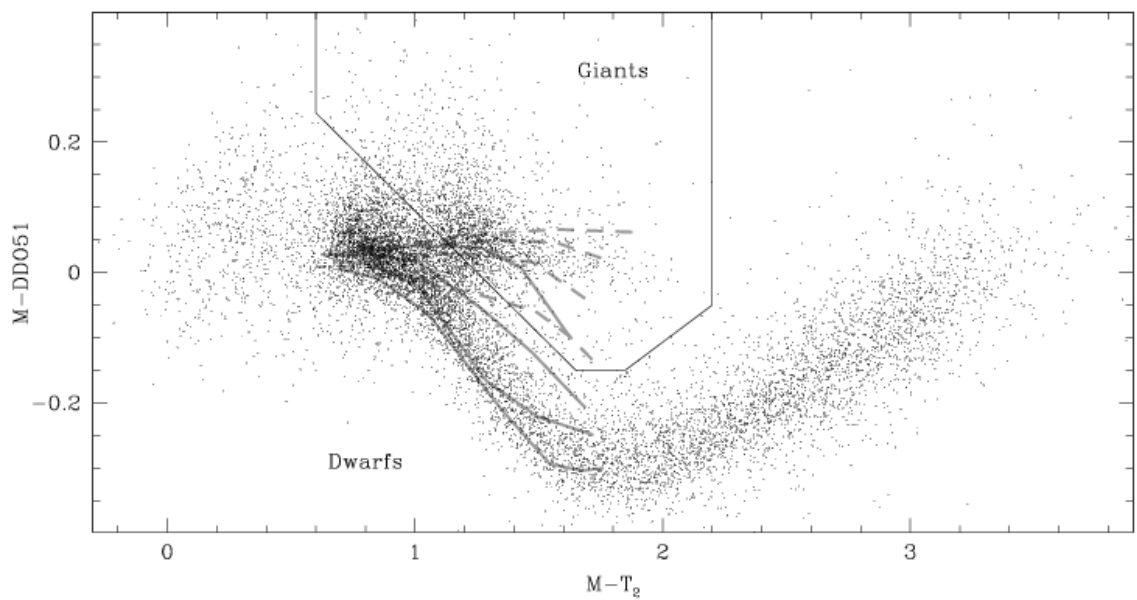} 
\caption{The two-color diagram (2CD) of all stellar objects in our survey
  that fall within the magnitude and profile shape limits described in \S2.
  The limits of our 2CD selection of giants are plotted as solid straight
  black lines; the limits were selected to lie along, but above the dwarf
  2CD locus.  Isometallicity lines for dwarfs ({\it solid grey lines})
  and for giants ({\it dashed grey lines}) are plotted for reference
  (from Figure 2b of \citeauthor*{pI}).  In increasing
  ($M - DDO51$) color, these lines correspond to [Fe/H] $= 0.0, -1.0, -2.0,
  -3.0$ for dwarfs and [Fe/H] $= -1.0, -1.5, -2.0, -3.0$ for giants.}
\label{fig:2color}
\end{center}
\end{figure}

\begin{figure}
\begin{center}
\plotone{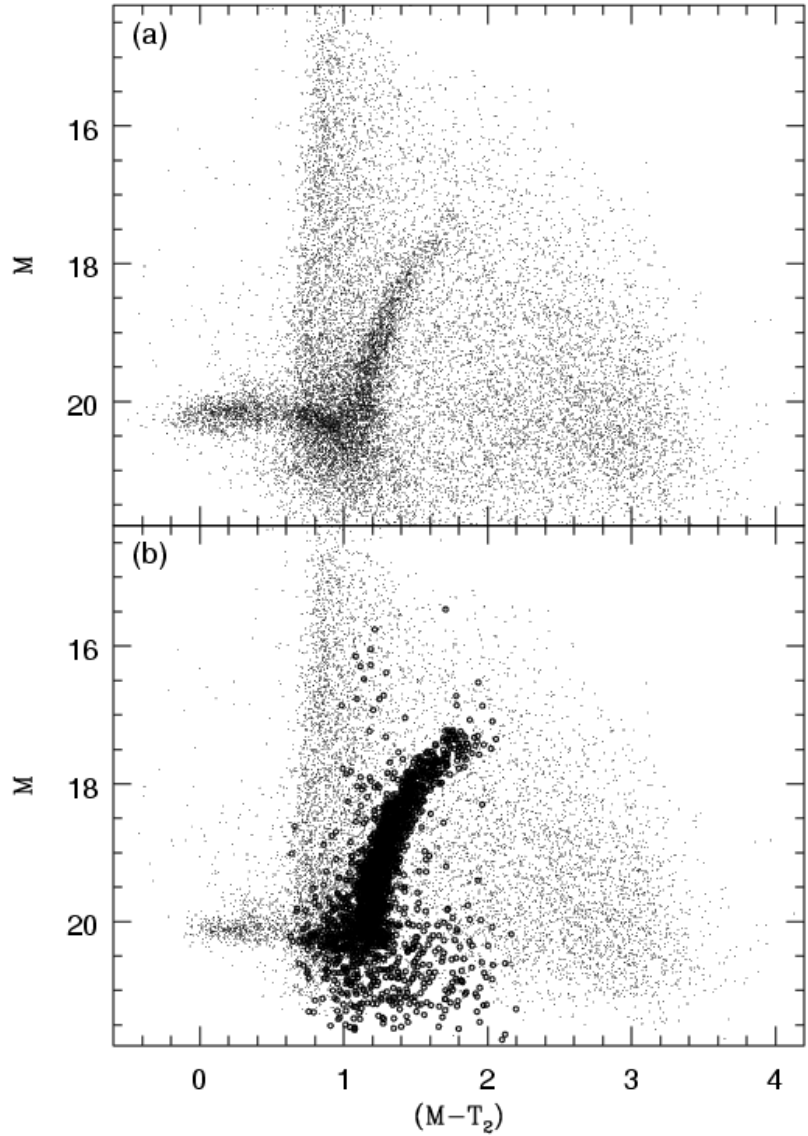} 
\caption{The color-magnitude diagrams (CMDs) of (a) all the star-like
  objects (as determined by PSF-fitting parameters) in our survey and (b)
  those objects with magnitude errors within the limits defined in \S2.
  Also, in (b), those stars selected to be giant star candidates in the 2CD
  (Figure \ref{fig:2color}) are plotted as open circles.}
\label{fig:cmd1}
\end{center}
\end{figure}

\begin{figure}
\begin{center}
\plotone{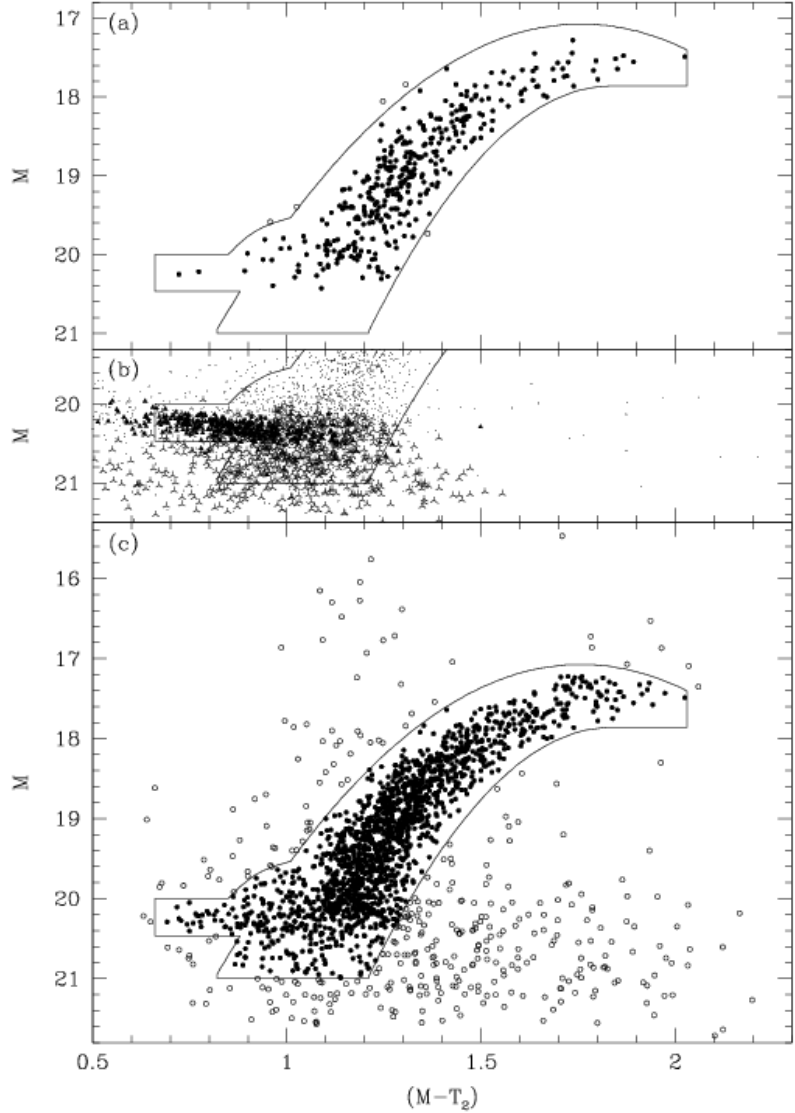} 
\caption{The CMD of stars selected as giant star candidates.  Panel (a)
  shows the CMD of the giants found to have Sculptor proper motion
  membership probabilities greater than 80\% by \citet{schweitzer95}.
  These stars are used to define the RGB/RHB selection to $M\sim20.5$.  As
  described more fully in the text, (b) displays the RHB stars
  ({\it triangles}) and the lower portion of the RGB ({\it three-pointed
    crosses}) as selected from the ($B-V, V$) CMD in \citet{m99}.  All
  stars photometered both in our study and in \citet{m99} ({\it points})
  are shown in (b).  This plot was used to define the RGB/RHB selection to
  $M\sim21.0$.  Finally, in (c), we demonstrate the application of our
  RGB/RHB selection to all giant star candidates found in our survey.  In
  (a) and (c), those giant candidates falling inside our RGB/RHB selection
  are represented by filled circles, while those outside are left as open
  circles.}
\label{fig:cmd2}
\end{center}
\end{figure}

\begin{figure}
\begin{center}
\plotone{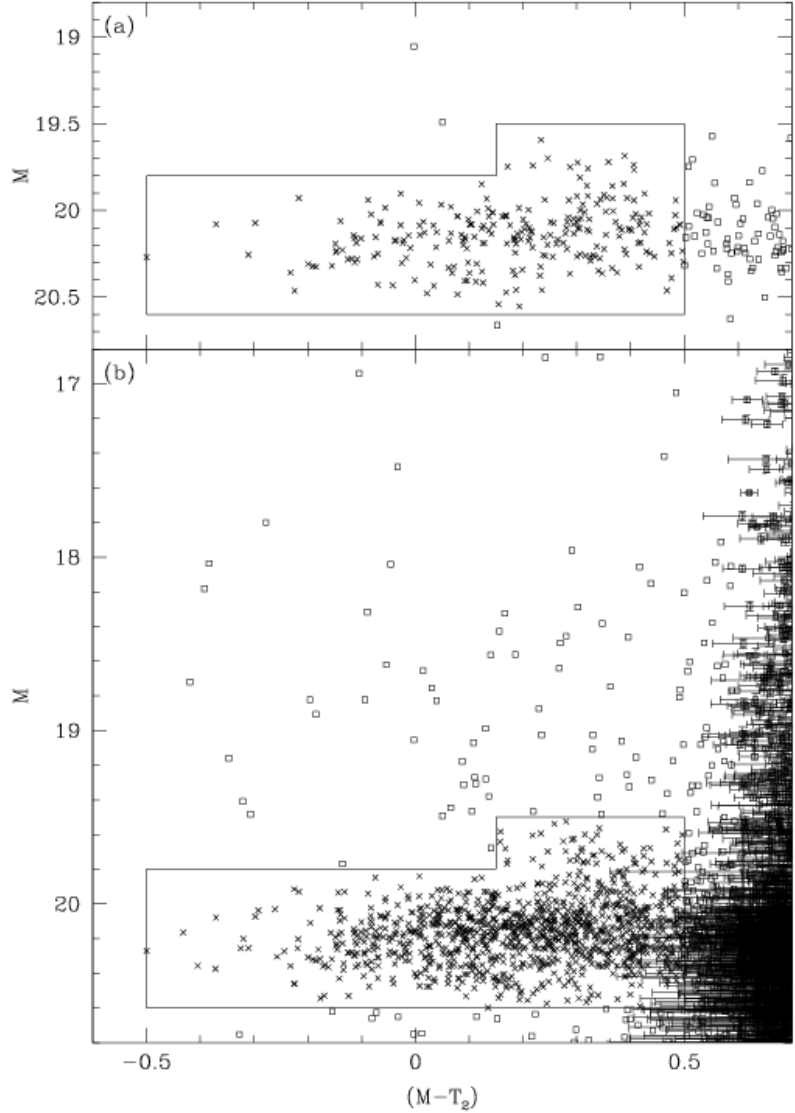} 
\caption{The CMD of stars in the color range of the BHB of Sculptor.  Both
  (a) and (b) plot all stellar objects regardless of magnitude errors.  The
  definition of the BHB selection is done in (a), which shows only those
  stars with Sculptor proper motion membership probabilities greater than
  80\%.  In (b), we demonstrate the application of this selection to all
  the stellar objects in our survey.  In both plots, those objects selected
  to be Sculptor BHB candidates are shown as crosses and those left
  unselected are open squares.  Finally, (b) displays error bars in both
  color and magnitude for objects with $(M-T_2)>0.6$ (the approximate blue
  edge of the MW field population) to demonstrate that there is little
  likely dwarf star contamination in this sample. }
\label{fig:hb}
\end{center}
\end{figure}

\begin{figure}
\begin{center}
\plotone{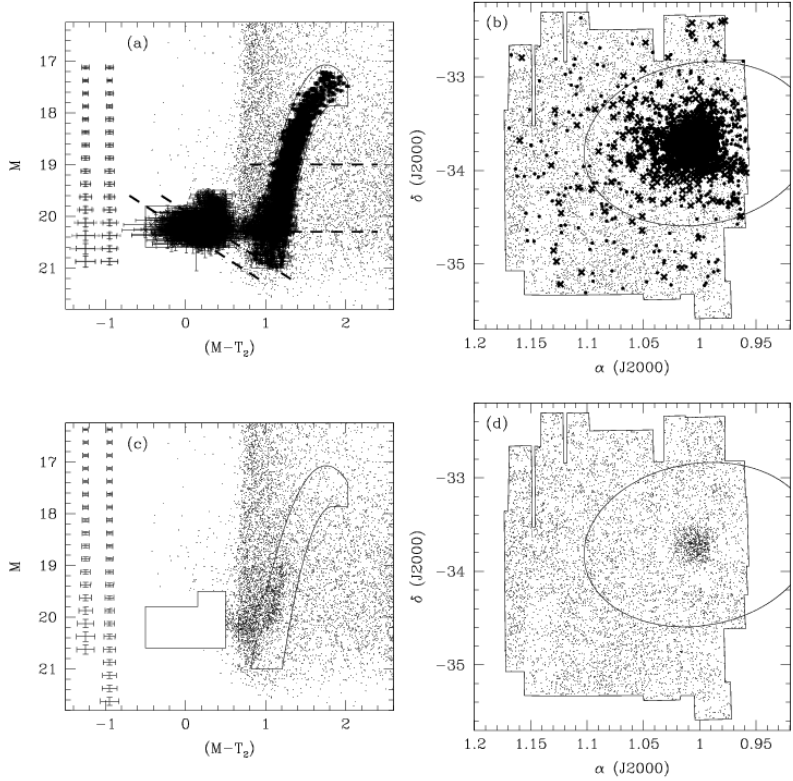} 
\caption{The distribution of all stars from our magnitude error and
  morphology-limited detected sample in the CMD ({\it left panels}) and on
  the sky ({\it right panels}).  The {\it top panels} include the stars
  that meet our Sculptor candidate criteria shown with larger symbols
  ({\it filled circles} for RGB/RHB and {\it crosses} for BHB, including
  individual error bars) while the
  bottom panels show only those objects that were eliminated by our
  selection criteria.  The boundaries of our CMD selection are shown in
  panels (a) and (c).  Dashed lines in panel (a) show the magnitude limits
  adopted in our analysis of the structure of the dSph.  Also, we show two
  sets of average color and magnitude errors in 0.25 magnitude bins at the left
  of panels (a) and (c).  The set to the left is for stars at a radius of
  $\leq0.5r_{lim}$, while the one to the right is for stars beyond this radius.
  The average errors are calculated for the Sculptor candidates in (a) and for
  all de-selected, field stars in (c).  Average error bars are determined both
  with ({\it grey}) and without ({\it black}) the BHB population in (a).
  Panel (b) demonstrates the general correspondence of the BHB and RGB/RHB
  distributions as well as the significant number of stars outside the
  nominal limiting radius determined by \citeauthor*{IH95} (plotted ellipse
  in panels (b) and (d)).  By comparison with Figure \ref{fig:uncutfield}
  it may be seen that some of the outermost Sculptor candidates appear to
  lie along individual CCD field boundaries (where the net depth of the
  survey is deeper).  We account for this effect in our analysis by
  choosing magnitude limited samples.  Our conservative Sculptor selection
  criteria have actually de-selected some likely Sculptor stars, as
  evidenced by the higher density of stars near the bottom of the RGB in
  panel (c) and the concentration of stars at the spatial center of
  Sculptor in panel (d).}
\label{fig:findb}
\end{center}
\end{figure}

\begin{figure}
\begin{center}
\plotone{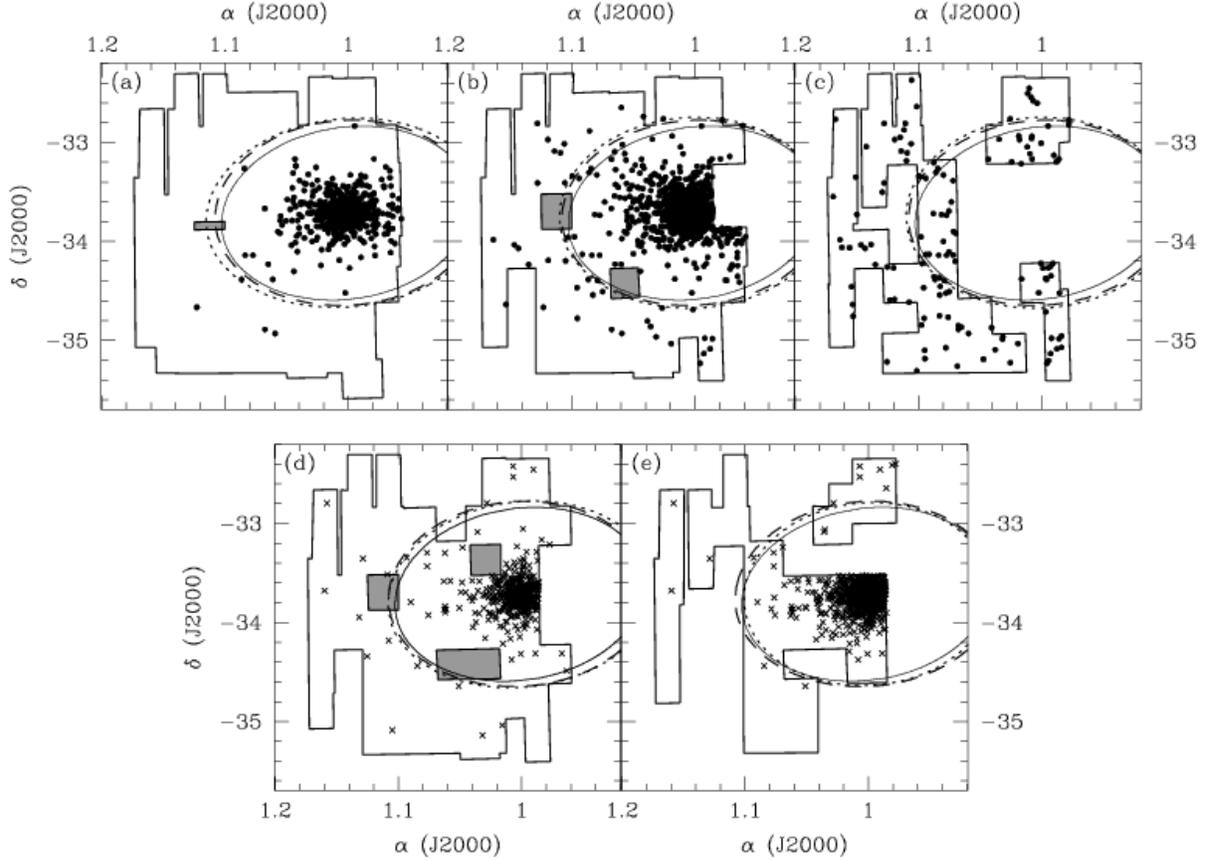} 
\caption{The spatial plots of all magnitude limited samples and their
  respective field boundaries or survey holes ({\it grey}): (a)
  $M\leq19.0$, (b) $M\leq20.3$, (c)$M\leq21.0$, (d) $T_2\leq19.9$, and (e)
  $T_2\leq20.3$.  Each panel gives for its subsample the limiting radius
  corresponding to the best-fit King profile ({\it dotted ellipse}; see
  Table \ref{tab:rpkingfits}), the limiting radius corresponding to the
  {\it mean} King profile parameters across all subsamples (Table
  \ref{tab:rpkingfits}; {\it dashed ellipse}), and the \citeauthor*{IH95}
  ({\it solid ellipse}) King limiting radius (\S5.2).  The ellipses in (c)
  are the same as those in (b) because the $M\leq21.0$ sample was not fit
  to a radial profile (see text).  The four derived King limiting radii
  (for the $M\leq19.0$, $M\leq20.3$, $T_2\leq19.9$, and $T_2\leq20.3$
  samples) are only slightly different from one another and the same within
  the fitting errors. }
\label{fig:sclmaglimdist}
\end{center}
\end{figure}

Sculptor candidates thus far selected by application of the 2CD and CMD
criteria provide a useful database for follow-up spectroscopy and
exploration of Sculptor dynamics; however, a study of the morphological
structure of Sculptor requires {\it homogeneously} selected samples that
account for variable magnitude limits across the survey area.  To do this,
we analyze the spatial distributions of various subsamples of Sculptor
candidates after imposing specific magnitude limits (those adopted for the
RGB/RHB samples are shown as dashed lines in Figure \ref{fig:findb}a) and
eliminating survey regions with imaging depths shallower than these limits
from consideration.  For the RGB/RHB sample, the various adopted magnitude
limits trade off the balance between sky coverage and depth.  An
$M\leq19.0$ limit reaches the vast majority of our fields (Figure
\ref{fig:histmaglim}) and it likely gives us the purest sample of Sculptor
RGB stars, albeit at the expense of a faint survey depth.  A second,
$M\leq20.3$ limit gives the maximum depth that includes most of the CCD
fields near the center of Sculptor.  A final, $M\leq21.0$ limit plumbs the
deepest fields in our survey, but over rather limited areas.  The latter
sample is put to limited use because it does not include the central fields
and is partially limited by poorer $T_2$ magnitudes.  As discussed in
\S3.2.2, the magnitude errors allowable for the BHB sample have been
loosened in order to reclaim a larger number of BHB stars.  Magnitude
limits that are adjusted for this different error sample (not shown in
Figure \ref{fig:histmaglim}) are used to define sample magnitude limits of
$T_2 \leq 19.9$ and $T_2 \leq 20.3$ for our BHB samples (see Figure
\ref{fig:findb}a).  Also, any field that did not have a corresponding $M$
depth that allowed sampling of the full BHB {\it color range} (i.e.
requiring $19.5 \leq M \leq 20.6$ to match the two respective BHB $T_2$
limits adopted above) is automatically rejected.  In the following
discussions, the above defined subsamples will be referred to by their
imposed magnitude limits (i.e. the RGB/RHB limited to $M\leq19.0$ is the
$M\leq19.0$ sample and the BHB limited to $T_2\leq19.9$ is the
$T_2\leq19.9$ sample, etc.).  The spatial distribution and field limits of
each sample are shown in Figure \ref{fig:sclmaglimdist}.

\section{SPECTROSCOPY}

The reliability of our photometrically selected Sculptor candidates
(RGB/RHB and BHB) can be assessed via radial velocity membership checks.
A complete census would require $\sim2700$ spectra; limited allotted
observing time and poor conditions during that allotted time has precluded
us from fulfilling this task.  However, we have obtained spectra for
$\sim5\%$ of our photometrically selected Sculptor candidates (mainly for
$\sim10\%$ of our RGB/RHB sample) and this allows us to at least gauge the
statistical reliability of our selection methodology in the magnitude range
probed by the obtained spectroscopy.  This section
describes our spectroscopic observations and an assessment of the
contamination of our candidate sample.
 
\subsection{CTIO HYDRA Spectroscopy}

Spectra for this study were obtained using the HYDRA multi-fiber system at
the Blanco 4-m telescope at the Cerro Tololo Inter-American Observatory
(CTIO) during UT 10 -- 12 November 2000 and UT 7 -- 10 October
2001\footnote{These allocations were also used for observations of Carina
  dSph stars presented in \citeauthor*{pVI} where a more lengthy discussion
  of the observations and analyses for these data are given.  We only
  summarize this previous presentation here; readers can see
  \citeauthor*{pVI} for more detail.}.  The Loral 3K$\times$1K CCD and the
790 lines/mm KPGLD grating (in first order) were used for observations
during November 2000 and yield a resolution of $\sim2600$ (or 2.6\AA\ per
resolution element) with spectral coverage from 7000--9250\AA.  During
October 2001, additional observations using the SITe 4K$\times$2K CCD and
the 380 lines/mm grating (in first order) were made.  A 200 $\mu$m slit
plate was placed after the fibers to improve the resolution to $\sim7600$
(1.2\AA\ per resolution element) with a spectral coverage from
7750--8700\AA.  During both runs, calibration lamp (Penray HeNeArXe)
exposures were taken in every HYDRA fiber setup for wavelength calibration.
Each standard radial velocity (RV) calibrator was observed through nearly a
dozen individual fibers in each HYDRA setup; 6 -- 13 of these calibrators
were observed during each run.
 
As with the imaging frames, routines from the IRAF {\ttfamily imred.ccdred}
package were used to reduce the raw CCD multi-fiber data.  Images were bias
subtracted, overscan corrected, trimmed, and then corrected for
pixel-to-pixel sensitivity variations and chip cosmetics by applying
``milky flats'' as described in the CTIO Hydra manual by N.
Suntzeff\footnote{http://www.ctio.noao.edu/spectrographs/hydra/hydra-nickmanual.html}.
Subsequent extraction and calibration of the individual fiber spectra were
completed with a local IRAF code incorporating functions from the
{\ttfamily imred.hydra} package as described in \citeauthor*{pVI}.
 
The RV of each observed star is determined via a modified version of the
cross-correlation (XC) technique developed by \citet{td79}.  Spectra are
Fourier-filtered to remove low-order (continuum) variations and then
cross-correlated against a master RV template spectrum.  The master is a
normalized spectrum that has all spectral regions that contribute little
more than noise to the XC function masked (i.e. set to one).  Masking of
the spectra eliminates all but the most vertical parts of the
uncontaminated spectral lines, which contains the most information for the
XC function.  The foundations and application of this XC methodology are
described more fully in \citet{sgr2} and \citeauthor*{pVI}.  

Each derived RV is assigned a numerical quality index \citep[with $Q=7$
being the highest quality, $Q=1$ the lowest; see][]{sgr2} gauged by the
strength (as compared to secondary peaks) and symmetry of the XC peak.  It
is found that RVs with $Q>3$ yield measurements that are acceptable for
membership discrimination, but the RV errors improve with higher $Q$.  We
set $Q=8$ for stars that have average velocities from multiple
observations\footnote{These stars are not necessarily of higher quality
  than some individual $Q=7$ observations; the $Q=8$ value just serves to
  distinguish an RV that is averaged over multiple measures from an RV from
  a single spectrum. However, in all but one case at least one $Q=7$
  measure contributes to the weighted averaged value}.  The dozens of
single observations of RV standards show typical standard deviations of
$5\,{\rm km\,s^{-1}}$ for the November 2000 data and $2\,{\rm km\,s^{-1}}$
for the higher resolution October 2001 data.  We adopt typical errors that
are twice the standard deviations determined above for our Sculptor target
stars due to the lower $S/N$ of these observations compared to the RV
standard observations.  These adopted values are consistent with
comparisons shown in Table \ref{tab:rvmean} (see below for the Table
description) for multiply observed stars and reasonable for the higher
quality ($Q>5$) spectra.  Although five different HYDRA setups were used
for Sculptor observations, mediocre to poor observing conditions meant that
not all targeted stars yielded usable spectra; therefore, the number of
useful spectra are significantly less than the maximum possible with
Hydra's 64 or 128 available fibers on these observing runs, despite
exposure times of up to 4.5 hours.  On the other hand, the number of Hydra
fibers actually exceeds the density of Sculptor targets, so that some
unassigned fibers could be placed on stars other than those in our
selection sample at no cost.  Thus we were able to target some additional
(brighter) stars with photometric properties placing them outside our
selection criteria; these stars are useful to get some idea of the
incompleteness of our Sculptor candidate sample.
 
\subsection{Magellan MIKE Spectroscopy}

Spectra were also obtained with the Magellan Inamori Kyocera Echelle (MIKE)
slit spectrograph \citep{bernstein03} mounted on the Magellan II (Clay)
6.5-m telescope at the Las Campanas Observatory on the nights of UT 27 --
28 January 2004 and 30 -- 31 December 2004.  
Reduction of the data from raw detector output to stellar
$v_{\rm hel}$ used fundamentally the same tools as the CTIO HYDRA
reductions described above, with the addition of one step to account for
slit centering errors (described below).  A more extensive discussion of
the reduction of these MIKE data is given (in the context of the Carina
dSph stars observed on the same observing run) in \citeauthor*{Mu05b}.  For
XCs to obtain $v_{\rm hel}$, we used a single red order (at $R=19000$) of
the double echelle spectrograph.  This order covers the region 8468 --
8693\AA\ and includes the strong \ion{Ca}{2} infrared triplet as well as
just over a dozen other lines that contribute usefully to the XC.  Two
other orders were cross-correlated separately but these did not improve
the precision of the derived $v_{\rm hel}$, even when all three of the XC
functions were combined.  RV standards were observed both at the beginning
and ending of the night while a ThAr calibration lamp was observed on
average once per hour throughout the night (though the instrument proved
extremely stable over the night).  The RV standards are used to measure
small induced offsets in the derived radial velocities that are particular
to the adopted radial velocity XC template.  Bias frames, exposures of the
quartz dome lamp, and ``milky flats'' were observed on a nightly basis to
correct for large-scale as well as pixel-to-pixel throughput variations.
Because all of the stars were observed through the $0\farcs9$ slit,
whereas the typical seeing was about $0\farcs7$, significant fractional
errors in derived RVs may arise from slit centering errors.  To measure the
velocity shifts that result from this effect, we independently
cross-correlate the telluric absorption features in each order for each
star against those in a set of observed radial velocity standards as well
as in dusk spectra (see discussion in \citeauthor*{pX}).

Given the resolution and average quality of data, the MIKE observations
should have a velocity precision of $\lesssim1\,{\rm km\,s^{-1}}$.  In
practice, however, the derived $v_{\rm hel}$ have a precision of closer to
$1.5\,{\rm km\,s^{-1}}$, the degradation arising primarily due to the
precision of the literature-quoted RV values of the standards.  The
velocity precision is also somewhat limited by the slit illumination
errors, which, in fact, vary with wavelength due to atmospheric dispersion.
This atmospheric dispersion introduces a wavelength dependent term in the
slit illumination that is hard to characterize without multiple
observations with MIKE.  The errors for spectra taken in January 2004,
$\pm1.5\,{\rm km\,s^{-1}}$, are slightly lower than the
$\pm2.5\,{\rm km\,s^{-1}}$ assessed for the December 2004 data due to
difficulties in the reduction of the latter.
These errors are more than suitable for evaluation
of kinematic membership and acceptable for assessing velocity dispersion,
$\sigma_v$, measurements for Sculptor.

\subsection{Radial Velocity Summary}

In total, 194 high- to mid-quality spectra were obtained in the three runs
(49 in 2000, 111 in 2001, 22 in January 2004, and 12 in December 2004), of
which 157 were of unique
stars.  Based on our photometric selection, 11 of these stars were expected
to be field stars, two were expected to be Sculptor BHB stars, and the rest
were expected to be Sculptor RGB/RHB stars.  As in \citeauthor*{pVI}, for
stars observed multiple times we calculate weighted averages,
$\langle v_{\rm hel} \rangle$, of the individual measured velocity values,
$v_i$, following
\begin{equation}
\langle v_{\rm hel} \rangle = \sum_{i} (\omega_{i}^2 v_i) /  \sum_{i} (\omega_{i}^{2}).
\label{eq:meanrv}
\end{equation}
Weights used in Equation \ref{eq:meanrv} are defined by
\begin{equation}
\omega_{i} = \omega_{{\rm quality,} i} \times  (1/ \epsilon_{v_i})
\end{equation}
and account for varying $S/N$ and resolution in the spectral observations.
As discussed in the previous sections, we assume
$\epsilon_{v_i}=10\,{\rm km\,s^{-1}}$,
$\epsilon_{v_i}=4\,{\rm km\,s^{-1}}$,
$\epsilon_{v_i}=1.5\,{\rm km\,s^{-1}}$, and
$\epsilon_{v_i}=2.5\,{\rm km\,s^{-1}}$ for representative relative velocity
errors for the November 2000, October 2001, January 2004, and December 2004
observations, respectively.
The previous use of our XC methodology has shown the
magnitude of the XC peak ($CCP$) to be correlated with the velocity
precision (i.e. a stronger CCP yields a better velocity measurement);
thus we include a ``quality'' term, $\omega_{{\rm quality,} i}$,
in weighting each RV.  Measurements of $CCP$ in the ranges
($<0.3)/0.3-0.5/0.5-1.0/(\geq$1.0) are assigned $\omega_{{\rm quality,} i}
= 0.5/1.0/2.0/3.0$, respectively (\citeauthor*{pVI}).

The individual observations and adopted mean velocities for the 36 stars
with multiple observations are listed in Table \ref{tab:rvmean}.  The
first line of each entry gives the calculated $\langle v_{\rm hel}
\rangle$ and standard deviation, $s(v_{\rm hel})$, of the individual
measures of $v_{\rm hel}$.  The Table also serves to show that for stars
observed twice in the same run, the adopted uncertainties are generally
consistent with $s(v_{\rm hel})/\sqrt{N}$ ($N$ is the number of individual
measures of the velocity).  However, we recognize the low
numbers in the statistics of this agreement.  
For the stars 7015937, 31000262, and 31000811,
$s(v_{\rm hel})$ is significantly different from the adopted
$\epsilon_{v_i}$ due to largely disparate velocities between observations.
This is almost certainly due to an incorrect determination of $v_{\rm hel}$
and/or $Q$ for at least one of the spectra of these stars, not an intrinsic
change
in the velocity of the observed star (and we note that one or more of the RVs
is near the systemic RV of Sculptor in each case).  We reobserved 31000262
and 31000811 in December 2004 with the MIKE echelle to clarify these as 
members/non-members.  The
star 31000262 is confirmed as a member while 31000811 is not; however, their
$s(v_{\rm hel})$ are still large.  
For four other stars
(1000844, 1008239, 28000278, and 28000376) the difference in velocities
are larger than $\sqrt{N}\times$ the adopted error estimates; though
binary stars may play a role here, with a normal distribution of errors we
would of course expect greater than 1-$\sigma$ differences to occur one
third of the time --- or for about 12 stars in Table \ref{tab:rvmean}.
That we obtain half this number overall likely reflects that we have
conservatively overestimated the true RV uncertainties.  
Nonetheless, these
seven stars with large internal RV differences are omitted from measurement
of the Sculptor systemic velocity, $v_{\rm sys}$, and $\sigma_v$.

Along with internal consistency, we also check our derived $v_{\rm hel}$
against 10 stars with previous RV measurements \citep[][hereafter
\citeauthor*{QDP}]{ad86,QDP}\footnote{A larger number of overlap stars may
  be checked against the \citet{tol04} database but tabulated data are not
  readily available.}.  The comparison between the literature values for these
stars and those obtained for the current study are listed in Table
\ref{tab:rvcomp}.  As stated earlier, stars with weighted velocity averages
from our own study
have been given $Q=8$ to distinguish them from single observations.  The
Table shows general consistency between previous observations and those
velocities determined for this study.  More specifically, our observations
more closely match those from the more recent \citeauthor*{QDP} study.
 
Finally, in Table \ref{tab:rvfull} we present all of the available radial
velocities for the Sculptor dSph.  For the sake of internal consistency,
our determined $v_{\rm hel}$ are used instead of literature observations
when both are available.  The 22 literature stars that were not reobserved
for this study are also given.  When both literature sources had
observations, the error-weighted mean velocity and its associated error are
given\footnote{We note that \citeauthor*{QDP} argue that their observations
  of H185 and H512 are significantly different from \citet{ad86}. Star H512
  was not observed by us.  Both of our observations (see Table
  \ref{tab:rvmean}) for H185 (our 1002429) are consistent with that found
  by \citeauthor*{QDP} but not \citet{ad86}.  If a binary, it is possible
  that we happened to observe H185 at the same phase as \citeauthor*{QDP}.
  Nevertheless, according to our spectroscopic membership criterion given
  below, all observations of both stars are consistent with them being
  Sculptor members, but, for consistency with \citeauthor*{QDP}, H185 is
  omitted from our systemic velocity and velocity dispersion measurements
  (even though doing so has an insignificant effect on the results).}.
Along with RV statistics, the Table also includes color-color-magnitude
measures and the elliptical distance
for each star assuming the mean ellipticity for the samples fit in \S5.2.
Finally, the last column allows for gauging the accuracy of our photometric
selection technique (see below).  Those stars satisfying the defined
criteria are denoted with a ``Y'' while those that do not are denoted with
a ``N''.  So, for example, a star that satisfies the photometric criteria
(P) but not the radial velocity criterion (V; defined below) has a ``P/V''
entry of ``Y/N''.  Several special cases are highlighted by footnotes to
the Table.

Figure \ref{fig:rvdist} gives the distribution of the spectroscopically
observed stars from Table \ref{tab:rvfull} both in (semi-major axis)
distance--velocity space and on-sky.  From Figure \ref{fig:rvdist}a, it is
immediately apparent that the vast majority of the observed stars are
kinematically similar.  Given the observations are nearly in the direction
of the South Galactic Pole (SGP), typical MW field dwarfs (mainly from the
disk at these magnitudes) and even the bulk of halo field giant
contaminants will be at lower heliocentric velocity than Sculptor so that
the dSph members should be fairly distinct.  For a population of stars
with net zero revolutionary velocity about the Galactic center viewed in 
the direction of Sculptor, the mean heliocentric velocity will be about
35 km s$^{-1}$; for higher, net positive revolutionary speeds, the mean
heliocentric velocity {\it decreases} towards 0 km s$^{-1}$. 
To determine the velocity of Sculptor, the mean of all available
velocities was calculated using a 3$\sigma$ rejection iterated until points
were no longer rejected to obtain
$\overline{v_{\rm hel}}=107.96\pm0.76\,{\rm km\,s^{-1}}$, in good agreement
with the previously determined Sculptor $v_{\rm sys}$ (\citeauthor*{QDP}:
$v_{\rm sys}=109.9\pm1.4\,{\rm km\,s^{-1}}$).
The 3$\sigma$ rejection algorithm essentially sets the RV membership
criterion to $80\leq v_{\rm hel} \leq 135$, very similar to the limits set
by \citet{tol04} for their Sculptor member search.  
Stars falling within
this velocity range we henceforth consider to be kinematically associated
with the Sculptor dSph.  A histogram of the distribution of stars with
observed RVs is shown in Figure \ref{fig:rvdist}b and appears to follow a
normal distribution centered on the determined mean velocity.  Of
particular interest in this set are the probable member stars outside our
derived King limiting radius (\S5.2), one of which is a selected BHB star.

\subsection{Preliminary Dynamical Information from the Radial Velocity
            Survey}

Using the 134 stars leftover after pruning (1) stars with large $s$ in
Table \ref{tab:rvfull} (e.g., possible binaries, including H185, or
problem RV measures),
and (2) observations from November 2000 (due to the large velocity
uncertainties for these spectra), Figure \ref{fig:rvprof} gives the
azimuthally averaged $v_{\rm hel}$ and $\sigma_v$ profiles.  The mean
velocity ($\langle v_{\rm hel} \rangle$), velocity error
($\epsilon_{\langle v_{\rm hel} \rangle}$), and $\sigma_v$ are derived by
the same formulae given by \citet{mateo91}.  The weighting scheme for
calculation of $\langle v_{\rm hel} \rangle$ here ($w_i =
1/\epsilon_{v_i}^2$ with $\omega_i^2 = w_i$ in Equation \ref{eq:meanrv})
assumes the adopted, run specific $\epsilon_{v_i}$ values given above for
the individual spectra.  For the multiply observed stars in Table
\ref{tab:rvmean}, the error as calculated for an error weighted mean
($\epsilon_{v_{\rm hel}} = 1/\sqrt{\sum(w_i)}$) is used; these values are
tabulated in Table \ref{tab:rvfull}.  The error in the velocity dispersion,
$\epsilon_{\sigma_v}$, is obtained from
\begin{equation}
\epsilon_{\sigma_v}^2 = \frac{\sigma_v^4 + \langle \epsilon_{v_i} \rangle ^4}{2N} \frac{1}{\sigma_v^2 - \langle \epsilon_{v_i} \rangle ^2},
\label{eq:sigerr}
\end{equation}
where $\langle \epsilon_i \rangle$ is the average error in the velocity
measurement and $N$ is the number of stars\footnote{Note that $\sigma_v$,
  $\epsilon_{\langle v_{\rm hel} \rangle}$, and $\langle \epsilon_{v_i}
  \rangle$ given here are, respectively, $\mu_{int}$, $\sigma_{\langle v
  \rangle}$, and $\langle \sigma_i \rangle$ in the nomenclature of
  \citet{mateo91}.}.  Equation \ref{eq:sigerr} results in an indeterminable
$\epsilon_{\sigma_v}$ when $\sigma_v^2\leq \langle \epsilon_i \rangle ^2$;
an indeterminable $\sigma_v$ results when $N\epsilon_{\langle v_{\rm hel}
\rangle}^2\leq \langle \epsilon_i \rangle ^2+\epsilon_{\langle v_{\rm hel}
\rangle}^2$.  Only bins that contain nonzero, determinable $\sigma_v$
{\it and} $\epsilon_{\sigma_v}^2$ are shown in Figure \ref{fig:rvprof}.  To
demonstrate possible biases caused by binning, the results for the data are
supplied for three different bin sizes.

The velocity profile of Sculptor is, in general, constant with semi-major
axis distance.  We have confirmed that this is the case for velocity as a
function of either equatorial coordinate as well.  The velocity profile is
no better fit with a linear relationship between $v_{\rm hel}$ and
$\alpha$, $\delta$, or $a$ than it is with a constant value regardless of
the bin size.  Therefore, the data suggest Sculptor is not rotating at
least at the level of the observed dispersion in the velocities
($\sqrt{N}\epsilon_{\langle v_{\rm hel} \rangle}\sim 9\,{\rm km\,s^{-1}}$).
However, we note that our RV sampling of Sculptor candidate members becomes
poor beyond $a\sim30\arcmin$; velocity trends beyond this radius are not
well constrained by our present data.  Using the trimmed sample of stars
presented in Figure \ref{fig:rvprof} and the above error weighting scheme,
the mean velocity of Sculptor is $\langle v_{\rm hel} \rangle =
110.43\pm0.79\,{\rm km\,s^{-1}}$.

We find a global dispersion of $\sigma_v=8.8\pm0.6\,{\rm km\,s^{-1}}$ for
the 134 stars in Figure \ref{fig:rvprof}.  While this is 40\% higher than
determined by \citeauthor*{QDP}, these authors only probe out to
$\sim10\arcmin$; as may be seen in Figure \ref{fig:rvprof}, our dispersion
over this same radius is in fact comparable to that of \citeauthor*{QDP}.
On the other hand, our global $\sigma_v$ appears to be comparable to the
global value given in the recent letter by \citet{tol04}, who actually
survey a similar radial extent as we do\footnote{\citet{tol04} derive
  different dispersions for Sculptor stars separated into metallicity
  groups, one with [Fe/H]$<-1.7$ and another having [Fe/H]$>-1.7$.  The
  global $\sigma_v$ set by the present data is intermediary to those
  \citeauthor{tol04} report for these metallicity-separated populations.
  In \S5.4 we make an approximately similar metallicity separation of
  Sculptor stars on the basis of photometry, however, this separation is
  not as reliable as the spectroscopic separation of \citeauthor{tol04}.
  We also have fewer overall Sculptor velocities than \citeauthor{tol04}
  with which to estimate reliably distinct dispersions by metallicity.}.
This correlation of derived dispersions with radial coverage points to the
likely existence of a {\it radial gradient} in Sculptor velocity
dispersion, with a hotter dispersion found at larger radius.  Such a trend
is suggested by Figure \ref{fig:rvprof}, where, in general, colder
dispersions are found at smaller radii, particularly when larger bin sizes
are adopted to improve the errors on the dispersion.  Clearly one would
prefer better statistics at larger radii, but we note that a radial
gradient is implied by the results of \citet{tol04}, who note that hotter
dispersions attach to the more metal-poor Sculptor stars and that these
stars have a larger radial extent than metal-rich Sculptor stars.  Our own
analysis of Sculptor shows a similarly distinct radial distribution of
``metal-poor'' and ``metal-rich'' populations (see discussion in \S5.4).

It seems difficult to escape the conclusion that the Sculptor dispersion
profile is, at minimum, flat to a significant fraction of the King limiting
radius, and could even be rising at large radius (at least there seems to
be a rise in the dispersion past $\sim0.4r_{\rm lim}$).  This is a
trend that is being found in other dSph systems (e.g.,
\citet*{mateo97, kleyna02, kleyna04}; \citeauthor*{Mu05b}).
Predominantly bound stellar systems are generally expected to produce
{\it declining} dispersion profiles because the Keplerian orbits of stars
within the system should produce lower velocities at their apocenters.  The
implications of flat/rising profiles have been discussed by, e.g.,
\citet{kroupa97} and \citet{kleyna99}, who interpret such trends as an
important signature of tidally disrupting systems.  A flat/rising
dispersion trend is also found in Ursa Minor to large radius (see Paper IX).  
Ironically, after
measuring a similarly flat profile in the Draco system, \citet{kleyna02}
have appealed to an enormous dark matter halo, $(M/L)_{\rm tot}=440\pm240$,
as an alternative means to keep the dispersion hot to large radius.  On the
other hand, \citeauthor*{Mu05b} show that prosaic models of tidally
disrupting satellites with more modest mass and on very radial orbits can also
give reasonable fits the similar $\sigma_v$ trend ({\it including} a centrally {\it colder}
dispersion) {\it and} radial profile for the Carina dSph.  It is reasonable
to expect similar models will work for Sculptor, which is at a similar
Galactocentric radius and orbital phase as Carina.  It is certainly
compelling that the Sgr system also shows a flat/rising $\sigma_v$ trend
\citet{sgr2, Ma05b}.  The structural and {\it dynamical} similarity of Sgr
to other dSphs strongly begs the question of whether the disrupting Sgr is
more dSph {\it paradigm} than exception \citep{Ma05b}.  It is also worth
noting that while the velocity distribution of stars in the inner parts of
Sculptor is Gaussian, the distribution in the outer parts becomes fairly
platykurtic --- a distribution unexpected for a fully bound population of
stars.  These issues are explored further in \citeauthor*{pIX}, \citet{Ma05b}, and
\citeauthor*{Mu05b}.  The main goal of this subsection is to demonstrate the
apparent flat dispersion profile in Sculptor and point out the consistency
of this trend with the interpretation that the Sculptor radial density
break population we measure below (\S5) may constitute unbound, tidal
debris.  

\subsection{Photometric Selection Accuracy}

\begin{figure}
\begin{center}
\plotone{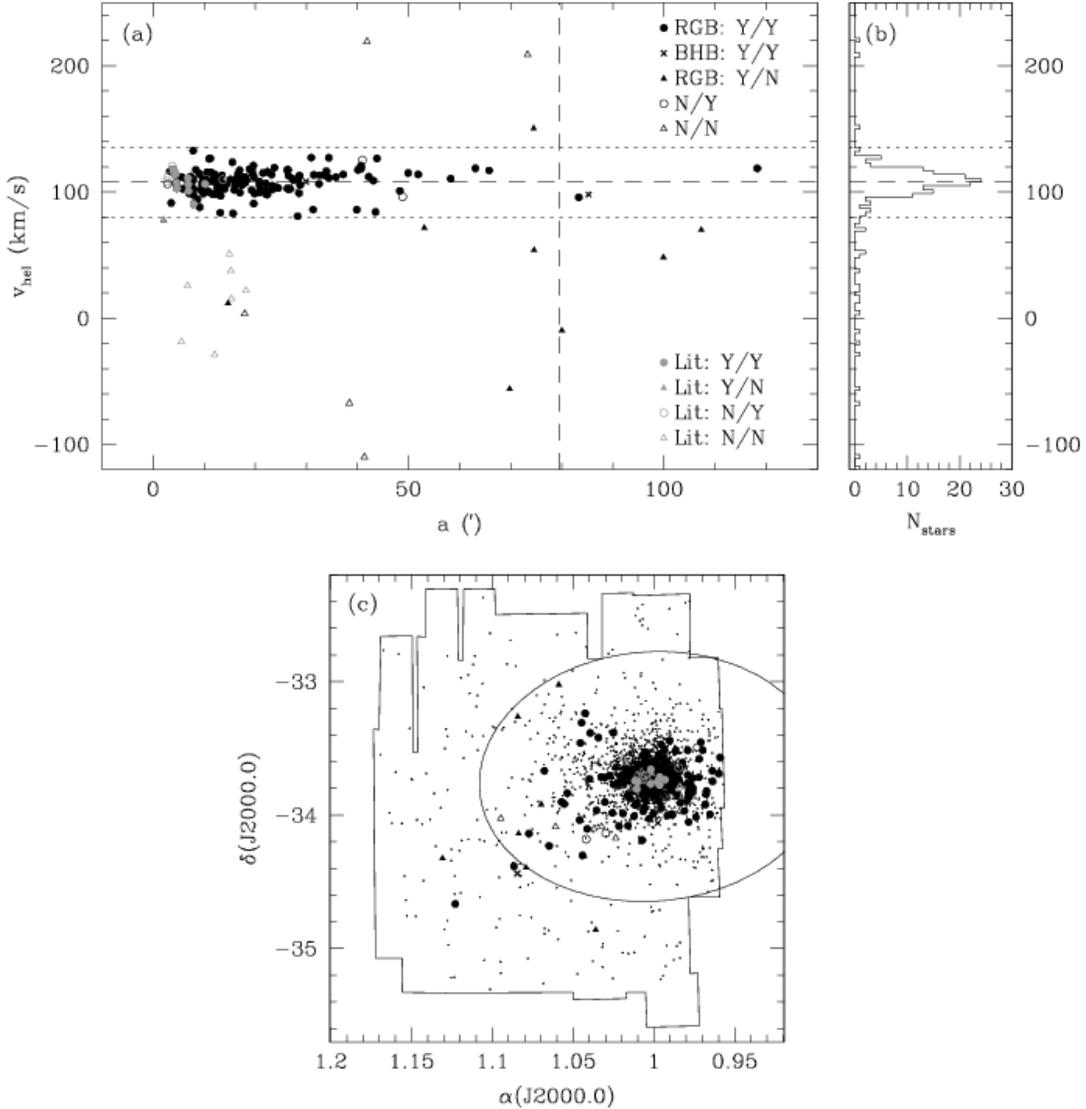} 
\caption{The (a) heliocentric radial velocity as a function of elliptical
  distance, $a$, from the derived cluster center, (b) histogram of the
  number of stars, $N_{\rm stars}$, at each velocity, and (c) spatial
  distribution of spectroscopically observed stars.  The legend in (a)
  gives the definition of each point symbol according to stellar type and
  photometric/velocity member status.  The distribution of points within
  $80\leq v_{\rm hel}\leq135$ (within the {\it horizontal dotted lines})
  is roughly symmetric about the derived
  $\overline{v_{\rm hel}}=107.96\pm0.76$ ({\it horizontal dashed line}).  The
  small points in (c) are stars photometrically selected to be Sculptor
  candidates (BHB and RGB/RHB) but have not yet been observed
  spectroscopically.  In particular, note (a) and (c) show six stars
  observed outside our derived mean King limiting radius (a: {\it vertical
    dashed line}, c: {\it solid ellipse}).  Three of these stars are both
  photometrically and spectroscopically selected as Sculptor members.}
\label{fig:rvdist}
\end{center}
\end{figure}

A primary motivation for the RVs in the present study (including those with
the lower velocity precision from observations in 2000) was to evaluate 
whether our photometric techniques to select Sculptor candidates are accurate 
so that we have confidence in the reliability of Sculptor structural properties derived 
from such data.  Figure \ref{fig:rvdist} summarizes the results of
this program.  The different symbols defined in the legend of Figure
\ref{fig:rvdist}a reflect the various categories in column 10 of Table
\ref{tab:rvfull}, where we correlate our photometric (\S3) and RV (\S4.3)
membership criteria.  Figure \ref{fig:rvselect} displays the Sculptor field
2CD and CMD to illustrate the selection/de-selection of stars having
derived RVs (where symbols are the same as in Figure \ref{fig:rvdist}).
Figure \ref{fig:rvselect} illustrates the efficacy of our
photometric selection technique at isolating individual, highly probable
Sculptor members; creating such reliable target lists is particularly
helpful for ensuring cost-effective spectroscopic observations of the
low-density regions of the dSph, where contamination by MW field stars
becomes a more significant problem (see Figure \ref{fig:bgdemon} and \S5.2.4).

Table \ref{tab:pvcomp} summarizes the accuracy of the photometric selection
criteria for the subsample of spectra obtained to date.  
Given that most of the spectra obtained to date are for $M \lesssim 19$ 
candidates, the statistics are properly interpreted to represent our
success for stars to these magnitudes, but do lend confidence that we
can expect improved selection efficiency at fainter  magnitudes over
selection methodologies based on only CMD-position of candidates.
The values given
in the final column are the percentage of photometric selections that agree
with the spectroscopic selection out of the total number of
spectroscopically observed stars.  Among all stars (BHB or RGB) we have
selected to be photometric members that have RVs, a combined 94\% are found
to be spectroscopic members (i.e., only 6\% are found to be false
positives).  Figure \ref{fig:rvdist} shows evidence for a decrease
in our false positive rate with radius from the center (i.e., with lower dSph
density); beyond in 0.5$r_{\rm lim}$, 71\% of our photometric members are
confirmed spectroscopic members.
This success rate can be compared to the 77\%
(308/401) velocity member success rate \citet{tol04} found over their
entire survey (which probes to similar magnitudes), a rate that declines to 
$\lesssim45\%$ for their stars observed beyond 0.5$r_{\rm lim}$.\footnote{To be
fair, it is not clear from the \citet{tol04} paper whether their results include
some stars that were only observed because they had spare spectrograph fiber optic
cables to place on stars that these authors would ordinarily consider to be poor
candidates, though the bulk of their targets are clustered about the Sculptor RGB.}
Based on these trends, we predict that
further study at even lower density regions of the dSph will show an even
wider disparity in success rates between CMD-only and $DDO51+$CMD methods for picking
candidate Sculptor stars.

Among our 22 photometrically-selected, {\it non}-members having derived
RVs, nine have RVs consistent with Sculptor membership.  That our
photometric selection is missing some Sculptor members is consistent with
our overall philosophy of sacrificing completeness for the sake of sample
reliability.  For purposes of mapping dSph structure, incompleteness
removes signal whereas unreliable candidates add background.  Our strategy
is to aim for near-zero background to make sure diffuse components are
mapped reliably, and this has driven adoption of rather conservative
selection criteria.  For example,
four of the nine ``false negatives'' in Table \ref{tab:pvcomp} actually
satisfy {\it both} the 2CD and CMD selection criteria in Figure
\ref{fig:rvselect}; these Sculptor giants were lost from our sample
{\it because they failed our point source image shape criteria}.  Thus,
these four stars (marked with footnote ``f'' in Table \ref{tab:rvfull}) are
technically not failures of the Washington+$DD051$ giant selection method
{\it per se}; nevertheless, we conservatively treat them as ``selection
failures'' in our accounting.  With the presently small spectroscopic
sample of non-selected stars it is difficult to estimate our true sample
incompleteness, especially given that it varies across the
color-color-magnitude volume as a function of mainly stellar temperature
and photometric error.  Figure \ref{fig:rvselect} in fact shows that the
other five false negatives tend to lie just outside one of our selection
limits, as might be expected.  The resulting high degree of reliability of
our photometric methodology is not unique to this study of Sculptor but is
also true for our similar studies of Ursa Minor, Draco, Leo I, Andromeda I and
Andromeda III \citep[\citeauthor*{pIV}; \citeauthor*{pIX}; \citeauthor*{pX}; ][]{ost05}.

\subsection{Review of Possible Sources of Contamination}

Our spectroscopic results show that our photometric samples are of a high
degree of reliability, and that our method confers a higher success rate
for identifying {\it bona fide} Sculptor stars than simple CMD-based selections
(e.g., compare Figure 11a to Figure 4 of \citep*{tol04}; see also
Papers VI and IX).
However, it
is worthwhile to investigate the origin of the remaining few 
sample interlopers found among the spectroscopic sample of Sculptor giant
candidates. 

First, our derived magnitudes are not perfect, so that there
is some photometric scatter across our selection boundaries.  Inspection of
Figure \ref{fig:rvselect} shows that a number of our false positives and
false negatives are near the borders of our selection regions.  The color
and magnitude error bars in Figure \ref{fig:rvselect} demonstrate that some
of the false classifications can be explained by photometric errors.  Table
\ref{tab:meanerr} gives the mean color and magnitude errors for the various
samples used here for the analysis of the structure of Sculptor.  The
tabulated means show where potential vulnerabilities to photometric
contamination are highest (e.g., on the lower RGB).

There are also some objects that lie more squarely within our selection
limits that are not kinematically associated with Sculptor.  These objects
likely fall in one of the following categories: point-like galaxies (which
are dominated by the light from giant stars or may have their magnesium
features redshifted out of the $DDO51$ band), metal-poor field subdwarfs
(which also have suppressed magnesium lines), or MW field giant stars.  Our
means of selecting against galaxies relies on morphological parameters
output by PSF-fitting, but point-like galaxies like quasars and AGN may
pass these criteria unabated.  In fact, among our sample of RGB/RHB
candidates we spectroscopically observed one quasar at ($\alpha$,$\delta$)
= ($01^{\rm h}03^{\rm m}33\fs71, -34\arcdeg55\arcmin49\farcs2$).  But 
any galaxies in the sample would likely be revealed by extreme, non-Milky 
Way-like RVs.

According
to the isometallicity lines given in Figure \ref{fig:2color}, metal-poor
([Fe/H]$\leq-2.5$) sub-dwarfs may also pass our 2CD selection.  Following
arguments given in \citeauthor*{pII}, we should find $\lesssim18$ such
subdwarfs in our {\it entire} survey area (i.e., at a density of
$\Sigma\lesssim 0.6 \times 10^{-3}$arcmin$^{-2}$), and we certainly 
have fewer than this limit.  Our selection
criteria allow for selection of field giant stars that happen to have
colors and magnitudes similar to the Sculptor RGB/RHB.  Again, the
contaminant surface density here is small given estimates from our control
fields (\S5.1).  It is worth noting that Milky Way stars orbiting {\it 
prograde} about the Milky Way will, in general, have 
$v_{hel} \lesssim +35$ km s$^{-1}$ in this direction of the sky ($[l,b]
\sim [288, -83]^{\circ}$).  This can be seen, e.g., by the distribution of 
velocities for non-Sculptor stars in Figure 4 of \citet{tol04}, which
has a concentration of stars with these velocities.
Only dynamically hot stars ---
stars with either retrograde orbits or with significant motions in the 
negative Galactic $Z_{GC}$ direction --- will be expected to have higher 
$v_{hel}$.  Several of our false positives are in this category.
In addition, we find that for those false positives having MIKE echelle
spectroscopy, the equivalent widths of the calcium infrared triplet lines 
are generally comparable to those of both Sculptor and Carina dSph members 
observed on the same run but less than other, more metal-rich stars also
observed in the Carina field (more details are presented in \citeauthor*{Mu05b}).
These velocity and chemical traits suggest that some halo stars are
contributing to the false positives. 

Still yet another explanation for at least some of the false positives is that
they are actually Sculptor stars.  We have already noted (\S4.4) that 
the velocity dispersion of the dSph appears to be growing with radius, 
yet we used the {\it same} $\sigma_v$ rejection at all radii to exclude
likely non-members.  However, of the nine false positives in our spectroscopic
sample we note that three lie within 15 km s$^{-1}$ of our Sculptor velocity
membership limits, six lie within 32 km s$^{-1}$ and all but one of these
are beyond 50 arcmin from the Sculptor center.
At these velocities and considering their $(l,b)$ these six ``false positives" 
must either be halo stars that are highly retrograde and/or that have a 
large Galactic $Z$ motion perpendicular to the disk; yet there are not
comparable numbers of false positive stars with the {\it opposite} properties,
as might be expected for a randomly populated, hot halo.  It seems conceivable
that at least some of these stars might be {\it bona fide} Sculptor giants,
consistent with the notion that the velocity dispersion is growing with 
radius.

No matter their origin, in the end, all of these contamination 
effects will be accounted for as (a small)
``background'' in our analysis of the Sculptor radial density fits (\S5.2),
and we expect our high $S/B$ Sculptor giant candidate lists to 
provide a reliable means by which to map the structural
properties of the Sculptor dSph.  While we do not have a similar
spectroscopic testing of the BHB samples, we shall show (\S5.3) that these
stars yield very similar Sculptor morphological properties to the RGB
samples.  

\section{DISTRIBUTION OF CANDIDATE SCULPTOR MEMBER STARS}

\subsection{Background Determination}

A proper assessment of the background is critical to proper determination
of the density distribution of the low density dSph galaxies.  This rings
particularly true for the Sculptor dSph given the reviewed (\S1) history of
attempts to determine its physical extent.  Our goal here has been to
attack the problem anew with a methodology aimed at substantially
{\it reducing} the background level so that fractional errors in its
estimation have significantly less effect on our morphological
parameterization of the system, particularly in its tenuous outer parts.
We will make a quantitative assessment of the resultant $S/B$ improvement
below (\S5.2.4), but a qualitative impression of our success is given by
the CMD of {\it non-}selected stars shown in Figure \ref{fig:findb}c, which
demonstrates the amount of (mostly) foreground material along the Carina
RGB that has been removed from consideration, while the spectroscopic
assessment of our contamination level (Table \ref{tab:pvcomp}) shows that
our residual contamination rate is very low to at least $M \sim 19$ ($\sim6\%$; \S4.5),
and likely very good to even fainter magnitudes.
Nevertheless, our contamination rate is not zero, and we now attempt to clarify how much
residual ``background'' level remains so that we can establish the radial
profile of Sculptor to the lowest possible densities.

In our previous papers (e.g., \citeauthor*{pII,pIV}), estimation of the
background to our RGB samples has been based on the assumption that, to
first order, the densities of MW halo giants and HB stars roughly follow
$R_{\rm GC}^{-3}$ laws, so that their contribution along a line of sight
is roughly flat in magnitude.  Under these circumstances, were one to shift
the RGB/RHB and BHB CMD selection criteria in magnitude (in the case of the
RGB/RHB samples, this is done through the CMD of stars already selected to
be giant candidates in the 2CD), the background contribution of these field
stars should be approximately constant.  For other classes of background
contaminant (e.g., metal-poor disk disk subdwarfs with an exponential 
density law, halo dwarfs which are close enough that heliocentric
distance is not a reasonable proxy for $R_{\rm GC}^{-3}$, and other
disk stars with large photometric errors) the
$R_{\rm GC}^{-3}$ assumption will be less correct, but the above method
provides a reasonable first order estimate of the background 
level.\footnote{Indeed, for some classes of contaminants --- e.g., blue, 
main sequence turnoff dwarfs from the Galactic disk and thick disk which
are not expected to lie in our Sculptor CMD selection region because of the
extremely large implied distances from the Galactic plane --- we will be
{\it overestimating} their representation by shifting the CMD-selection
region to brighter magnitudes.  This situation counteracts the shallowing 
of the $R_{\rm GC}^{-3}$ assumption for halo MSTO dwarfs.}
Table \ref{tab:cmdbg} gives the results of shifting our adopted CMD selection
regions in magnitude.  The brightest magnitude offset (i.e. largest
negative $\Delta M$ listed) is set by the saturation limit of the CCD
frames; shifting the region to fainter magnitudes is precluded because of problems
with survey incompleteness in this direction.
When determining a background estimate with this method, one must
``clear'' those small $\Delta M$ offsets where Sculptor stars themselves
contribute; for each sample in Table \ref{tab:cmdbg} the CMD offsets where
background stars dominate are those below the horizontal line.  Taking the
average counts below these lines and dividing by the total area covered by
each sample yields the estimated background number density,
$\Sigma_{\rm CMD}$, given in the final line of Table \ref{tab:cmdbg} along
with its Poissonian error.

A second method by which we can estimate the background is by determining
the number of ``Sculptor giants'' selected by our 2CD and CMD criteria but
at large distances from the Sculptor core.  Unfortunately, our Sculptor
radial surface density plots never converged to a constant background level
even as we continued to add spatial coverage at ever larger radii.  To
remedy this problem we observed four background ``control'' fields
significantly separated (by $5\arcdeg$) from the center of the dSph in each
of the cardinal directions (Figure \ref{fig:uncutfield}a).  At such remote
locations with respect to Sculptor (with $r_{\rm lim}$ estimated at
$76\farcm5$ by \citeauthor*{IH95}) it is less likely that we will see a
significant contribution from Sculptor stars.
 These four fields were taken
during the 23 October 2001 observing run (see Table \ref{tab:obs}) and are
of similar quality to our normal survey fields, though their magnitude
limits vary.  
Table \ref{tab:fieldbg}
gives the background densities in our complete list of giants (just the 2CD
selection) and in each of our magnitude limited samples for each 353.85
arcmin$^2$ control field; no data are given when the fields were insufficiently
deep for proper comparison with a particular, magnitude-limited sample.
  Confidence limits for a Poissonian probability
distribution are given at the 90\% and 68\% levels.  Unfortunately, three
of the four control fields were insufficiently deep in $T_2$ to
characterize accurately the background number density of our BHB samples.
The final column gives the average control field background level,
$\Sigma_{\rm CF}$.

Table \ref{tab:fieldbg} illustrates several things.  First, the background
densities derived by this independent method are also extremely low; only
0, 1, or 2 stars were found for any sample in any field. This reflects both
the efficacy of the Washington$+DDO51$ technique to throw out foreground
dwarfs as well as the high, $b = -83^{\circ}$ latitude of the Sculptor system,
which limits the number of MW contaminants
overall.  The 2CD giant selection throws out 95-98\% of the total stars in
the control fields and 98-100\% of the total is thrown out when both the
2CD and CMD selections are imposed.  Second, the background densities
estimated from the control fields are similar to those found with the
``CMD offset'' method.  Although the background densities derived with the
latter method are slightly smaller than those determined by the control
fields, they are well within the confidence limits (though we acknowledge
that the confidence limits on the control field densities are large
relative to the actual values because of the small number statistics from
which they are derived).  Finally, the background estimates obtained with
either the method illustrated in Table \ref{tab:cmdbg} or that in Table
\ref{tab:fieldbg} are consistent (within Poissonian errors) with the
spectroscopic determination of a small ($\sim$5\%) ``false positive" rate in
our RGB samples to either $M=19.0$ or $M=20.3$ (Table \ref{tab:pvcomp};
\S4.5).

\subsection{Radial Plots}

\subsubsection{Fitting Functions}

We fit our Sculptor density distribution with both a single component King
model and with a power-law with a flat core (PLC) model, the two functions
used by \citet{kleyna98} in their study of the Ursa Minor dSph.  The
fitted King functions define the stellar surface density,
$\Sigma_{\rm K}(Q)$, as
\begin{eqnarray}
\Sigma_{\rm K}(Q) = \left\{ \begin{array}{ll}k\left\{(1+Q^{2})^{-\frac{1}{2}} - \left [ 1 + \frac{r_{\rm lim}^{2}}{a^{2}(1-\varepsilon)}\right ]^{-\frac{1}{2}}\right\}^{2} + \Sigma_{b} &, ~{\rm if}  ~Q^{2} \le \frac{r_{\rm lim}^{2}}{a^{2}(1-\varepsilon)} \\
\Sigma_{b} &, ~{\rm otherwise}
\end{array} \right.  ~,
\label{eq:kingprof}
\end{eqnarray}
where
\begin{eqnarray}
Q^{2} &=& \frac{X^{2}}{a^{2}} + \frac{Y^{2}}{a^{2}(1-\varepsilon)^{2}}, \\
X &=&  (\xi - \xi_{0})\sin{\theta} + (\eta - \eta_{0})\cos{\theta}, \\
Y &=& -(\xi - \xi_{0})\cos{\theta} + (\eta - \eta_{0})\sin{\theta},
\end{eqnarray}
$k$ is the normalization constant, $r_{\rm lim}$ is the King limiting
radius, $a(1-\varepsilon)^{\frac{1}{2}}$ is the semi-major core radius,
$\varepsilon = (1-b/a)$ is the system ellipticity, $\theta$ is a rotation
angle on the sky (where the normally defined position angle, $PA$, is given
by $90-\theta$), $\xi$ and $\eta$ are the stellar coordinates deprojected
to the tangent plane to the sky, and $\Sigma _b$ is the fitted background
density.  The PLC model is given by
\begin{equation}
\Sigma_{\rm PLC}(Q) = \frac{M(\nu -1)}{\pi a^{2}(1-\varepsilon)(1+Q^{2})^{\nu}} + \Sigma_{b},
\label{eq:PLCprof}
\end{equation}
where $M$ is the total number of stars, $\nu>1$ is the power-law fall off,
and the other parameters carry the same definition as in the King model.
Best-fitting parameters were calculated using Bayesian and maximum
likelihood techniques while errors are estimated using a Bayesian approach
in conjunction with a Markov Chain technique.  The fitting algorithms are
explained in greater detail by \citet{ost02} and \citet{ost05}.

\subsubsection{Initial Profile Fitting Results}

The best-fitting King profile parameters to our different magnitude-limited
samples are given in Table \ref{tab:rpkingfits} along with previous
literature values for comparison.  The best-fitting PLC profile parameters
are given in Table \ref{tab:rpplcfits}.  Tables \ref{tab:rpkingfits} and
\ref{tab:rpplcfits} also give the error-weighted means of the model
parameters derived from the four different stellar samples analyzed.  Note
that the error on the quoted mean values will be somewhat under-estimated
because the different samples are not completely independent of one
another.  The best-fit model profiles are shown against the stellar
densities in Figure \ref{fig:rpbgfit}.  A formal model is not fit to the
$M\leq21.0$ sample because the center of the galaxy is not represented in
this sample.  Instead, this sample is modeled by the functional parameters
fit to the $M\leq20.3$ sample but scaled to the same average density of
stars as the $M\leq20.3$ sample in fields where both samples are
represented; the result is overplotted on the $M\leq20.3$ sample as open
circles.

Table \ref{tab:rpkingfits} shows the considerable agreement of all our
samples in each derived morphological parameter and their general
consistency with most of the literature values (though see \S5.2.4).
Figure \ref{fig:sclmaglimdist} shows the King limiting radius for the
various samples, and the differences are nearly imperceptible.  The
consistency between the model parameters found for the RGB/RHB and BHB
samples is especially gratifying given the differences in the selection
methodology used for each sample and the expected relative purity of the
BHB sample.  The only parameters with variations inconsistent with the
formal errors are the center coordinates.  However, given the flatness of
the surface density in the Sculptor core and our uneven sampling of
Sculptor with position angle, the profile center is not well constrained;
the observed $\sim1\farcm5$ center coordinate variations are, therefore,
not surprising.  To check for possible correlations between parameters
presumably more susceptible to the uneven $PA$ coverage of the data, the
fitting program was also run with the center coordinates, $PA$, and
$\varepsilon$ fixed to their tabulated mean values; no significant changes
in the derived $r_c$ and $r_{\rm lim}$ from the best-fit values in Table
\ref{tab:rpkingfits} were found.

\subsubsection{Background Level Effects on the Radial Profile}

Comparison of Tables \ref{tab:rpkingfits} and \ref{tab:rpplcfits} reveals
an important difference between the King and PLC fits: The model-estimated
backgrounds are significantly larger in the King fits than they are in the
PLC fits, and they are larger than those found by the background
estimations described in \S5.1.  The explanation lies in the mismatch
between the actual distribution of Sculptor stars and a King profile. This
difference is portrayed by (1) the distributions of stars in Figure
\ref{fig:sclmaglimdist}, which show an ``extratidal'' population with
declining radial density {\it outside} the best-fit King limiting radii,
and (2) by the poor fit of the King profile to the density distribution at
the largest radii in Figure \ref{fig:rpbgfit}.  The issue of whether a King
profile is {\it physically} meaningful for a dSph system
aside\footnote{Because the crossing time/two-body relaxation time for a
  dSph galaxy is long compared to globular clusters, it is not clear that
  a King profile, which was designed to describe these compact, higher
  density systems, {\it should} be an appropriate physical model for dSphs.
  We return to the physical interpretation of the Sculptor density profile
  in \S6.}, the adopted, one component King
function (Equation \ref{eq:kingprof}) simply is not {\it mathematically}
suited to accommodating the approximately $r^{-2}$ decline of the Sculptor
surface density at large radii.  When allowed to ``find'' the background on
its own, the King profile fit averages the ``extratidal'' Sculptor stars
into the mean background level.  On the other hand, the PLC fit, which is
{\it designed} to accommodate a power law decline in density, obtains
background estimates much closer to those estimated from the control fields
and the ``CMD offset'' method.  On the other hand, the PLC models fail to
account for the apparent inflection point, or ``break'', in the radial
profile near $60\arcmin$, where the slope of the profile apparently
changes.

To remedy this problem, we next attempt King function fits where the
backgrounds are fixed to our measured (\S5.1) values.  Since the only
measurable background density in the control fields is for the $M\leq20.3$
and $M\leq21.0$ samples, only for these do we use $\Sigma_{\rm CF}$ to
estimate the background; for the other samples we use $\Sigma_{\rm CMD}$.
We are then, in effect, always using the {\it larger} of the two 
background estimates.
As expected, however, the results are unsatisfactory in that they now
result in clearly exaggerated tidal radii (Figure \ref{fig:fixbgking}); in
attempting to compensate
for an observed density profile that
now clearly shows a significant density at large radii, the fitting code
must inflate $r_{\rm lim}$.

On the other hand, the density profile {\it within} the break radius
{\it is} well described by the initial King profile fit, which is largely
insensitive to the low density outskirts of our maps due to the overestimated
background level.  This is demonstrated
by Figure \ref{fig:allsamprp}, which compares the original King profile
fits to the the true density distributions accounting for the {\it
  measured} background levels.  The measured background levels are
subtracted from the points plotted and, as in Figure \ref{fig:rpbgfit}, are
shown to highlight where our data reach $S/B=1$.  In the case of the BHB
stars the $S/B=1$ limit is apparently {\it beyond the limits of our main
Sculptor survey}.  As Figure \ref{fig:allsamprp} demonstrates, the central
regions of the dSph are, in fact, well described by the King profiles in
Table \ref{tab:rpkingfits} down to approximately three orders of magnitude
below the central density; for this reason, we feel it is justified to
adopt the parameters found in Table \ref{tab:rpkingfits} as appropriate
descriptors for the central Sculptor populations.  However, a diffuse
``break'' population dominates the density beyond the King limiting radius.
Like other dSphs we have studied in a similar manner
(\citeauthor*{pII,pIV,pIX,pX}), Sculptor can be described by a
``King+Break'' (K+B) profile --- a characteristic seen in models of
tidal-disrupting satellite galaxies \citep[][see \S6]{johnston98, mayer01}.
This break is compared to a least-squares fitted and $\Sigma \propto r^{-2}$
power-laws in Figure \ref{fig:allsamprp}.

From Table \ref{tab:rpplcfits} we see that the PLC profile parameters are
fairly consistent except in the case of the $T_2\leq19.9$ sample where the
small sample size is apparently causing difficulty in the fit.  A notable
characteristic of the PLC fits is their $\nu \sim 2$ index, a value for
which the PLC model reduces to a projected Plummer model.  As we pointed
out above and show in Figures \ref{fig:rpbgfit} and \ref{fig:allsamprp},
while the PLC fits seem to give better matches to the data {\it over all
  radii}, they do not account 
for the profile break and increasingly
fail to match the measured density at large radii.  The $\nu \sim 2$
indices derived with the PLC fits correspond to $\sim r^{-4}$ density
declines --- faster than the $\sim r^{-2}$ declines that provide a better
fit to the outer densities (e.g., see left panels in Figure
\ref{fig:allsamprp}).  This particular failing of the PLC models is
illustrated, for example, by the fact that in the $T_2\leq19.9$ case, the
PLC-fit ``background'' is nearly an order of magnitude higher than that
calculated (Table \ref{tab:cmdbg}).

To put a more quantitative measure on the relative quality of the fits using
the different backgrounds and surface density profiles, we have calculated
a simple $\chi^2$ statistic\footnote{Note that Ostheimer's (2002) fitting
  routine does not minimize this $\chi^2$ to determine best-fitting sets of
  parameters.  The {\it a posteriori}-calculated $\chi^2$ statistic is used
  simply as a relative figure of merit to intercompare models.}
using the data presented in Figures
\ref{fig:rpbgfit} and \ref{fig:allsamprp}; the results are tabulated in Table
\ref{tab:fitchi}.  For the ``King+break'' (``K+B") profiles, the break population is modeled by
both the least-squares fitted and an adopted $r^{-2}$ power-law.  The starting radius
of the power law (given in Table \ref{tab:fitchi}),
$r_{\rm break}$, is that which yields the best
$\chi^2$ value;
the normalization is determined by the value of the model King profile at $r_{\rm break}$.  
In general, when compared to the simple King profile (without a break
component), the PLC model proves to be a better fit when using either the
model or measured
$\Sigma_b$.  However, in two out of the four cases determined using
the measured $\Sigma_b$, a K+B profile
provides a better fit to the data than does the PLC profile.  Moreover, the
K+B profile provides a better fit when compared to a simple King
profile {\it even when the inflated, model $\Sigma_b$ is used}; the one
exception is the $M\leq19.0$ sample, which has only one point that could be
confidently considered part of a break population.  The overall
consistency of $r_{\rm break}$ across all samples is encouraging given
that it is independently determined for each sample.

While providing comparable fits to the overall distribution, the failings
of the PLC models to match the observed features (density and inflection
point) at large radii are the primary reasons why,
in the end, we prefer the K+B
profile description for the Sculptor morphology.  But for either the PLC
or K+B description, a clear and relevant point is that Sculptor
extends to the edge of our main survey field ($150\arcmin$ from the
Sculptor center), within which there does not seem to be a
``Sculptor-free'', background region.

\subsubsection{Comparison to Previous Results}

Our derived King profile fit parameters (Table \ref{tab:rpkingfits}) --- in
particular $PA$, $\varepsilon$, and $r_c$ --- are in general agreement with
previously measured values for Sculptor.  The most important difference
appears to be that we consistently find King limiting radii near the upper
end of the previously reported range and twice that of the lower estimates,
most notably the recent CCD study by \citeauthor*{W03}.  Only the
\citet{e88b} value for the limiting radius is larger than ours.  We find a
large King limiting radius despite our adoption of the King parameters
derived with the artificially inflated, {\it model-fit} backgrounds.  As
found in previous attempts to fit the Sculptor radial profile (see \S1),
we also find that use of the {\it measured} background level increases our
derived King limiting radius and, also, reduces the quality of the fit at
smaller radii as shown in Figure \ref{fig:fixbgking}.  Moreover, we find a
sizable excess density {\it beyond} the King limiting radius when using
either the measured or the model-fit background.  As \citet{tol04}
have done, we have also spectroscopically confirmed Sculptor velocities for 
several of these outliers. 

The origin of the difference in our measured King radius compared to that
found in other studies is not likely due to differences in the depth of the
data.  Our actual imaging limit is comparable to that of all of the
previous studies, except the \citeauthor*{W03} study, which reaches
$\sim2.5$ magnitudes {\it deeper} than our data.  Instead, the difference
in model fits probably lies in the significantly better $S/B$ delivered by
our {\it methodology}, which allows us to explore the dSph to substantially
larger radius with much greater sensitivity.  All previous Sculptor surveys
employed simple statistical starcounts to map the dSph.  Figure
\ref{fig:bgdemon} demonstrates the significant improvement gained by our
approach: We compare the equivalent Sculptor radial $S/B$ (given by
$\Sigma_{\rm K}/\Sigma_{b}-1$) profiles delivered by our data when we (1)
adopt the traditional, simple starcounting approach to measure the radial
density (i.e., all stars detected are used to determine both
$\Sigma_{\rm K}$ and $\Sigma_{b} $), (2) mimic a ``CMD-filtered'' starcount
analysis\footnote{As discussed in \S1, this has become a popular method of
  mapping stellar systems to large radii
  \citep[see, e.g.,][]{gfiq95,oden01,oden02,rock02}.  One filters the
  starcounts to those sources that lie near the primary CMD locations
  occupied by stars in the stellar system.} whereby only stars meeting our
CMD criterion for RGB/RHB selection are used to measure the density, and
(3) adopt {\it both} our 2CD and CMD selection criteria to find only the
most reliable Sculptor giant candidates, as discussed in \S3.  In all
cases, we use the $M\leq20.3$ sample.  As may be seen, making use of
color-magnitude information to tune the starcount selection to the Sculptor
RGB/RHB does improve the $S/B$ of a survey by about a factor of $\sim4$
over simple star counting techniques (as used by, e.g., \citeauthor*{IH95}
and \citeauthor*{W03}).  But even this CMD filtering falls far short of the
nearly 200 times lower background achieved by using the two-criteria,
Washington$+DDO51$ technique to weed out foreground dwarf stars first.  The
radius at which the profile density sinks to the associated background is
$\sim20\arcmin$ using simple starcounts and $\sim40\arcmin$ using the CMD
selection for RGB/RHB stars, but expands to beyond $100\arcmin$ for the 2CD
{\it and} CMD-filtered, giant star candidate sample.  Note, because
Sculptor is situated near the SGP, it represents the {\it best} situation
with regard to foreground dwarf contamination; thus, use of the
Washington$+DDO51$ approach should have a greater impact on the study of
lower latitude systems.

The actual situation with regard to previous Sculptor studies is actually
worse than demonstrated in Figure \ref{fig:bgdemon}.  For example, in both
\citeauthor*{IH95} and \citeauthor*{W03} the authors did not discriminate
between stars and background galaxies, but counted {\it all} sources in
their survey.  However, as demonstrated by, e.g., \citet{rm93}, at
$M\gtrsim 20$ galaxies actually outnumber stars at the (North) Galactic
Pole by approximately 5:1.  In the Sculptor field near the SGP, the
background in our star counts profile becomes 2.1 or 1.6 times greater if
we make no image morphology cuts or do not select against likely galaxies
(large $\chi$ at faint $M$), respectively.  Were one to probe to deeper
magnitudes in search of more dSph-rich regions of the CMD (e.g., beyond the
MSTO), one might be thwarted by the fact that the galaxy contamination rate
not only rises faster than the contribution from field star contaminants,
but it rises faster than the typical luminosity function of the dSphs
themselves \citep[see, e.g., the luminosity functions for old stellar
populations summarized in Figure 5 of][]{rm93}.  Obviously, such
contamination will have its greatest impact on the lowest density regions
of the dSph, but even the core regions of the \citeauthor*{IH95} and
\citeauthor*{W03} studies have been significantly impacted by the lack of
galaxy pruning.  The quoted source counts of IH95 are about {\it an order
of magnitude} greater in the Sculptor core than we find from the sum of all
our stellar detections after removing morphologically-discriminated
galaxies. Yet the \citeauthor*{IH95} study only probes slightly fainter.
On the other hand, the \citeauthor*{W03} survey probes 2.5 mag deeper than
ours, so they would be expected to have a higher source density in the
core.  However, assuming the luminosity function of Sculptor is
approximately like that of a globular cluster \citep[see, e.g.,][]{e88a},
we find an expected factor of $\lesssim20$ in Sculptor stars gained by the
increased depth of \citeauthor*{W03}, which cannot account for the W03
factor of $\sim45$ greater central surface density compared to our
morphologically-discriminated counts in the Sculptor center.

We can also directly compare the $S/B$ of our Washington$+DDO51$ filtered
counts to those {\it adopted by} the \citeauthor*{IH95} and
\citeauthor*{W03} studies.  Using Table 3 from \citeauthor*{IH95} and
Figure 5 from \citeauthor*{W03}, we find peak values of $S/B\sim20$ and
$S/B\sim1000$ for these studies, respectively.  In contrast, we have a peak
$S/B\sim8000$ for our data (Figure \ref{fig:bgdemon}).  These $S/B$
comparisons are slightly misrepresentative because the backgrounds in these
previous surveys were actually calculated at smaller radii, where Sculptor
stars contribute.

Finally, it is worth pointing out that tracing Sculptor by its BHB stars
yields results with comparable $S/B$ to those from our Washington$+DDO51$
giant star mapping technique.  Given the expense of obtaining the
narrow-band $DDO51$ photometry, it would seem that simply mapping BHB stars
with broadband photometry may be a more effective means to explore the
extended structure of the Sculptor dSph.  However, we stress that our
long-term goal is not only to determine the {\it spatial} characteristics
of Sculptor and other dSph systems, but to derive their large-scale
{\it dynamics}, and giant stars are much more accessible spectroscopic
targets than are BHB stars at the same distance.  In this vein, Figure
\ref{fig:bgdemon} has important bearing on the prospects for spectroscopic
studies of the Sculptor dSph at very large radii: By a semi-major axis
radius of $\sim 200\arcmin$ (i.e., 2.5 King limiting radii), less than 5\%
of $M\leq20.3$ stars in the RGB part of the CMD are actually part of
Sculptor, which has a density of $\lesssim4$ deg$^{-2}$ (Figure
\ref{fig:allsamprp}).  Thus, even with wide field, multifiber spectroscopy,
it will not be possible to explore Sculptor RGB stars at such large radii
efficiently if the input sample is selected only as ``RGB-like'' stars in
the CMD.  (Note that while they are easier to find as ``needles in the
haystack'', BHB stars have a sky density an order of magnitude
{\it smaller} than that of the RGB, and, of course, they are fainter.)  It
is in this very low regime, where our relatively reliably identified
samples of $(M, T_2, DDO51)$-selected RGB stars are therefore particularly
valuable.

\subsection{Two-dimensional Structure}

Break populations are just one feature associated with satellites
undergoing tidal disruption.  Discernible tidal tails and internal
substructure (e.g., lumpiness and {\ttfamily S}-shapes) are others.
\citeauthor*{W03} point out the possibility of a tidal tail extending to
the southeast at a semi-major axis radius of $30\arcmin$ to $40\arcmin$
(note that this is ``extratidal'' for them, but well inside our newly
determined King limiting radius).  Moreover, both \citet{e88c} and
\citeauthor*{IH95} report an increasing Sculptor ellipticity with radius,
which is an effect one would observe if tidal arms were forming
\citep{jcg02}.

Our high $S/B$ Sculptor maps offer the opportunity to look for these
previously noted features.  Figure \ref{fig:all.cont} plots contours of the
inner regions of all but our $M\leq21.0$ sample.  We also over-plot the
$M\leq20.3$ and $T_2\leq19.9$ samples (Figure \ref{fig:all.cont}e) to show
the strong correspondence between the spatial distribution of these two
samples where the spatial coverage overlaps.  Finally, to give our most
complete representation of Sculptor, we have plotted contours derived from
the combined $M\leq20.3$ and $T_2\leq19.9$ samples (Figure
\ref{fig:all.cont}f).

Figure \ref{fig:all.cont} displays interesting aspects of the central
distribution of Sculptor stars.  An increase in the ellipticity of Sculptor
with radius is apparent from both the $M\leq19.0$ and $M\leq20.3$ contour
plots: While the dSph is more or less round within about $20\arcmin$, it
becomes extended in the east-west direction at larger radii, as has been
previously noted (e.g., \citeauthor*{IH95}).  Moreover, while these maps
show a slightly higher density toward the northeast in our RGB/RHB samples,
with none of our samples can we confirm a Sculptor extension to the
southeast as reported by \citeauthor*{W03}.  As one possible explanation
for the difference we draw attention to the comment made by
\citeauthor*{W03} that the interpretation of possible tidal extensions in
their data might have been influenced by the fact that their contour maps
made were made without benefit of star/galaxy discrimination in their
source catalog; we have already discussed (\S5.2.4) how significant a
contribution galaxies are likely to be to the \citeauthor*{W03} source
counts.  On the other hand, the Sculptor maps of \citeauthor*{IH95} (see
their Figure 1f) also suggest an extension of Sculptor to the northeast
albeit at lower significance than we see in our maps.  It is interesting
that this extension is along the direction of the Sculptor proper motion as
determined by \citet{schweitzer95}, as might be expected for tidal debris.

As a further demonstration of the increasing ellipticity with radius and to
assess the apparent northeast excess (which would imply isophotal
twisting), Figure \ref{fig:elong} maps the azimuthal trends of the mean
Sculptor density for different radial annuli for both the $M\leq19.0$ and
$M\leq20.3$ samples.  For this analysis, the radial annulus sectors are
shaped according to the ellipticity parameters derived from the mean fit to
the entire galaxy (line 5 in Table \ref{tab:rpkingfits}) and radial plots
were determined for azimuthal bins.  If the model ellipse is a perfect fit
at a particular semi-major axis radius, there should be no azimuthal trend
in density and all sectors should have the model surface density.  If, on
the other hand, the galaxy is {\it more} or {\it less} elliptical than the
model at a given radius, then the density will show variations from the
model: In the case of ellipticity changing at a fixed PA, one will find an
{\it increase} in the density over the model along the semi-major axis
$PA$s and {\it deficits} along the semi-minor axes, whereas if it is
{\it rounder} it will show the opposite trend.  Any isophotal twisting (as,
for example, one might see with the onset of well-defined tidal tails) will
be evidenced as azimuthal shifts of such features with radius.  The radial
bins in Figure \ref{fig:elong} are logarithmically and coarsely spaced to
improve the signal-to-noise of the points.  Our ability to interpret this
figure is complicated by the small area represented by annuli at short
radii and incomplete survey coverage at larger radii.  Nevertheless, Figure
\ref{fig:elong} seems to confirm the impression that the galaxy changes
shape from rounder than our global fitted shape to more elongated than our
global fitted shape between $r=22\farcm7$ and $r=40\farcm8$.  The data do
not show obvious signs of isophotal twisting, but are of insufficient
quality to do so reliably.

\subsection{Spatial Metallicity Variation}

The idea of a metallicity spread within Sculptor has long been recognized
from photometric studies of its giant branch,
\citep{kd77,nb78,sd83,dac84,e88a}, via studies of its RR Lyrae population
\citep{g93,kaluzny95}, and through spectroscopic analysis
\citep{tol01,shet03,tol03,tol04}.  \citet{m99} have argued that this spread
is really a {\it bimodal} population distribution described by populations
with [Fe/H] $\sim -2.2$ and $\sim -1.4$; it was suggested to possibly be
trimodal by \citet{grv94}.  Spatial gradients in the Sculptor populations
(whether due to a true gradient or as a change in mixing ratio of a bimodal
metallicity distribution) have also previously been explored
\citep{light88,dc96,hk99,m99,harbeck01,tol04}.  Given our larger area
coverage of Sculptor than in previous studies and the more reliable
catalog of giant stars we have created, it is worthwhile to re-evaluate
the spatial distributions of stars with apparently different metallicities.

Since we have no better indicator of intrinsic metallicity for the bulk of
our giants other than their position in the CMD, we divide our RGB
populations into ``metal-rich'' (red) and ``metal-poor'' (blue) halves by
fitting a fourth order polynomial to the center of the CMD locus of the RGB
as shown in Figure \ref{fig:splitcmd}.  This division into metallicity
halves is reasonable whether or not Sculptor's metallicity distribution
function is bimodal or more continuous.  Figure \ref{fig:cumdist}a shows
the cumulative radial distributions of these two RGB halves for our
$M\leq19.0$ sample, where the stars have been binned along the abscissa in
terms of their equivalent semi-major axis distance from the Sculptor
center.  To make these distributions, we have fixed the Sculptor center,
$PA$, and $\varepsilon$ to that of the mean of our best-fitted King models
(Table \ref{tab:rpkingfits}).  The separation of the blue and red RGB
distributions is clear, with the more metal-rich Sculptor population being
more centrally concentrated (as has been found by previous studies).  The
difference in these two distributions is quantified using a
Kolmogorov-Smirnov (K-S) test and the results are given in Table
\ref{tab:ks}.  This table provides the effective number of data points
($N_e$), the K-S statistic ($D$), and the K-S significance level (KS\%), as
calculated by the algorithms provided by \citet{nr}.  The results of this
test show a strong disagreement with the null hypothesis that these two
spatial distributions are the same.

We repeat the analysis with deeper probes of the RGB, but here we meet with
a population mixing problem due to contamination of the ``blue'' RGB by the
Sculptor RHB.  The RHB population properly belongs to the {\it red}
(metal-rich) RGB \citep[e.g.,][]{m99}, and therefore, a simple RGB division
will tend to {\it wash out} the difference in the distributions between the
red and blue populations.  We have removed the CMD regions of our
$M\leq20.3$ sample that are clearly dominated by RHB stars, but have left
in the sample that part of the CMD where the RHB and the blue RGB overlap
(compare Figures \ref{fig:cmd2} and \ref{fig:splitcmd}).  This residual RHB
contamination of the blue RGB sample will cause a modest dilution of the
red/blue spatial distinction.  Indeed, while we find the distributions of
the blue and red stars in Figure \ref{fig:cumdist}b to be significantly
different, they are slightly less so than in the $M\leq19.0$ case (note the
difference in K-S probabilities in Table \ref{tab:ks}).

Obviously, we expect the BHB sample, which belongs to the more metal-poor
Sculptor population, to track the spatial distribution of the blue RGB
sample.  We test this by comparing the $T_2\leq20.3$ sample to the two
$M\leq20.3$ samples in Figure \ref{fig:cumdist}c.  To do this fairly, we
have excluded sky regions of the $M\leq20.3$ samples that are not included
in the $T_2\leq20.3$ sample.  It is hard to distinguish between the BHB and
blue RGB samples, as might be expected (Figure \ref{fig:cumdist}c), and the
distributions {\it are identical} at the 99\% level as given by the K-S
test.  On the other hand, the BHB distribution is significantly different
than that of the red RGB sample.

To further illustrate their different distributions, the blue and red RGB
distributions have been individually fit with King model profiles.  The
best-fitting parameters are given in Table \ref{tab:rpsplitfits}, the
individual radial profiles illustrated in Figure \ref{fig:splitrp} and the
sky distributions shown in Figure \ref{fig:sclsplitdist} together with
their best-fitting King limiting ellipses.  The background,
$\Sigma_{\rm CMD}$, for each radial profile has been determined using the
``CMD offset'' method for each half of the RGB.  Again, the greater
concentration of the metal-rich population is apparent by comparing, for
example, the differences in the core/limiting radii in Table
\ref{tab:rpsplitfits}.  However, our new, wide area data now show this
spatial difference to be maintained to the nominal $80\arcmin$ limiting
radius of the system and beyond.  While both metallicity populations
apparently contribute to the ``break'' population beyond this radius, the
metal-poor population density is greater than the metal-rich population
there by an order of magnitude.  The radial gradient in the make-up of
Sculptor is also demonstrated in Figure \ref{fig:splitrp} by the variation
of the fractional population balance parameter, $(B-R)/(B+R)$, where $B$
and $R$ are the number of blue and red RGB stars within an elliptical
annulus.  Interestingly, the small east-west extension of the metal-rich
population as seen in Figure \ref{fig:sclsplitdist} leads to the finding of
a {\it greater} ellipticity compared to that for the metal-poor population.

\section{SUMMARY AND DISCUSSION}

We have undertaken the most sensitive morphological survey of the Sculptor
dSph system to date.  Our ability to increase the signal-to-noise of the
measured Sculptor structure is due to our orders of magnitude improved
signal-to-background, $S/B$, over that in previous work.  By using the
Washington$+DDO51$ photometric system to create maps of Sculptor giant
stars, and by taking advantage of the strong contrast of Sculptor BHB stars
to Galactic field stars of the same color, we are able to reduce substantially the
backgrounds, $B$, in our maps and thereby track the radial
profile of very high probability Sculptor stars to the largest angular
extents ($\sim150\arcmin$) explored for this system thus far.  At the edge
of our survey area, the primary limit on our ability to resolve detailed
{\it two-dimensional} structure on the scale of tens of arcminutes is the
quantization noise of individual BHB and RGB sources at low densities.

These data have been applied to a variety of analyses to obtain the
following results:

\begin{enumerate}

\item Through spectroscopic follow-up of 157 stars we verify that our
  Washington$+DDO51$ methodology to select candidate Sculptor RGB stars
  actually produces true Sculptor members with a 94\% reliability rate.
  Thus our Sculptor RGB candidates can be trusted to give an accurate
  tracing of the structure of the Sculptor system.  Admittedly, this
  reliability rate comes at some expense in terms of completeness, and we
  have identified some radial velocity members of Sculptor with
  color-color-magnitude combinations that place these stars {\it just
    outside} our conservative selection criteria.
 
\item The Sculptor structural parameters are derived anew. Both simple King
  models and Power Law + Core (PLC) models provide unsatisfactory
  descriptions of the full Sculptor density distribution over the full
  range of our survey.  Neither can properly account for a ``break'' to a
  shallower, approximately $r^{-2}$ radial density fall-off near a
  semi-major axis radius of $\sim60\arcmin$.  On the other hand, the
  Sculptor radial profile is well-fit by a King profile {\it within} this
  radius, so that, as we have found for other dSph systems we have explored
  \citep[\citeauthor*{pII,pIV,pX}; ][]{sgr1}, Sculptor is well-described
  by a ``King + Power Law'' function.  Once adopting this prescription to
  account for the extended break population around Sculptor, we find
  central King model parameters that are within the range of those found
  previously by other studies, but an $r_{\rm lim}= 79\farcm6\pm3\farcm3$
  King limiting radius that is at the high end of those previously
  reported; most notably, our limiting radius is nearly twice that found by
  the recent, deep CCD starcount survey by \citeauthor*{W03}.  We argue
  that such discrepancies arise from substantially higher contamination by
  both foreground stars and background galaxies in the previous surveys,
  which limited their effective signal-to-noise, particularly in the more
  diffuse outer regions of Sculptor.  As a cautionary measure, we assess
  the dependence of our own derived King radius on adopted background.  We
  consistently find $r_{\rm lim}\approx 80\arcmin$, despite allowing our
  fitting program to use an inflated background that is at least a factor
  of two, and up to an order of magnitude, larger than the background
  {\it measured} in control fields $5\arcdeg$ from the Sculptor center.
  
\item We confirm the existence of an extended component to the Sculptor
  system beyond the $\sim80\arcmin$ King limiting radius, and map this
  break population to the $150\arcmin$ extent of our main survey area.  An
  exorbitantly large background density would need to be adopted to
  eliminate this excess population at large radii.  Instead, two
  independent methods used to measure our background levels agree that they
  are very low, and demonstrate that we detect the Sculptor break
  population at a $S/B\sim3$ at the $150\arcmin$ limit of our survey, and
  with $S/B\sim10-40$ (depending on which sample of stars we use) near the
  break radius.  Given this degree of significance, there can be no doubt
  that this break population is real.  Moreover, the break population is
  seen with both RGB/RHB {\it and} BHB stars, and we have 
  at least three (see \S4.6) stars in
  the break population spectroscopically confirmed \citep[][ report an
  additional two]{tol04}.  We stress this point
  because our previous detection (\citeauthor*{pII}) of a break population
  around the Carina dSph using the same methodology has been questioned by
  \citet{morrison01} and \citeauthor*{W03} (but has been shown to be real
  in \citeauthor*{pVI} and again in \citeauthor*{Mu05b}).

\item The contrast of Sculptor RGB stars in its break population with
  respect to other field stars at the same color and magnitude are so low
  that simple strategies to pick such stars out of the CMD for
  spectroscopic follow-up will result in relatively low rates for
  uncovering actual Sculptor giants --- for example, less than 5\% at 2.5
  King radii.  Plowing through large numbers of stars with multifiber
  spectroscopy is not necessarily the best solution to probing the dynamics
  of Sculptor RGB stars at large radii, since the density of $M\leq20.3$
  RGB stars is $\lesssim4\,{\rm deg^{-2}}$ at 2.5 King limiting radii.  In
  contrast, at the same radius our Washington$+DDO51$ Sculptor RGB sample
  is expected to increase the success rate for finding {\it bona fide}
  Sculptor stars by more than an order of magnitude, and at a density
  efficiently probed with single slit spectroscopy.  We hope to undertake
  this type of follow-up work, and have presented a successful proof of
  concept with results using the Magellan telescope and MIKE spectrograph
  here.
  
\item The two-dimensional structure of Sculptor shows evidence for a change
  in ellipticity with radius, from rounder to more elliptical, as
  previously reported by \citet{e88c} and \citeauthor*{IH95}.  Our deeper
  data sets also show an extension from Sculptor to the northeast along the
  direction of the \citet{schweitzer95} proper motion vector.  This
  feature, lying between $20\arcmin$ and $40\arcmin$, is also seen at
  lower significance in the contour plot by \citeauthor*{IH95}.  We do not,
  however, confirm the extended feature to the southeast reported by
  \citeauthor*{W03}.
 
\item Preliminary dynamical results using a total of 134 Sculptor stars
  ranging out to $\sim0.6$ King limiting radii show a mean Sculptor
  velocity of $\langle v_{\rm hel}\rangle=110.43\pm0.79\,{\rm km\,s^{-1}}$,
  a global velocity dispersion of $\sigma_v=8.8\pm0.6\,{\rm km\,s^{-1}}$,
  and no evidence for rotation.  The radial trend of the velocity dispersion
  for our Sculptor stars
  is found to be relatively flat, however, as found in other dSph systems
  (e.g., \citeauthor*{Mu05b}), Sculptor may show a rise in velocity
  dispersion beyond $\sim0.4r_{\rm lim}$.  In \citeauthor*{Mu05b} we show how
  such a profile is one signature of tidally disrupting satellites on very
  radial orbits.
  
\item When the Sculptor RGB is divided into red and blue halves,
  representing predominantly more metal-rich and more metal-poor Sculptor
  populations, a clear difference in spatial distributions is observed.
  The more metal-rich giant stars are more centrally concentrated and,
  interestingly, more elliptically-distributed, than the metal-poor giants.
  The distribution of BHB stars closely follows the distribution of blue
  RGB (metal-poor) stars, as expected.  These metallicity-dependent
  differences in spatial distribution echo previous findings
  \citep*{light88,dc96,hk99,m99,harbeck01,tol04}.

\end{enumerate}

Spatial and dynamical characteristics of the Sculptor system reported here
(e.g., the clearly detected break population, signs of increasing
ellipticity with radius, and a rising/flat velocity dispersion profile) are
classic signatures of a dwarf satellite undergoing tidal disruption.  In
this interpretation, the Galactic gravitational tides are stretching
Sculptor at large radii internal to tidal radius, and stripping stars that
extend into a break population beyond the nominal King limiting radius
(which might then actually correspond to the tidal radius of the system).
K+B density profiles are consistent with model expectations for
the appearance of a tidally disrupting dwarf galaxy in a MW-like potential
\citep{johnston98,mayer01}, where the break population represents unbound
stars.  Unbound stars at all radii can inflate the velocity dispersion and
lead to the observed flat velocity dispersion profile
\citep[e.g.,][ \citeauthor*{Mu05b}]{kroupa97}.

On the other hand, breaks {\it within} the tidal radius of a disrupting
system have also been found in some N-body models explored by
\citet*{jcg02}.  Thus, Sculptor could be tidally disrupting, and still have
the observed break population be bound.  On the other hand, if one
calculates where the tidal radius of Sculptor should be from its mass
relative to that of the MW one finds that it is close to, or less than, the
observed King limiting radius assuming a Sculptor mass of 1.5$\times 10^7$
M$_{\sun}$, consistent with an assumed $M/L = 10.9$ in solar units
(IH95)\footnote{Using the current Sculptor Galactocentric distance as an
  approximate orbital semi-major axis distance
  \citep[$a=79\pm4\,{\rm kpc}$; ][]{mateo98} and assuming a low orbital
  eccentricity ($e\lesssim.1$), Equation 1 by \citet{burk97} yields a
  tidal radius $r_t\sim65\arcmin$ for a Sculptor mass of 1.5$\times 10^7$
  M$_{\sun}$.  A higher eccentricity ($e\sim0.5$) orbit yields a smaller
  tidal radius, $r_t\sim40\arcmin$.  The \citet{schweitzer95} orbit
  determined for Sculptor implies $e=0.25$, between these two values.}.  A
characteristic of {\it all} of the numerous orbiting satellite models
presented in the \citet*{jcg02} study is an increase in the isophotal
ellipticity as a function of radius.  This increase in ellipticity is
greater for those satellites on more circular orbits.  \citet*{jcg02} give
observational signatures that help in determining the physical process
behind the development of the structures seen in their models.  They define
the radius where the ellipticity of an originally circular model satellite
becomes $>0.02$ as $r_{\rm distort}$ and the innermost radius at which the
change in the slope of the density power-law becomes $>0.2$ as
$r_{\rm break}$.  Their simulations show that if
$r_{\rm break}/r_{\rm distort}>1$ and the slope of the power law beyond
$r_{\rm break}$ ($\gamma_{xt}$) is greater than $-3$, the likely cause of
the distortions and break population is tidal stripping.  We estimate
$r_{\rm distort}\sim30\arcmin$ from Figures \ref{fig:all.cont}b and
\ref{fig:elong} and $r_{\rm break}\sim60\arcmin$ and $\gamma_{xt}=-2$ from
Figure \ref{fig:allsamprp}.  These numbers satisfy the above conditions for
tidal stripping set by the simulations.

\begin{figure}
\begin{center}
\plotone{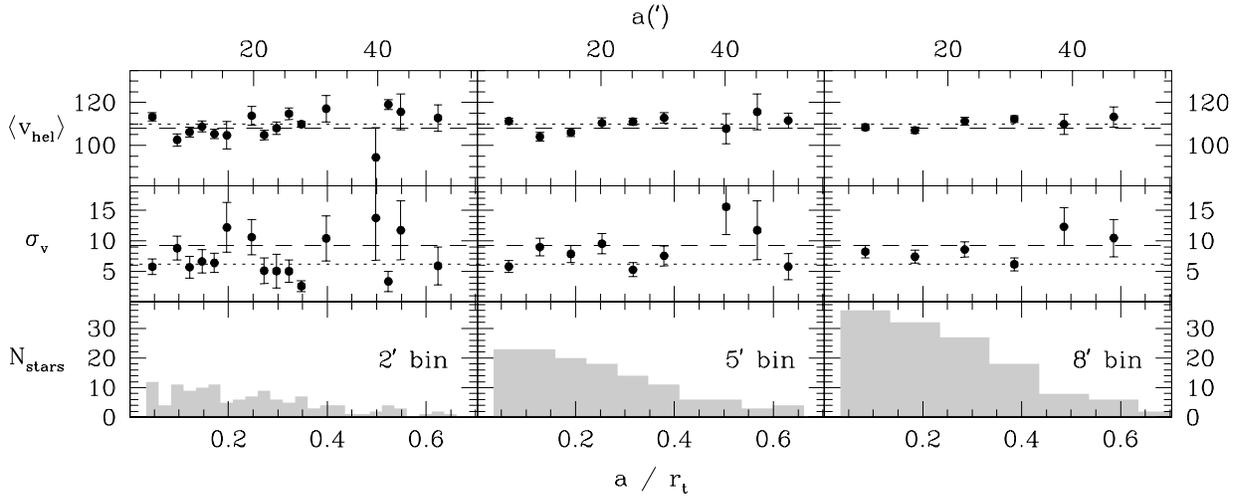} 
\caption{The azimuthally averaged $v_{\rm hel}$ (${\rm km\,s^{-1}}$;
  {\it top row}), $\sigma_v$ (${\rm km\,s^{-1}}$; {\it middle row}), and
  number of stars per bin, $N_{\rm stars}$, ({\it bottom row}) profiles.
  Points along the abscissa are marked both in terms of the semi-major axis
  distance, $a$, and the ratio of this distance to the mean of the derived
  King limiting radius found in \S5.2. The results are given for three
  different bin sizes: $2\arcmin$ ({\it left column}), $5\arcmin$
  ({\it center column}), and $8\arcmin$ ({\it right column}).  The values
  of $v_{\rm sys}$ and the global $\sigma_v$ as derived by
  \citeauthor*{QDP} ({\it dotted line}) and this study ({\it dashed line})
  are also shown for reference.  Note the value determined by
  \citeauthor*{QDP} was determined from stars with $a\lesssim10\arcmin$;
  however, we extend the line here for comparison at all radii.}
\label{fig:rvprof}
\end{center}
\end{figure}

\begin{figure}
\begin{center}
\plotone{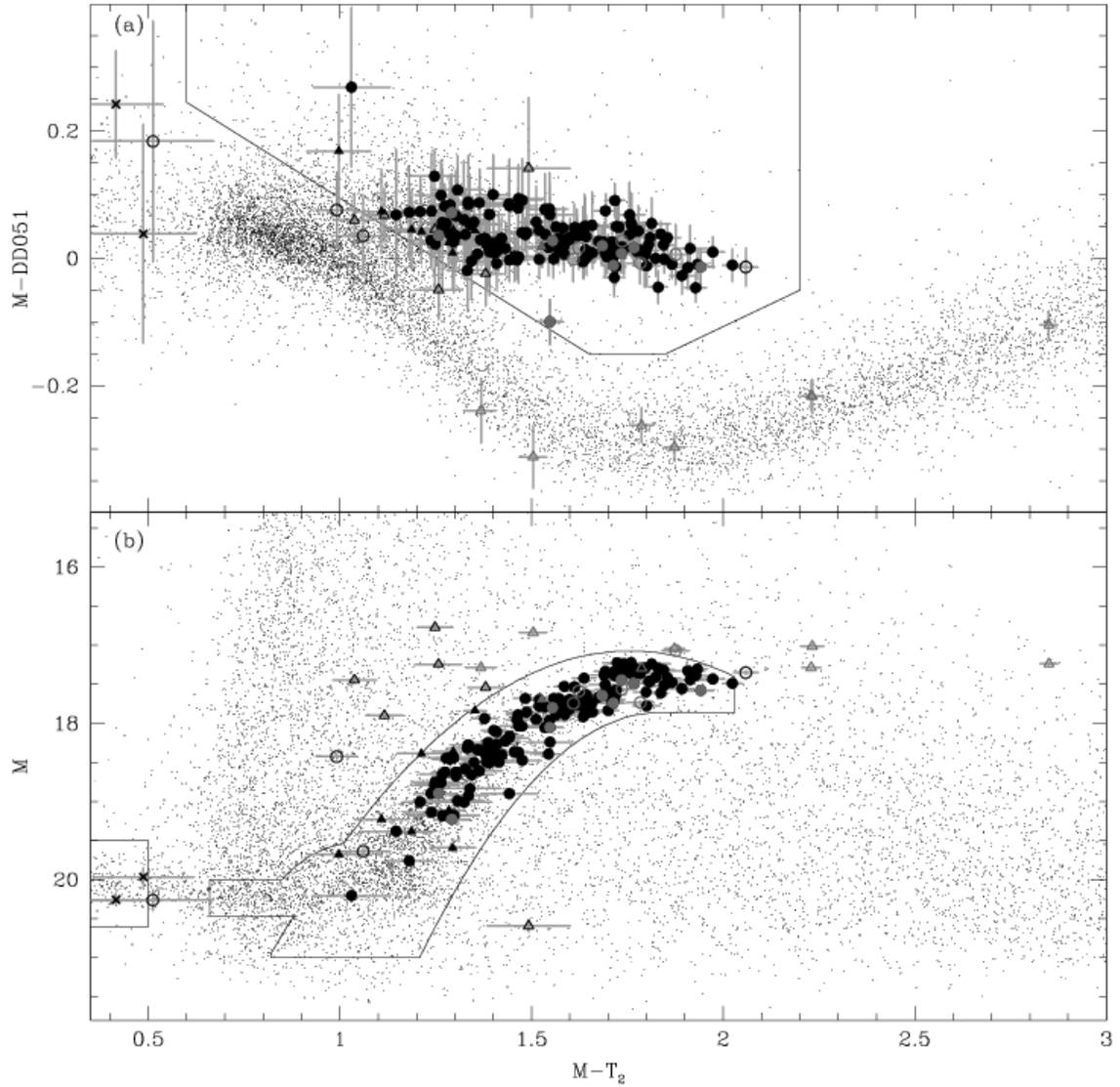} 
\caption{The distribution of all stars passing our photometric error and
  morphological limits in (a) the 2CD and (b) the CMD.  The stars that
  have been observed spectroscopically are marked in the same way as
  Figure \ref{fig:rvdist} and include error bars ({\it grey}).  See Table
  \ref{tab:pvcomp} for a summary of the accuracy of our photometric
  selection technique as evidenced by our spectroscopic data. }
\label{fig:rvselect}
\end{center}
\end{figure}

\begin{figure}
\begin{center}
\plotone{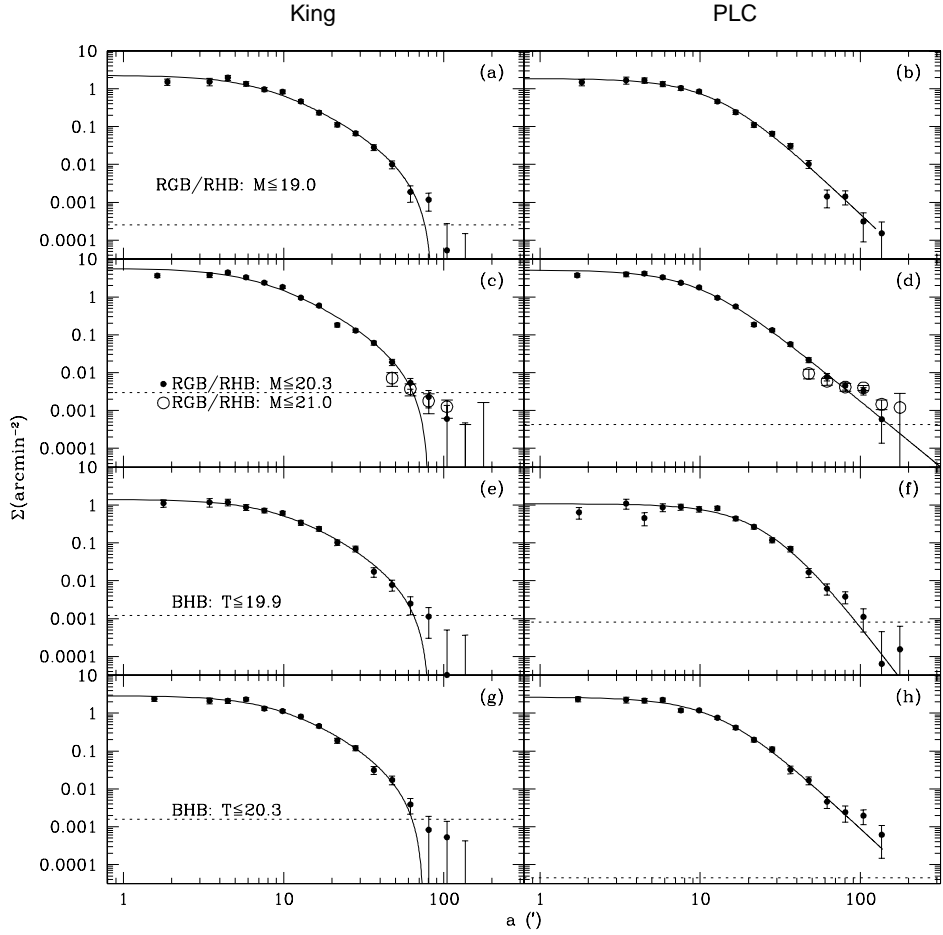} 
\caption{Initial radial surface density fits using the King ({\it left})
  and PLC ({\it right}) models.  The fits are shown for each sample: (a)
  and (b), the $M\leq19.0$ sample; (c) and (d), the $M\leq20.3$ and
  $M\leq21.0$ samples; (e) and (f), the $T_2\leq19.9$ sample; and (g) and
  (h), the $T_2\leq20.3$ sample.  The abscissa is given in units of the
  semi-major axis. The best-fitting models ({\it solid lines}) are
  overplotted on the resulting observed density points, which have error
  bars based on Poissonian statistics.  Since our $M\leq21.0$ sample did
  not contain the central parts of the galaxy, this sample is fit using
  model parameters derived for our $M\leq20.3$ sample, scaled to densities
  appropriate for comparison (overplotted in panels (c) and (d) as open
  circles).  (In detail, the $M\leq21.0$ data points are scaled by the
  ratio of the number of stars in this sample to the number of stars in the
  $M\leq20.3$ sample limited to the area covered by both samples.)  The
  background densities {\it as determined by our fitting program} are
  subtracted from the observed density and overplotted ({\it dotted line})
  in each panel to show where the $S/B=1$.  The background for the
  $M\leq21.0$ is taken to be the same as the $M\leq20.3$ background, again
  scaled by the factor defined above (they are, therefore, equivalent on
  the plot).  Note the significantly lower backgrounds determined by the
  PLC fits, whereas the King fits tend to place the background levels
  higher than the density of the break population.}
\label{fig:rpbgfit}
\end{center}
\end{figure}

\begin{figure}
\begin{center}
\plotone{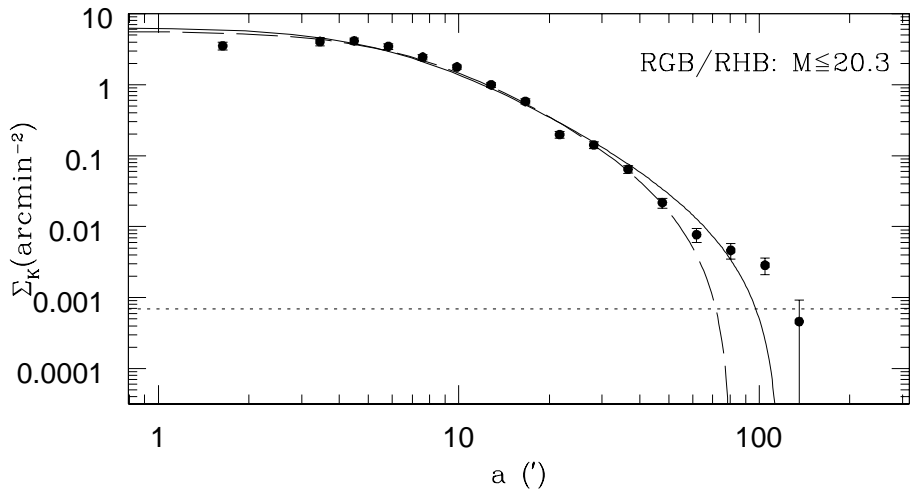} 
\caption{A demonstration of an attempt to fit our observed density
  distributions when fixing the background to the observed value.  We plot
  the observed density distribution of the $M\leq20.3$ sample that has been
  background subtracted according to the background density seen in our
  control fields ({\it dotted line}).  The initial fit to this profile is
  shown ({\it dashed line}) along with the model derived after fixing the
  background density to our observed value ({\it solid line}).  All
  parameters are nearly identical to those derived in the initial fit
  except for the King limiting radius which increases from $r_{\rm lim}
  \sim80\arcmin$ to $r_{\rm lim}\sim115\arcmin$.}
\label{fig:fixbgking}
\end{center}
\end{figure}

\begin{figure}
\begin{center}
\plotone{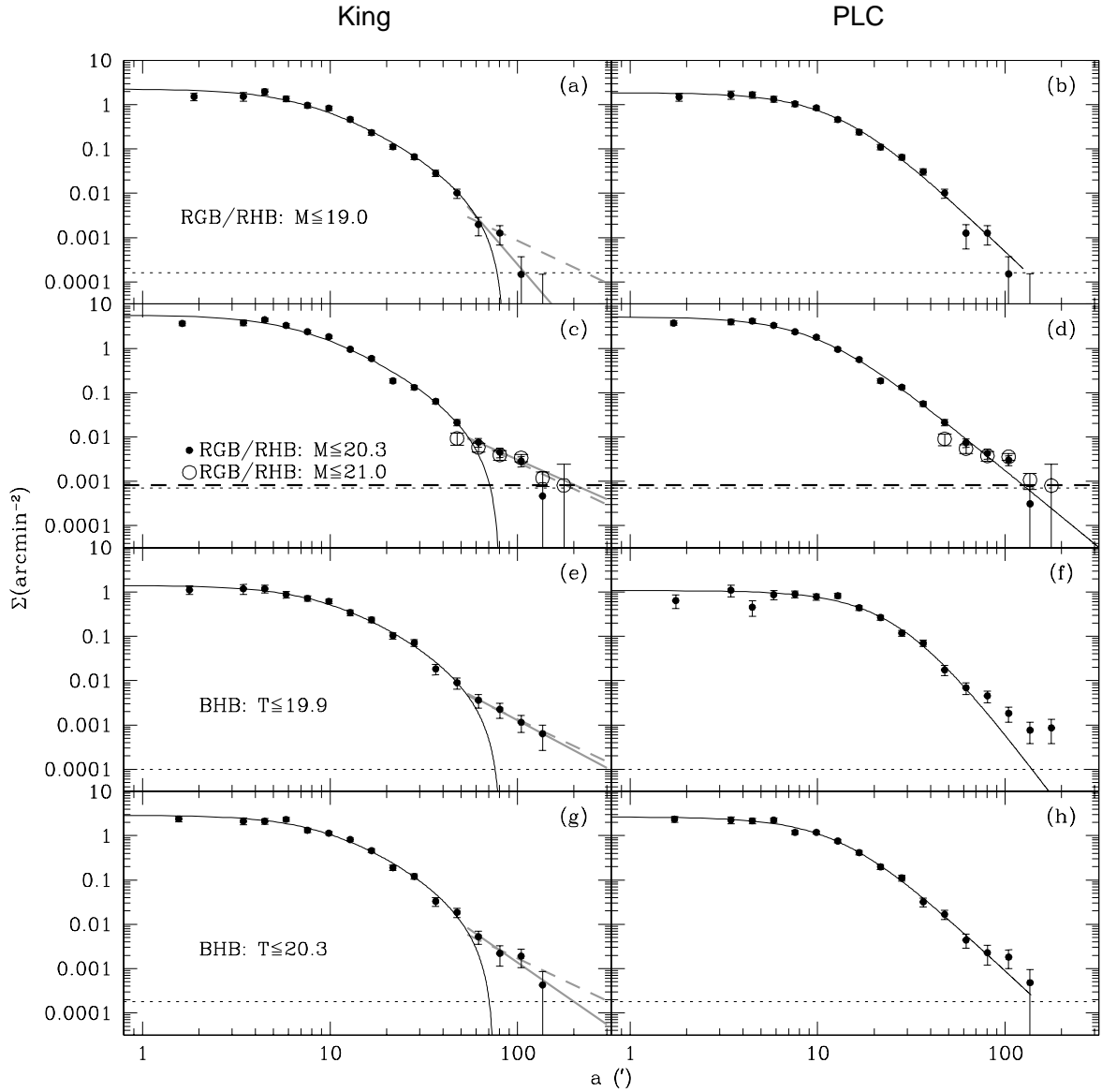} 
\caption{The same as Figure \ref{fig:rpbgfit} except now the initial fits,
  which characterize the {\it inner} parts of Sculptor well, are compared
  to data corrected by our independently determined background levels.  The
  background used is the larger of either $\Sigma_{\rm CF}$ or
  $\Sigma_{\rm CMD}$.
  Since we are now using backgrounds that are different between the
  $M\leq20.3$ and $M\leq21.0$ samples, both background levels are plotted
  in panels (c) and (d) (with {\it dotted} and {\it dashed lines},
  respectively).  Points considered to be part of the ``break'' population
  are modeled with a least-squares fit power law ({\it solid grey line}) in the
  left panels.  The power laws are $\Sigma\propto r^{-4.9}$, $r^{-1.9}$,
  $r^{-2.3}$, and $r^{-2.9}$ in panels (a), (c), (e), and (g),
  respectively.  We also show a simple $\Sigma\propto r^{-2}$ law
  ({\it dashed grey line}) for comparison.  Goodness-of-fit values useful
  in comparing these fits are given in Table \ref{tab:fitchi}.}
\label{fig:allsamprp}
\end{center}
\end{figure}

\begin{figure}
\begin{center}
\plotone{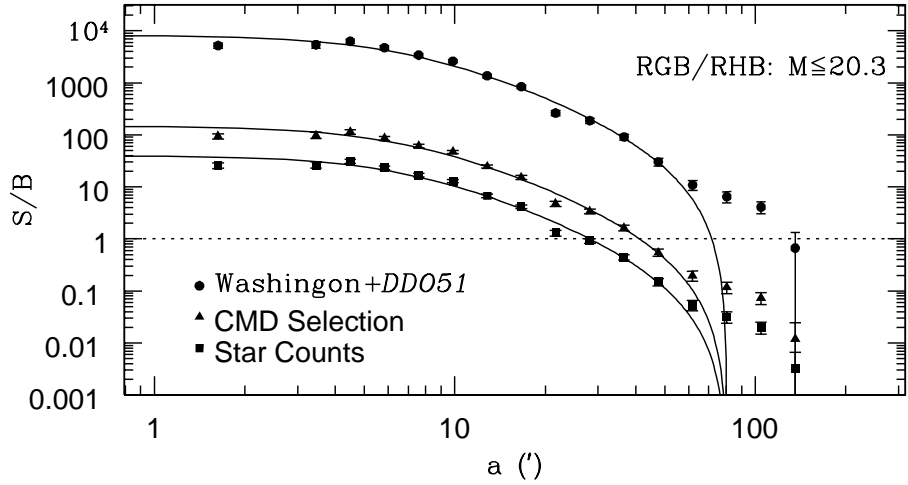} 
\caption{A comparison of $S/B$ for various strategies implemented to
  analyze our photometric catalog, including traditional starcounts
  (i.e., all stars within our error limits after removing galaxies;
  {\it squares}); a CMD-filtered scheme (making use of our CMD selection
  region from Figure \ref{fig:cmd2}; {\it triangles}); and our complete
  CMD and 2CD selection using Washington$+DDO51$ photometry
  ({\it circles}).  The results for our $M\leq20.3$ sample presented in
  Figure \ref{fig:allsamprp} are normalized to the expected $S/B$ for each
  of these three cases.  This normalization is estimated by taking the
  ratio of the counts in the core of the dSph ($a<3\arcmin$) to those
  expected to be due to a background as measured by our control fields.
  Both the points and the model ({\it solid line}) are normalized by this
  ratio.  As may be seen, the $S/B$ is improved by 0.6 order of magnitude
  in modifying the technique from simple starcounts to CMD-filtered
  starcounts, and improved by another 1.7 orders of magnitude in moving
  from CMD-filtered starcounts to our Washington$+DDO51$, CMD+2CD selection
  (\S3).}
\label{fig:bgdemon}
\end{center}
\end{figure}

\begin{figure}
\begin{center}
\plotone{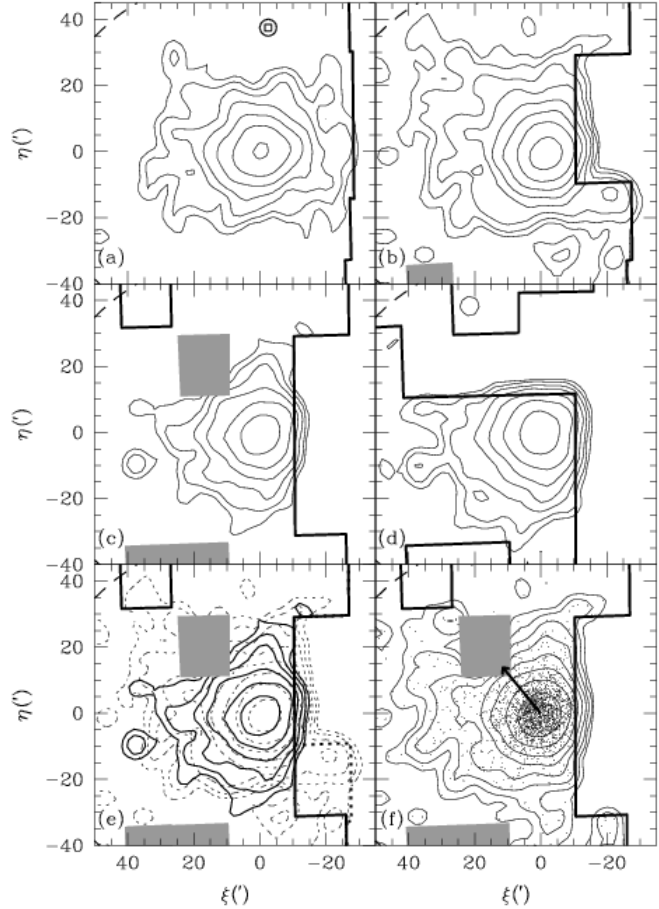} 
\caption{Contour plots of the (a) $M\leq19.0$, (b) $M\leq20.3$, (c)
  $T_2\leq19.9$, and (d) $T_2\leq20.3$ samples in the central regions of
  Sculptor.  A section of the ellipse associated with the derived mean
  King limiting radius is also shown ({\it dashed line to upper left}).
  Panel (e) overlays the $M\leq20.3$ ({\it dotted lines}) and $T_2\leq19.9$
  ({\it solid lines}) contours to show the strong correspondence between
  these two distributions.  Panel (f) combines the two samples in (e) to
  give our most complete two-dimensional representation of Sculptor.  In
  (f), we also plot the proper motion vector determined by
  \citet{schweitzer95}, which has a position angle of $\sim40\arcdeg$.  The
  error in the position angle of the proper motion is indicated by the grey
  sector at the base of the arrow.  Sections of the samples that are
  omitted due to insufficient
  depth are displayed with grey boxes as in Figure \ref{fig:sclmaglimdist}.
  For all contours, individual stars have been smoothed by a
  unit-normalized Gaussian kernel with $\sigma=2\farcm5$, truncated at
  3$\sigma$.  The one $\sigma$ extent of this kernel and the sampling bin
  ($2\arcmin\times2\arcmin$) size are shown at the top of (a).  The
  relative density levels are the same for each panel: 1, 2, 4, 8, 16,
  32, 64 and 128.  For a sense of what this translates to in terms of the
  actual star sample, the stars used to create the contours in (f) are
  over-plotted.}
\label{fig:all.cont}
\end{center}
\end{figure}

\begin{figure}
\begin{center}
\plotone{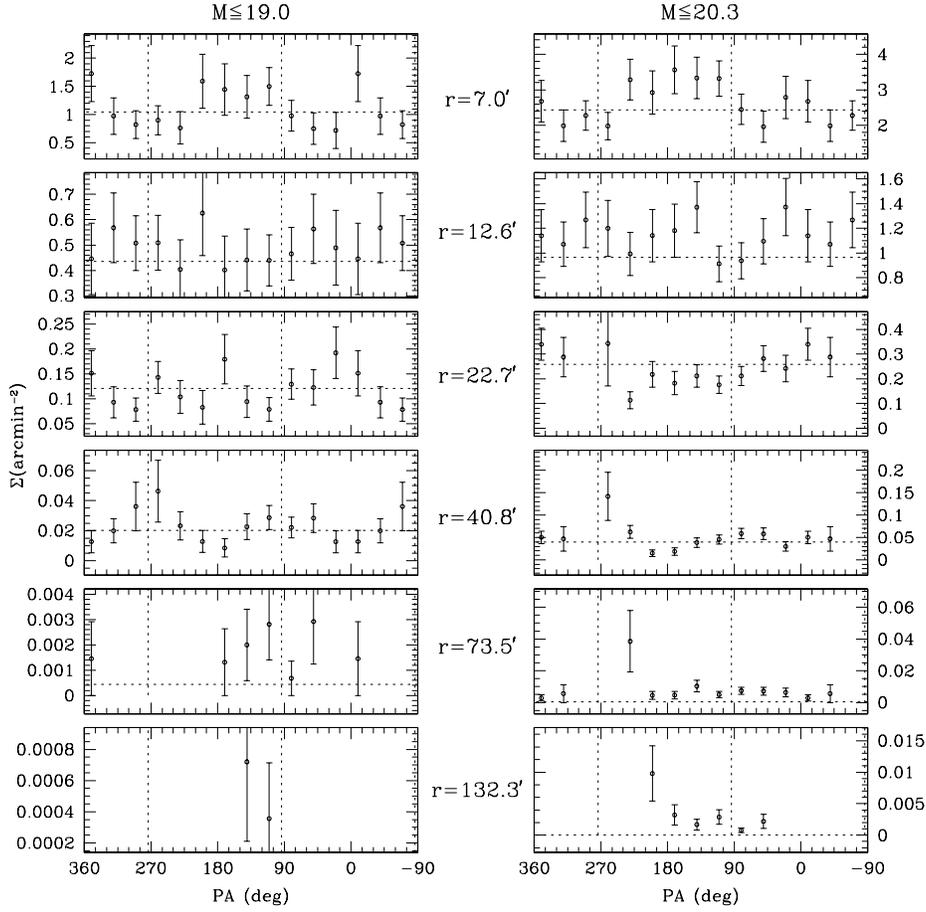} 
\caption{Measures of the variation in the surface density with azimuth in
  the $M\leq19.0$ ({\it left}) and $M\leq20.3$ ({\it right}) samples as a
  function of semi-major axis radius.  The marked radii are the bin centers
  just as in previous radial plots.  The {\it horizontal dotted line} shows
  for each elliptical annulus the density for that annulus derived from the
  global King profile fit to all radii.  Were there no variation in
  isophote shape with radius, the densities of individual annular sectors
  would be constant at all radii, and this constant value should match
  that of the global fit, to the degree that the fitted radial profile is a
  good match to the observed radial profile.  Changes in isophote shape
  from the nominal Sculptor ellipticity (line 5 of Table
  \ref{tab:rpkingfits}) will show up as systematic deviations in the
  densities of annular sectors at any radius.  The tendency for systematic
  deficits in density near the semi-major axes ({\it vertical dotted
    lines}) and larger densities near the semi-minor axes seen at smaller
  radii indicate that the central Sculptor profile is rounder than the
  nominal ellipticity from the global fit.  On the other hand, the
  opposite trend observed at larger radii (e.g., at $r=40\farcm8$) suggests
  that the outer contours of Sculptor are more elliptical than the global
  King profile fit, with the ellipticity along the same $PA$ as the global
  fit.  No strong evidence for isophotal twisting is seen, but incomplete
  survey coverage at intermediate and large radii leaves great uncertainty
  in this conclusion.}
\label{fig:elong}
\end{center}
\end{figure}

\begin{figure}
\begin{center}
\plotone{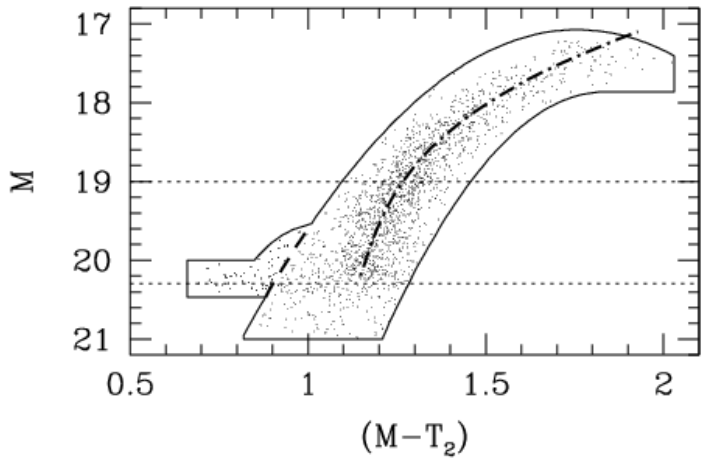} 
\caption{Demonstration of our division of the Sculptor RGB into
  ``metal-poor'' (blue) and ``metal-rich'' (red) halves.  A fourth-order
  polynomial has been fit to the distribution of stars within our RGB/RHB
  selection ({\it dot-dashed line}).  This division of the RGB into blue
  and red parts is done only for the $M\leq19.9$ and $M\leq20.3$ samples
  whose limits are shown by dotted lines.  Because the RHB population is
  expected to track the spatial distribution of the red half of the RGB, we
  trim (with the {\it dashed line}) the RHB-dominated part of our selection
  region away from the blue half of the RGB.  This will reduce, but not
  totally eliminate the ``dilution'' of the distinct blue RGB spatial
  distribution due to the the overlap of the RHB with the blue RGB in the
  CMD (e.g., Figure \ref{fig:cmd2}). }
\label{fig:splitcmd}
\end{center}
\end{figure}

\begin{figure}
\begin{center}
\plotone{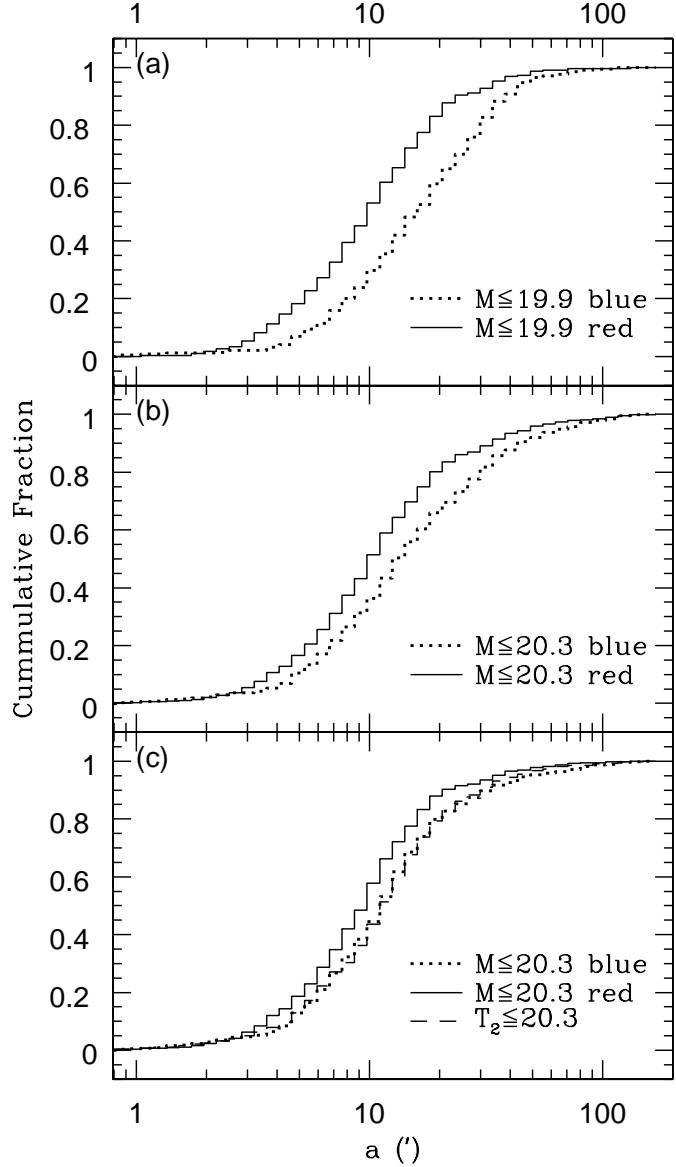} 
\caption{Cumulative distributions of the red ({\it solid lines}) and blue
  ({\it dotted lines}) populations (see Figure \ref{fig:splitcmd}) for (a)
  the $M\leq19.0$ sample, (b) the $M\leq20.3$ sample, and (c) the
  $M\leq20.3$ sample only over the area corresponding to the $T\leq20.3$
  sample ({\it dashed line}).  In all three cases the red population is
  much more concentrated than the blue population.  In (c), the BHB
  population is shown to closely follow the blue $M\leq20.3$ sample.  A
  similar test could be done for the RHB, which should follow the red
  population; however, we are unable select a reasonably pure sample of RHB
  stars from our CMD with a large enough sample size. }
\label{fig:cumdist}
\end{center}
\end{figure}

\begin{figure}
\begin{center}
\plotone{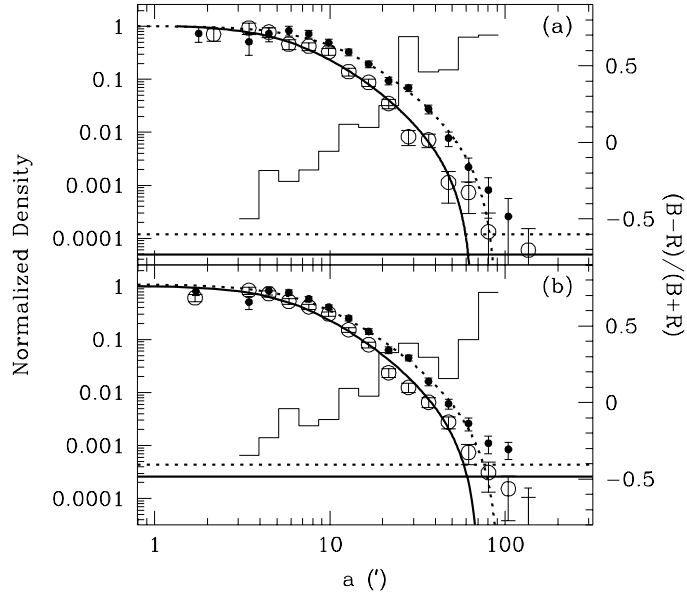} 
\caption{Radial surface density distributions for the blue ({\it filled
    circles}) and red ({\it open circles}) RGB populations in the (a)
  $M\leq19.0$ and (b) $M\leq20.3$ samples.  The best-fitted King profiles
  are overplotted for the blue ({\it dotted line}) and red ({\it solid
    line}) populations along with the background determined by the ``CMD
  offset'' method.  As illustrated in Figure \ref{fig:cumdist}, the blue
  population is much more extended than its red counterpart.  This gradient
  is further demonstrated by the overlayed color index $(B-R)/(B+R)$ (shown
  by the {\it histogram}), where $B$ and $R$ are, respectively, the number
  of ``blue'' and ``red'' stars within a given annulus. }
\label{fig:splitrp}
\end{center}
\end{figure}

\begin{figure}
\begin{center}
\plotone{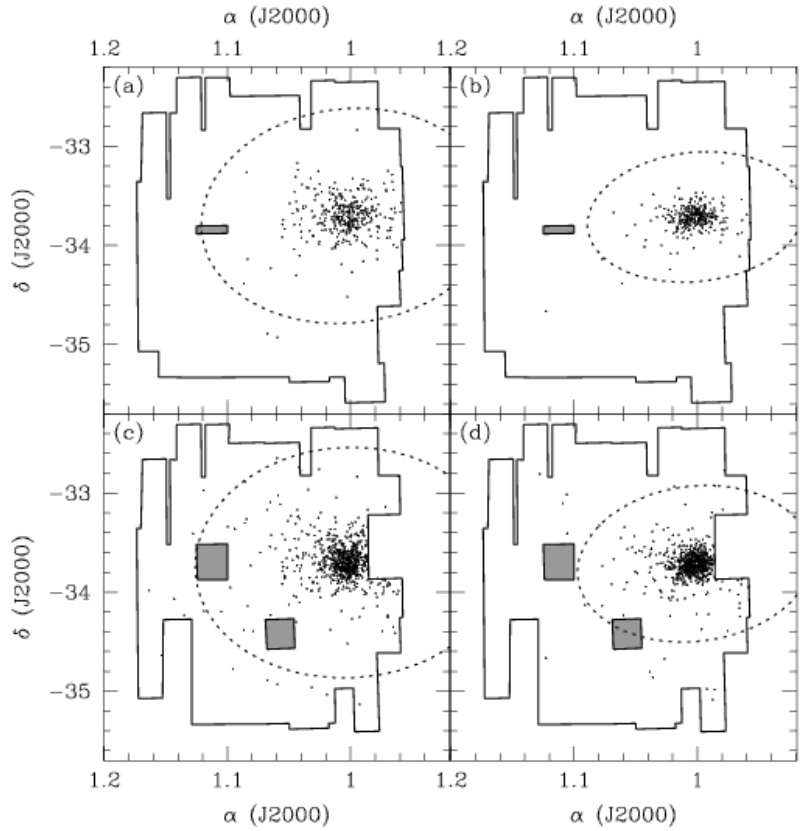} 
\caption{The spatial distribution of the ``blue'' ({\it left}) and ``red''
  ({\it right}) RGB populations for the $M\leq19.0$ ({\it top}) and
  $M\leq20.3$ ({\it bottom}) samples.  The derived King limiting radii
  ({\it dotted line}) are also shown.  The red population is much more
  concentrated than the blue population; but, interestingly, it is also
  {\it more} elliptical. }
\label{fig:sclsplitdist}
\end{center}
\end{figure}

Yet another explanation for the break population is that it is a
dynamically hotter, but {\it bound} component to the Sculptor system.  If
Sculptor has a significant, extended dark matter component, as has recently
been reported for the Draco dSph \citep{kleyna02}, a bound stellar
component extending to twice the King limiting radius might be
accommodated.  However, Sculptor is at the low end of the dSph $M/L$ range,
with an inferred central $M/L\sim10\,M_{\sun}/L_{\sun}$,\footnote{Although
  we have redefined the structural parameters of the Sculptor system here,
  our core radius ($7\farcm14\pm0\farcm33$) is only slightly larger than
  that found previously by \citeauthor*{IH95} ($5\farcm8\pm1\farcm6$) and
  values of our velocity dispersion at small radii are insignificantly
  different from that of \citeauthor*{QDP} ($6.2\pm1.1\,{\rm km\,s^{-1}}$;
  see Figure \ref{fig:rvprof}).  Since $M/L\propto \sigma_v^2/r_c$, a
  recalculation of $M/L$ using our new values would yield a 15--20\%
  smaller value than previously determined ($M/L=10.9$;
  \citeauthor*{IH95}); an insignificant difference considering the large
  errors associated with the calculation.} compared to the central Draco
$M/L$ of $\sim100\, M_{\sun}/L_{\sun}$ \citep[or even larger ---
$440\,M_{\sun}/L_{\sun}$ --- as recently reported by ][]{kleyna02}, so it
seems less likely that the break population of Sculptor stars is sheltered
within a massive dark matter halo of the inferred size of Draco's.  On the
other hand, the extraordinary inflation of the Draco $M/L$ by
\citet{kleyna03} was a direct result of an observed flat velocity
dispersion, as we have seen here\footnote{We note \citet{wilk04} have
  recently suggested a fall in the outer velocity dispersion for both Draco
  and Ursa Minor.  However, our own work on Ursa Minor (Paper IX) is
  inconsistent with the presence of a such a ``cold points" at large radii,
  and \citet{lokas} have also raised concerns about the
  \citet{wilk04} cold population in the outer parts of Draco.}.  Given this
tendency for some modelers to accommodate the flat dispersions profiles by
adding mass to the dSph at large radii, it is likely that a convincing
dismissal of the ``large dark matter halo'' interpretation will require
tracking the break population, both photometrically and spectroscopically,
to even larger radii, to the point where obvious tidal tails should be
discernible.  However, because N-body models of disrupting satellites have
been found that can explain flat velocity dispersion
profiles {\it and} K+B radial density profiles (\citeauthor*{Mu05b}), we believe that
the weight of evidence for tidal disruption models is now greater, and so
explore this interpretation further here.

Assuming the break population is from tidal disruption, we can calculate
the fractional mass-loss rate of Sculptor based on our newly mapped radial
profile.  For satellites on circular orbits, the mass loss is due to
constant stripping of the satellite over time.  For more eccentric orbits,
the mass loss comes in more violent pulses when the satellite passes
perigalacticon.  \citet{jsh99} give two prescriptions for the derivation
of the mass-loss rate, which have been modified by the more recent analysis
of \citet*{jcg02}.  Using the corrected formula for the mass loss from the
above simulations, Table \ref{tab:dfdt} gives the resulting fractional
mass-loss rates for the various samples in this paper.  The last
``reliable'' point that defines $r_{\rm break}$ is taken as the last point
that is consistent with the $\Sigma\propto r^{-2}$ fall-off.  The mean of
the values in Table \ref{tab:dfdt} are $(df/dt)_1=0.042\,{\rm Gyr^{-1}}$
and $(df/dt)_2=0.083\,{\rm Gyr^{-1}}$.  This is a relatively mild
fractional mass-loss rate yet places it squarely on the trend of increasing
fractional mass-loss rate with inferred $(M/L)_{V,{\rm tot}}$ noted by
\citet[][ see also \citealt{m02,Ma05b}]{m03}.  \citet{Ma05b} discuss as one
explanation for this trend an increase in tidal disruption in systems that
have a larger {\it inferred $M/L$} (see also Majewski 2002).
That Sgr --- which
is clearly losing mass into tidal arms to create {\it its} break population
--- participates in this trend lends support for this notion
\citep{m03,Ma05b}.

If Sculptor is losing stars due to tidal stripping, our analysis of the
different stellar populations (\S5.4) suggests that Sculptor would be
contributing stars to the MW halo predominantly from its more metal-poor
population.  Extrapolated back in time, this differential mass loss would
suggest that Sculptor once had a larger metal-poor population.  Whereas the
two Sculptor populations --- as we have divided them --- are roughly equal
in proportion, if we extrapolate the Sculptor mass loss backwards for
$12\,{\rm Gyr}$ we find that the original metal poor population would have
been about 50\% larger.  A similar argument has been used to estimate the
original balance of populations in the Carina dSph (\citeauthor*{pII};
\citealt{m02}).

\acknowledgments
KBW would like to thank the Department of Astronomy at the University of
Virginia for its active support of undergraduate research.  Preliminary
portions of this work were used for completion of a senior thesis at the
University of Virginia.  Both SRM and KBW thank his Ph.D. thesis adviser, M. A.
Bershady, for patience while a large portion of this undergraduate project
was completed by KBW at the University of Wisconsin -- Madison.  We
acknowledge support from NSF grants AST-9702521, AST-0307851 and AST-0307417,
a Cottrell Scholar Award from The Research Corporation, NASA/JPL contract
1228235, the David and Lucile Packard Foundation, and the generous support
of The F. H.  Levinson Fund of the Peninsula Community Foundation.  PMF holds
a Levinson Graduate Fellowship awarded through the Peninsula Community
Foundation and is also supported by the Virginia Space Grant Consortium.
SRM is extremely grateful to the Carnegie Observatories for a Visiting
Associateship, which allowed access to the Swope telescope to undertake this
survey of the Sculptor dSph.  Finally, we appreciate useful conversations
with Ricardo Mu\~noz and Kathryn Johnston, and helpful comments from the
referee that helped improve the discussion.

\newpage
\begin{deluxetable}{ r c c c c }
\tabletypesize{\footnotesize}
\tablewidth{0pt}
\tablecaption{Imaging Observations}
\tablehead{ \colhead{UT Date} & \colhead{Seeing($\arcsec$)\tablenotemark{a}} & \colhead{Moon Illum.} & \colhead{$N_{\rm fields}$} & \colhead{$N_{\rm phot}$} }
\startdata
\cutinhead{SITe \#1 Chip}
26 Aug -- 02 Sep 1999 & 1.3 & 1.0 -- 0.5 & 16 &  \phn7 \\
26 Oct -- 30 Oct 1999 & 1.5 & 0.9 -- 0.5 & 11 &  \phn6 \\
19 Dec -- 24 Dec 1999 & 1.5 & 0.9 -- 0.9 & 14 &  \phn4 \\
30 Oct -- 08 Nov 2000 & 1.8 & 0.2 -- 0.9 & 18 &  \phn0 \\
\cutinhead{SITe \#3 Chip}
23 Oct -- 24 Oct 2001 & 1.6 & 0.5 -- 0.6 &  \phn4 &  \phn4 \\
16 Nov -- 20 Nov 2001 & 1.2 & 0.0 -- 0.3 & 33 & 33 \\
\enddata
\label{tab:obs}
\tablenotetext{a}{All seeing values are averages over all Sculptor
  observations for each run.}
\end{deluxetable}

\newpage
\begin{deluxetable}{ c r c c c r }
\tabletypesize{\footnotesize}
\tablewidth{0pt}
\tablecaption{Adopted Radial Velocities for Multiple Observations}
\tablehead{\colhead{ID} & \colhead{$v_{\rm hel}$\tablenotemark{a}} & \colhead{$CCP$} & \colhead{$Q$} & \colhead{$s(v_{\rm hel})$\tablenotemark{a}} & \colhead{UT Date} }
\startdata
 1000427 & {\bf 111.68} &  \nodata & \nodata &   \phn4.82  & \multicolumn{1}{c}{\nodata}  \\
 \nodata &  113.10 &  1.00    &       7 & \nodata & 13Nov2000  \\
 \nodata &  103.50 &  0.25    &       4 & \nodata & 07Oct2001  \\
 1000834 & {\bf 103.62} &  \nodata & \nodata &   \phn0.95  & \multicolumn{1}{c}{\nodata}  \\
 \nodata &  101.10 &  0.73    &       7 & \nodata & 13Nov2000  \\
 \nodata &  103.80 &  1.02    &       7 & \nodata & 07Oct2001  \\
 1000844 & {\bf 126.64} &  \nodata & \nodata &  14.04  & \multicolumn{1}{c}{\nodata}  \\
 \nodata &  110.10 &  1.32    &       7 & \nodata & 13Nov2000  \\
\enddata
\label{tab:rvmean}
\tablenotetext{a}{Units are ${\rm km\,s^{-1}}$.}
\end{deluxetable}

\newpage
\begin{deluxetable}{ r c c c c c }
\tabletypesize{\small}
\tablewidth{0pt}
\tablecaption{Comparison with Literature Radial Velocities}
\tablehead{ \colhead{ID} & \colhead{$v_{\rm hel}$\tablenotemark{a}} & \colhead{$\epsilon_{v_{\rm hel}}$\tablenotemark{a}$^,$\tablenotemark{b}} & \colhead{$CCP$} & \colhead{$Q$} & \colhead{Reference\tablenotemark{c}} }
\startdata
 1002429 & 106.33 &    3.71 & \nodata &       8 & 0 \\
    H185 & 125.50 &    4.70 & \nodata & \nodata & 1 \\
    H185 & 110.00 &    1.80 & \nodata & \nodata & 2 \\
 1003537 & 117.50 &    4.00 &    0.69 &       7 & 0 \\
    H241 & 116.10 &    4.70 & \nodata & \nodata & 1 \\
    H241 & 119.70 &    2.10 & \nodata & \nodata & 2 \\
 1014728 & \phn99.20 & 4.00 &    0.42 &       6 & 0 \\
    H394 & 109.70 &    3.30 & \nodata & \nodata & 2 \\
\enddata
\tablenotetext{a}{Units are ${\rm km\,s^{-1}}$.}
\tablenotetext{b}{For the current study, velocity errors provided
  here are either those adopted for the specific run or, when multiple
  observations were available, the error on the weighted mean velocity.}
\tablenotetext{c}{References: (0) This paper; (1) \citet{ad86};
  (2) \citeauthor*{QDP}. }
\label{tab:rvcomp}
\end{deluxetable}

\newpage
\begin{deluxetable}{ r r r r r r r r r r c r c}
\tabletypesize{\tiny}
\tablewidth{0pt}
\tablecaption{Heliocentric Radial Velocities of Selected Stars}
\tablehead{ \colhead{ID} & \colhead{$\alpha$(J2000.0)} & \colhead{$\delta$(J2000.0)} & \colhead{UT Date\tablenotemark{a}} & \colhead{$M$} & \colhead{$M-T_2$} & \colhead{$M-D$} & \colhead{$v_{\rm hel}$\tablenotemark{b}} & \colhead{$\epsilon_{v_{\rm hel}}$\tablenotemark{c}} & \colhead{$CCP$} & \colhead{$Q$} & \colhead{$a$ ($\arcmin$)} & \colhead{P/V\tablenotemark{d}}}
\startdata
 1000425 & 00 59 09.40 & -33 50 54.5 &13Nov2000 &   18.39$\pm$0.03 &    1.54$\pm$0.04 &    0.02$\pm$0.04 &   99.50 & \nodata &    0.93 &       7 &  17.0031 & Y/Y \\
 1000427 & 00 59 11.70 & -33 30 49.8 &  \nodata &   18.34$\pm$0.02 &    1.37$\pm$0.03 &    0.03$\pm$0.03 &  111.68 &    3.71 & \nodata &       8 &  19.7936 & Y/Y \\
 1000474 & 00 59 10.24 & -33 45 53.3 &07Oct2001 &   19.38$\pm$0.08 &    1.15$\pm$0.09 &    0.07$\pm$0.10 &  106.00 & \nodata &    1.00 &       7 &  12.6904 & Y/Y \\
 1000834 & 00 59 12.08 & -33 46 20.9 &  \nodata &   17.53$\pm$0.02 &    1.59$\pm$0.03 &    0.01$\pm$0.03 &  103.62 &    3.71 & \nodata &       8 &  12.6169 & Y/Y \\
 1000844 & 00 59 12.71 & -33 41 09.3 &  \nodata &   17.96$\pm$0.02 &    1.46$\pm$0.03 &   -0.00$\pm$0.03 &  126.64 &    3.71 & \nodata &       8 &  11.2082 & Y/Y \\
 1001140 & 00 59 14.55 & -33 40 39.8 &13Nov2000 &   17.70$\pm$0.02 &    1.72$\pm$0.03 &   -0.03$\pm$0.03 &  100.10 & \nodata &    0.93 &       7 &  10.9588 & Y/Y \\
 1001378 & 00 59 15.76 & -33 42 48.5 &13Nov2000 &   17.64$\pm$0.02 &    1.72$\pm$0.03 &   -0.01$\pm$0.03 &  109.70 & \nodata &    0.86 &       7 &  10.4815 & Y/Y \\
 1001547 & 00 59 16.93 & -33 40 10.5 &13Nov2000 &   17.30$\pm$0.02 &    1.83$\pm$0.02 &   -0.05$\pm$0.03 &  100.60 & \nodata &    1.11 &       7 &  10.6561 & Y/Y \\
\enddata
\label{tab:rvfull}

\tablenotetext{a}{Dates for stars multiply observed by us are given in
  Table \ref{tab:rvmean}.}

\tablenotetext{b}{Units are ${\rm km\,s^{-1}}$.}

\tablenotetext{c}{Errors from literature stars;
  1/$\sqrt{ \sum(1/\epsilon_{v_i}^2)}$ for multiply observed stars.  Both
  are given in ${\rm km\,s^{-1}}$.  Errors for stars observed by us on Nov
  2000, Oct 2001, Jan 2004, and Dec 2004 are adopted as 10, 4, 1.5, and 2.5
  ${\rm km\,s^{-1}}$, respectively. }

\tablenotetext{d}{Satisfaction of our photometric criteria (\S3) is denoted
  by a ``Y'' or ``N'' for P.  Satisfaction of our velocity criteria (\S4)
  is denoted similarly for V.}

\tablenotetext{e}{Star selected as giant candidate but fell redward of
  our RGB/RHB selection at the bright end.}

\tablenotetext{f}{Previously observed star that is consistent with our
  photometric candidate constraints but does not satisfy our $\chi$
  morphological constraint.}

\tablenotetext{g}{Photometrically selected BHB candidate.}

\tablenotetext{h}{Star fell redward of our BHB selection.}

\tablenotetext{i}{Star fell blueward of our 2CD selection limit.}

\tablenotetext{j}{Mean $v_{\rm hel}$ inconsistent with Sculptor but at
  least one individual observation was consistent.}

\end{deluxetable}

\newpage
\begin{deluxetable}{ l c c c}
\tabletypesize{\small}
\tablewidth{0pt}
\tablecaption{Summary of Photometric Selection Accuracy}
\tablehead{\colhead{} & \colhead{V: Yes} & \colhead{V: No} & \colhead{\% Correct} }
\startdata
P: Yes (RGB) &      137  &  8  &\phn94 \\
P: Yes (BHB) &\phn\phn2  &  0  &   100 \\
P: No        &\phn\phn4  &  6  &\phn60 \\
P: Yes (Lit) &\phn\phn9  &  1  &\phn90 \\
P: No  (Lit) &\phn\phn5  &  7  &\phn58 \\
\hline
P: Yes (Tot) &      148  &  8  &\phn94 \\
\enddata
\label{tab:pvcomp}
\end{deluxetable}

\newpage
\begin{deluxetable}{l c c c c c c c}
\tabletypesize{\small}
\tablewidth{0pt}
\tablecaption{ Mean Magnitude and Color Errors\tablenotemark{a} }
\tablehead{ \colhead{Sample\tablenotemark{b}} & \colhead{$N_{\rm fields}$} & \colhead{$N_{\rm stars}$} & \colhead{$\langle \epsilon_M \rangle$} &
\colhead{$\langle \epsilon_{(M-T_2)} \rangle$} & \colhead{$\langle \epsilon_{(M-DDO51)} \rangle$} &
\colhead{$\langle \epsilon_V \rangle$} & \colhead{$\langle \epsilon_{(B-V)} \rangle$} }
\startdata
Error Limit         & 96      & 11794 & 0.041$\pm$0.023 & 0.057$\pm$0.031 & 0.061$\pm$0.034     &         \nodata &         \nodata  \\
Giants              & 96      & \phn2018  & 0.051$\pm$0.023 & 0.067$\pm$0.029 & 0.073$\pm$0.033     &         \nodata &         \nodata  \\
RHB\tablenotemark{c}             & \nodata & \phn\phn386   & 0.104$\pm$0.023 & 0.129$\pm$0.024 & 0.156$\pm$0.030\tablenotemark{d} & 0.049$\pm$0.018 & 0.070$\pm$0.026  \\
Lower RGB           & \nodata & \phn\phn677   & 0.138$\pm$0.046 & 0.166$\pm$0.054 & 0.219$\pm$0.080\tablenotemark{d} & 0.064$\pm$0.028 & 0.100$\pm$0.044  \\
BHB                 & 96      & \phn1096  & 0.100$\pm$0.036 & 0.155$\pm$0.048 & 0.148$\pm$0.052\tablenotemark{d} &         \nodata &         \nodata  \\
$M\leq19.0$         & 95      & \phn\phn634   & 0.029$\pm$0.009 & 0.040$\pm$0.011 & 0.041$\pm$0.014     &         \nodata &         \nodata  \\
$M\leq20.3$         & 80      & \phn1410  & 0.047$\pm$0.023 & 0.060$\pm$0.025 & 0.066$\pm$0.031     &         \nodata &         \nodata  \\
$M\leq21.0$         & 53      & \phn\phn134   & 0.047$\pm$0.017 & 0.077$\pm$0.026 & 0.070$\pm$0.024     &         \nodata &         \nodata  \\
\enddata
\label{tab:meanerr}
\tablenotetext{a}{$\pm$ values are standard deviations.}

\tablenotetext{b}{Error Limit -- Stars satisfying error and
  morphological limits (\S2); Giants -- Stars satisfying the 2CD
  selection; RHB, Lower RGB -- Stars selected to be RHB or Lower RGB
  stars according to Equations \ref{eq:rhb} and \ref{eq:lrgb}; BHB --
  Full BHB sample selected in Figure \ref{fig:hb}.}

\tablenotetext{c}{Eight stars in this \citet{m99} sample did not have measured $M$, $T_2$, and $DDO51$ magnitudes.}

\tablenotetext{d}{These values are immaterial because the $(M-DDO51)$
  color was not used in the selection of this sample.}

\end{deluxetable}

\newpage
\begin{deluxetable}{ c c c c c c c }
\tabletypesize{\small}
\tablewidth{0pt}
\tablecaption{Counts In Offset CMD Selection Boxes}
\tablehead{ & \multicolumn{3}{c}{$M$ Limit} && \multicolumn{2}{c}{$T_2$ Limit} \\
\cline {2-4} \cline{6-7} 
\colhead{$\Delta M$} & \colhead{19.0} & \colhead{20.3} & \colhead{21.0} && \colhead{19.9} & \colhead{20.3} }
\startdata
\phm{$-$}0.00  &   \phn634   & 1410    &  134    &&  442 &  853 \\
$-0.25$  &   485   & 1207    &  111    &&  151 &  613 \\
$-0.50$  &   288   &  \phn945    &\phn   70    &&\phn   50 &  215 \\
$-0.75$  &   120   &  \phn698    &\phn   56    && \phn  18 &\phn   69 \\
$-1.00$  & \phn34   &  \phn445    &\phn   42    && \phn  13 & \phn  27 \\
\cline{2-2}
$-1.25$  &\phn\phn     6   &\phn  219    &\phn   32    &&\phn   11 &\phn   19 \\
$-1.50$  &\phn\phn     6   &\phn  102    &\phn   23    &&\phn   10 &\phn   14 \\
\cline{6-6}
$-1.75$  &\phn\phn     7   &\phn\phn   48    &\phn   20    &&\phn\phn    5 &\phn   10 \\
\enddata
\tablenotetext{a}{Units are square degrees.}
\tablenotetext{b}{Units are 10$^{-3}$ arcmin$^{-2}$.}
\label{tab:cmdbg}
\end{deluxetable}

\newpage
\begin{deluxetable}{ l c c c c c }
\tabletypesize{\footnotesize}
\tablewidth{0pt}
\tablecaption{Control Field Background Number Densities\tablenotemark{a}}
\tablehead{ \colhead{Sample\tablenotemark{b}} & \colhead{$\sum_{\mbox{North}}$} & \colhead{$\sum_{\mbox{South}}$} & \colhead{$\sum_{\mbox{East}}$} & \colhead{$\sum_{\mbox{West}}$} & \colhead{$\sum_{\mbox{CF}}$} }
\startdata
Giants        & $5.7^{+9.3}_{-5.7}\ ^{+4.5}_{-4.5}$ & $5.7^{+9.3}_{-5.7}\ ^{+4.5}_{-4.5}$ & $5.7^{+9.3}_{-5.7}\ ^{+4.5}_{-4.5}$ & $2.8^{+8.2}_{-2.8}\ ^{+3.7}_{-2.8}$ & $5.0^{+4.5}_{-2.6}\ ^{+2.2}_{-2.1}$ \\
$M\leq19.0$   & $0.0^{+6.5}_{-0.0}\ ^{+3.1}_{-0.0}$     & $0.0^{+6.5}_{-0.0}\ ^{+3.1}_{-0.0}$     & $0.0^{+6.5}_{-0.0}\ ^{+3.1}_{-0.0}$     & $0.0^{+6.5}_{-0.0}\ ^{+3.1}_{-0.0}$     & $0.0^{+3.3}_{-0.0}\ ^{+1.6}_{-0.0}$           \\
$M\leq20.3$   & $0.0^{+6.5}_{-0.0}\ ^{+3.1}_{-0.0}$     & $2.8^{+8.2}_{-2.8}\ ^{+3.7}_{-2.8}$ & $0.0^{+6.5}_{-0.0}\ ^{+3.1}_{-0.0}$     & $0.0^{+6.5}_{-0.0}\ ^{+3.1}_{-0.0}$     & $0.7^{+3.5}_{-0.7}\ ^{+1.6}_{-0.7}$ \\
$M\leq21.0$   & $2.8^{+8.2}_{-2.8}\ ^{+3.7}_{-2.8}$ & $2.8^{+8.2}_{-2.8}\ ^{+3.7}_{-2.8}$ & $0.0^{+6.5}_{-0.0}\ ^{+3.1}_{-0.0}$     & \nodata      & $1.9^{+4.4}_{-1.3}\ ^{+2.0}_{-1.3}$ \\
$T_2\leq19.9$ & \nodata             & \nodata             & $0.0^{+6.5}_{-0.0}\ ^{+3.1}_{-0.0}$     & \nodata             & $0.0^{+6.5}_{-0.0}\ ^{+3.1}_{-0.0}$           \\
$T_2\leq20.3$ & \nodata             & \nodata             & $0.0^{+6.5}_{-0.0}\ ^{+3.1}_{-0.0}$     & \nodata             & $0.0^{+6.5}_{-0.0}\ ^{+3.1}_{-0.0}$           \\
\enddata
\label{tab:fieldbg}
\tablenotetext{a}{Units are 10$^{-3}$ arcmin$^{-2}$. First and second error
  pairs give, respectively, the 90\% and 68\% Poissonian confidence
  limits.}
\tablenotetext{b}{See the text for definitions of the samples.}
\end{deluxetable}

\newpage
\begin{deluxetable}{ l l l l l l l l }
\tabletypesize{\scriptsize}
\tablewidth{0pt}
\tablecaption{King Radial Profile Fits}
\tablehead{ & \multicolumn{2}{c}{Center Coords (J2000.0)} & & & & & \\
\cline{2-3} \colhead{Sample\tablenotemark{a}} & \colhead{$\alpha$} & \colhead{$\delta$} & \colhead{$PA$($\arcdeg$)} &
\colhead{$\varepsilon$} & \colhead{$r_{c}$($\arcmin$)} & \colhead{$r_{\rm lim}$($\arcmin$)\tablenotemark{b}} & \colhead{$\Sigma_{b}$\tablenotemark{c}} }
\startdata
$M\leq19.0$ &  $1.001727\pm0.000006$ & $-33.71026\pm0.00081$ & $95.5\pm2.7$ & $\phm{>}0.33\pm0.03$ & $6.97\pm0.69$ & $\phm{>}85.14\pm\phn7.40$ & $0.26\pm0.21$ \\
$M\leq20.3$ &  $1.002043\pm0.000039$ & $-33.70979\pm0.00146$ & $92.4\pm1.8$ & $\phm{>}0.28\pm0.03$ & $6.50\pm0.46$ & $\phm{>}80.76\pm\phn5.98$ & $2.95\pm0.54$ \\
$T_2\leq19.9$ &  $0.999622\pm0.000052$ & $-33.71548\pm0.00214$ & $96.8\pm4.1$ & $\phm{>}0.31\pm0.05$ & $8.47\pm1.06$ & $\phm{>}81.53\pm10.19$    & $1.22\pm0.38$ \\
$T_2\leq20.3$ &  $1.000712\pm0.000028$ & $-33.71270\pm0.00308$ & $94.3\pm3.1$ & $\phm{>}0.26\pm0.04$ & $8.53\pm0.78$ & $\phm{>}75.06\pm\phn5.46$ & $1.56\pm0.54$ \\
Mean\tablenotemark{d} &  $1.001663\pm0.000006$ & $-33.71077\pm0.00066$ & $93.8\pm1.3$ & $\phm{>}0.29\pm0.02$ & $7.14\pm0.33$ & $\phm{>}79.59\pm\phn3.34$ & \multicolumn{1}{c}{\nodata}  \\
\tableline
1           & 1.002486 & $-33.69407$ & \multicolumn{1}{c}{\nodata}      & $>0\phm{.00\pm0.03}$          & \multicolumn{1}{c}{\nodata}     & (80 -- 120)    & \multicolumn{1}{c}{\nodata} \\
2           & \phn\phn\nodata  & \phn\phn\nodata   & $98\phd\phn\pm2$     & $\phm{>}0.35\pm0.05$ & \multicolumn{1}{c}{\nodata}     & $\phm{>}46\phd\phn\phn\pm\phn3\phd\phn\phn$ & \multicolumn{1}{c}{\nodata} \\
3           & 1.019497 & $-33.56441$ & \multicolumn{1}{c}{\nodata}      & $<0.3\phm{0\pm0.03}$         & 7.5         & $>75\phm{.00\pm10.03}$        & \multicolumn{1}{c}{\nodata} \\
\enddata
\label{tab:rpkingfits}
\tablenotetext{a}{References: (1) \citet{s38a,s38b}; (2) \citeauthor*{H61}; (3) \citet{DKK}; (4) \citet{e88b}; (5) \citeauthor*{IH95}; (6) \citeauthor*{W03}.}
\tablenotetext{b}{This parameter is described as the tidal radius, $r_t$,
  in most treatments, including the original discussion by \citet{king62}.
  We have adopted the name limiting radius, $r_{\rm lim}$, here to avoid
  confusion between this {\it fitting parameter} and the true tidal radius
  of the system.}
\tablenotetext{c}{Units are 10$^{-3}$ arcmin$^{-2}$.}
\tablenotetext{d}{Error weighted mean of the four samples.  Errors will be
  slightly under-estimated since the samples are not completely
  independent.}
\end{deluxetable}

\newpage
\begin{deluxetable}{ l c c c c c c c }
\tabletypesize{\scriptsize}
\tablewidth{0pt}
\tablecaption{PLC Radial Profile Fits}
\tablehead{ & \multicolumn{2}{c}{Center Coords (J2000.0)} & & & & &  \\
\cline{2-3} \colhead{Sample} & \colhead{$\alpha$} & \colhead{$\delta$} & \colhead{$PA$($\arcdeg$)} &
\colhead{$\varepsilon$} & \colhead{$r_{c}$($\arcmin$)} & \colhead{$\nu$} & \colhead{$\Sigma_{b}$\tablenotemark{a}} }
\startdata
$M\leq19.0$ &  $1.001751\pm0.000008$ & $-33.70764\pm0.00047$ & $89.81\pm0.12$ & $0.33\pm0.04$ & $10.73\pm1.03$ &  $2.03\pm0.13$  & $0.00\pm0.04$\tablenotemark{b} \\
$M\leq20.3$ &  $1.002271\pm0.000005$ & $-33.71114\pm0.00050$ & $89.43\pm0.38$ & $0.27\pm0.03$ & $ 8.60\pm0.69$ &  $1.74\pm0.10$  & $0.42\pm1.71$ \\
$T_2\leq19.9$ &  $0.991578\pm0.000015$ & $-33.71696\pm0.00205$ & $89.90\pm0.08$ & $0.49\pm0.03$ & $19.55\pm3.39$ &  $2.82\pm0.55$  & $0.80\pm0.59$ \\
$T_2\leq20.3$ &  $1.001197\pm0.000004$ & $-33.71297\pm0.00069$ & $89.90\pm0.03$ & $0.23\pm0.04$ & $11.77\pm1.41$ &  $1.99\pm0.19$  & $0.00\pm0.34$\tablenotemark{c} \\
Mean\tablenotemark{d} &  $1.001299\pm0.000003$ & $-33.71016\pm0.00030$ & $89.89\pm0.03$ & $0.34\pm0.02$ & $11.73\pm0.50$ &  $1.88\pm0.07$ & \nodata \\
\enddata
\label{tab:rpplcfits}
\tablenotetext{a}{Units are 10$^{-3}$ arcmin$^{-2}$.}
\tablenotetext{b}{Value: $\Sigma_b = 2.19\times10^{-8}$ arcmin$^{-2}$.}
\tablenotetext{c}{Value: $\Sigma_b = 2.70\times10^{-6}$ arcmin$^{-2}$.}

\tablenotetext{d}{Error weighted mean of the four samples.  Errors will
  be slightly under-estimated since the samples are not completely
  independent.}

\end{deluxetable}

\newpage
\begin{deluxetable}{ l c c c c c c c c c}
\tabletypesize{\footnotesize}
\tablewidth{0pt}
\tablecaption{Goodness-of-fit Estimates ($\chi^2$)}
\tablehead{ & \multicolumn{2}{c}{Model $\Sigma_b$} & & \multicolumn{6}{c}{Measured $\Sigma_b$} \\ \cline{2-3} \cline{5-10} \colhead{Sample} & \colhead{King} & \colhead{PLC} & & \colhead{King} & \colhead{PLC} & \colhead{K+B\tablenotemark{a}} & \colhead{$r_{\rm break}$\tablenotemark{a,b}} & \colhead{K+B\tablenotemark{c}} & \colhead{$r_{\rm break}$\tablenotemark{c,b}} }
\startdata
$M\leq19.0$   & 4.38 & 4.05 && 4.42 & 4.37 & 3.98 & 61 & 4.26 & 79 \\
$M\leq20.3$   & 8.77 & 5.67 && 9.58 & 5.54 & 7.81 & 61 & 7.74 & 61 \\
$T_2\leq19.9$ & 3.46 & 5.03 && 5.05 & 5.97 & 2.70 & 57 & 2.71 & 59 \\
$T_2\leq20.3$ & 5.06 & 4.06 && 5.64 & 3.96 & 4.36 & 55 & 4.46 & 60 \\
\enddata
\label{tab:fitchi}
\tablenotetext{a}{``King+Break'' profile using the least-squares fit power law to the density points beyond $r_{\rm break}$ as given in Figure \ref{fig:allsamprp}.  The least-squares fit for the $M\leq20.3$ did not include the final point (because it was below the background) so the $\Sigma \propto r^{-1.9}$ fit is actually worse than the $\Sigma \propto r^{-2}$ fit.} 
\tablenotetext{b}{Semi-major axis distance given in arcmin.}
\tablenotetext{c}{``King+Break'' profile using a $\Sigma \propto r^{-2}$ power law for the break population.}
\end{deluxetable}

\newpage
\begin{deluxetable}{ l c c l }
\tabletypesize{\footnotesize}
\tablewidth{0pt}
\tablecaption{Kolmogorov-Smirnov Statistics for Split RGB Populations}
\tablehead{ \colhead{Comparison} & \colhead{$N_e$} & \colhead{$D$} & \colhead{KS\%} }
\startdata
$M\leq19.0$ blue to red & 157.6656 & 0.2491 & \phn4.2483$\times10^{-7}$ \\
$M\leq20.3$ blue to red & 364.2317 & 0.1562 & \phn3.0136$\times10^{-6}$ \\
$T_2\leq20.3$ to $M\leq20.3$ blue & 339.8766 & 0.0233 & 99.196                \\
$T_2\leq20.3$ to $M\leq20.3$ red  & 360.7070 & 0.1522 & \phn8.9081$\times10^{-6}$ \\
\enddata
\label{tab:ks}
\end{deluxetable}

\newpage
\begin{deluxetable}{ l c c c c c c c }
\tabletypesize{\scriptsize}
\tablewidth{0pt}
\tablecaption{King Radial Profile Fits for Split RGB Populations}
\tablehead{ & \multicolumn{2}{c}{Center Coords (J2000.0)} & & & & & \\
\cline{2-3} \colhead{Sample} & \colhead{$\alpha$} & \colhead{$\delta$} & \colhead{$PA$($\arcdeg$)} &
\colhead{$\varepsilon$} & \colhead{$r_{c}$($\arcmin$)} & \colhead{$r_{\rm lim}$($\arcmin$)} & \colhead{$\Sigma_{\rm CMD}$\tablenotemark{a}} }
\startdata
Blue $M\leq19.0$ & $1.002102\pm0.000011$ & $-33.70122\pm0.00131$ & $96.3\pm4.7$ & $0.27\pm0.05$ & $9.57\pm1.31$ & $89.24\pm10.78$    & $0.12\pm0.07$ \\
Red $M\leq19.0$  & $1.001609\pm0.000007$ & $-33.71245\pm0.00091$ & $96.0\pm3.1$ & $0.41\pm0.04$ & $6.08\pm0.86$ & $65.74\pm\phn7.00$ & $0.05\pm0.04$ \\
Blue $M\leq20.3$ & $1.002319\pm0.000048$ & $-33.70173\pm0.00256$ & $92.6\pm2.1$ & $0.25\pm0.04$ & $7.47\pm0.78$ & $92.63\pm\phn9.96$ & $0.44\pm0.13$ \\
Red $M\leq20.3$  & $1.002213\pm0.000032$ & $-33.71280\pm0.00219$ & $96.9\pm3.0$ & $0.34\pm0.03$ & $5.95\pm0.59$ & $71.28\pm\phn5.48$ & $0.26\pm0.10$ \\
\enddata
\label{tab:rpsplitfits}
\tablenotetext{a}{Units are 10$^{-3}$ arcmin$^{-2}$ as derived from the
  ``CMD offset'' method.}
\end{deluxetable}

\newpage
\begin{deluxetable}{ l c c c c c c }
\tabletypesize{\footnotesize}
\tablewidth{0pt}
\tablecaption{Fractional Mass Loss Estimates\tablenotemark{a}}
\tablehead{ \colhead{Sample} & \colhead{$n_{\rm break}$\tablenotemark{b}} & \colhead{$\Sigma_{\rm break}$\tablenotemark{c}} & \colhead{$r_{xt}$($\arcmin$)} & \colhead{$n_{xt}$\tablenotemark{b}} & \colhead{$\left( \frac{df}{dt}\right)_1$\tablenotemark{d}} & \colhead{$\left( \frac{df}{dt}\right)_2$\tablenotemark{d}} }
\startdata
$M\leq19.0$   &\phn623.7 & 1.97 &\phn80.4 &\phn5.3 & 3.97 &\phn4.41 \\
$M\leq20.3$   &   1350.6 & 7.63 &   104.5 &   37.5 & 5.62 &\phn7.89 \\
$T_2\leq19.9$ &\phn422.4 & 3.61 &   135.8 &   17.6 & 4.84 &   11.95 \\
$T_2\leq20.3$ &\phn839.1 & 5.27 &   104.5 &   10.1 & 2.44 &\phn8.77 \\
Mean          & \nodata & \nodata & \nodata & \nodata & 4.22 &\phn8.26 \\
\enddata
\tablenotetext{a}{Using a value of $r_{\rm break}=61\farcm8$ for all
  samples.}
\tablenotetext{b}{Non-integer values due to background subtraction.}
\tablenotetext{c}{Units are 10$^{-3}$ arcmin$^{-2}$.}
\tablenotetext{d}{Units are 10$^{-2}\,{\rm Gyr^{-1}}$.}
\label{tab:dfdt}
\end{deluxetable}


\clearpage

\clearpage

\clearpage

\clearpage

\clearpage

\clearpage

\clearpage

\clearpage

\clearpage
\epsscale{0.7}

\clearpage
\epsscale{1.0}

\clearpage

\clearpage

\clearpage

\clearpage
\epsscale{0.8}

\clearpage
\epsscale{1.0}

\clearpage

\clearpage

\clearpage
\epsscale{0.9}

\clearpage
\epsscale{0.8}

\clearpage
\epsscale{1.0}

\clearpage

\clearpage

\clearpage

\end{document}